\documentclass[useAMS, usenatbib, usegraphicx]{mn2e}
\usepackage{amssymb, amsmath, longtable}  
\usepackage{longtable}
\usepackage{color}

\def \aap{A\&A}
\def \aaps{A\&AS}

\def \aj{AJ}

\def \apj{ApJ}
\def \apjl{ApJ}
\def \apjs{ApJS}
\def \araa{ARA\&A}

\def \mnras{MNRAS}

\newcommand{\AlII}{\hbox{{\rm Al}{\sc \,ii}}}

\newcommand{\CII}{\hbox{{\rm C}{\sc \,ii}}}
\newcommand{\CIII}{\hbox{{\rm C}{\sc \,iii}}}
\newcommand{\CIV}{\hbox{{\rm C}{\sc \,iv}}}

\newcommand{\FeII}{\hbox{{\rm Fe}{\sc \,ii}}}
\newcommand{\HI}{\hbox{{\rm H}{\sc \,i}}}

\newcommand{\Ha}{\hbox{{\rm H}$\alpha$}}

\newcommand{\OI}{\hbox{{\rm O}{\sc \,i}}}

\newcommand{\OVI}{\hbox{{\rm O}{\sc \,vi}}}
\newcommand{\SiII}{\hbox{{\rm Si}{\sc \,ii}}}
\newcommand{\SiIII}{\hbox{{\rm Si}{\sc \,iii}}}
\newcommand{\SiIV}{\hbox{{\rm Si}{\sc \,iv}}}

\newcommand{\lya}{\mbox{Ly$\alpha$}}
\newcommand{\lyb}{\mbox{Ly$\beta$}}

\newcommand{\kms}{\mbox{km~s$^{-1}$}}
\newcommand{\cmm}{\mbox{cm$^{-2}$}}

\newcommand{\kmsmpc}{\mbox{km~s$^{-1}$~Mpc$^{-1}$}}

\newcommand{\Deg}{\mbox{$^{\circ}$}}

\newcommand{\NHI}{\hbox{$N_{\rm \HI}$}}
\newcommand{\NCIV}{\hbox{$N_{\rm \CIV}$}}

\newcommand{\hMpc}        {\mbox{$h^{-1}$~Mpc}}

\newcommand{\Tbar}{\overline{T}}

\title[VLT LBG Redshift Survey II]{The VLT LBG Redshift Survey\thanks{ Based on observations collected at the European
    Organisation for Astronomical Research in the Southern Hemisphere,
    Chile, during programs 075.A-0683, 077.A-0612, 079.A-0442,
    081.A-0418 and 082.A-0494} II: Interactions between galaxies and the
  IGM at $z\sim3$} \author[N. H. M. Crighton et
  al.]{N. H. M. Crighton$^1$\thanks{email:neil.crighton@durham.ac.uk},
  R. Bielby$^{1,2}$, T. Shanks$^1$, L. Infante$^3$,
  C. G. Bornancini$^{4,5}$, \newauthor N. Bouch\'e$^6$,
  D. G. Lambas$^{4,5}$, J. D. Lowenthal$^{7}$, D. Minniti$^{3,8}$,
  S. L. Morris$^1$, \newauthor N. Padilla$^3$, C. P\'eroux$^9$,
  P. Petitjean$^2$, T. Theuns$^{10,11}$, P. Tummuangpak$^1$, \newauthor
  P. M. Weilbacher$^{12}$, L. Wisotzki$^{12}$ \& G. Worseck$^{13}$
  \\
  \\
  $^1$Department of Physics, University of Durham, South Road, Durham DH1 3LE, UK\\
  $^2$Institut d'Astrophysique de Paris, UMR7095 CNRS, Universit\'e Pierre et Marie Curie, 98 bis Bld Arago, 75014, Paris, France \\ 
  $^3$Departamento de Astronom\'ia y Astrof\'isica, Ponticia Universidad Catolica de Chile, Casilla 306, Santiago 22, Chile \\ 
  $^4$Instituto de Astronom\'ia Teo\'rica y Experimental, IATE, Observatorio Astron\'omico, Universidad Nacional de C\'ordoba,\\ 
  Laprida 854, X5000BGR, C\'ordoba, Argentina. \\ 
  $^5$Consejo Nacional de Investigaciones Cient\'ificas y T\'ecnicas, Avenida Rivadavia 1917, C1033AAJ,  Buenos Aires, Argentina. \\ 
  $^6$University of Santa Barbara, Broida, Santa Barbara, 93106, USA \\ 
  $^7$Department of Astronomy, Smith College, Northampton, MA 01063, USA \\ 
  $^8$Vatican Observatory, V00120 Vatican City State, Italy \\ 
  $^9$Laboratoire d'Astrophysique de Marseille, OAMP, Universite Aix-Marseill\`e \& CNRS, 13388 Marseille cedex 13, France\\ 
  $^{10}$Institute of Computational Cosmology, Department of Physics, University of Durham, South Road, Durham DH1 3LE, UK\\ 
  $^{11}$Universiteit Antwerpen, Campus Groenenborger, Groenenborgerlaan 171, B-2020 Antwerpen, Belgium\\
  $^{12}$Astrophysikalisches Institut Potsdam, An der Sternwarte 16, D-14482 Potsdam, Germany\\ 
  $^{13}$Dept of Astronomy and Astrophysics, UCO/Lick Observatory, University of California, 1156 High Street, Santa Cruz, CA 95064}
\begin{document}

\date{Accepted X December 15. Received X December 14; in
  original form X October 11}

\pagerange{\pageref{firstpage}--\pageref{lastpage}} \pubyear{X}

\maketitle

\label{firstpage}

\begin{abstract}
  We have measured redshifts for 243 $z\approx3$ quasars in nine VLT
  VIMOS LBG redshift survey areas, each of which is centred on a known
  bright quasar. Using spectra of these quasars, we measure the
  cross-correlation between neutral hydrogen gas causing the
  \lya\ forest and 1020 Lyman-break galaxies at $z\approx3$. We find
  an increase in neutral hydrogen absorption within
  $\approx5$~\hMpc\ of a galaxy in agreement with the results of
  Adelberger et al. (2003, 2005). The \lya-LBG cross-correlation can
  be described by a power-law on scales larger than 3~\hMpc. When
  galaxy velocity dispersions are taken into account our results at
  smaller scales ($<2$~\hMpc) are also in good agreement with the
  results of Adelberger et al. (2005). There is little immediate
  indication of a region with a transmission spike above the mean IGM
  value which might indicate the presence of star-formation feedback.
  To measure the galaxy velocity dispersions, which include both
  intrinsic LBG velocity dispersion and redshift errors, we have used
  the LBG-LBG redshift space distortion measurements of Bielby et
  al. (2010). We find that the redshift-space transmission spike
  implied in the results of Adelberger et al. (2003) is too narrow to
  be physical in the presence of the likely LBG velocity dispersion
  and is likely to be a statistical fluke.  Nevertheless, neither our
  nor previous data can rule out the presence of a narrow, real-space
  transmission spike, given the evidence of the increased
  \lya\ absorption surrounding LBGs which can mask the spike's
  presence when convolved with a realistic LBG velocity
  dispersion. Finally, we identify 176 \CIV\ systems in the quasar
  spectra and find an LBG-\CIV\ correlation strength on scales of
  10~\hMpc\ consistent with the relation measured at $\approx$Mpc
  scales.
\end{abstract}


\section{Introduction}
\label{sec:introduction}

The interaction between galaxies and the surrounding intergalactic
medium (IGM) is a crucial component of galaxy formation. The majority
of baryons at $z\sim3$ reside in the intergalactic medium
\citep[e.g.][]{Petitjean93, miralda-escud_ly_1996,
  schaye_model-independent_2001}.  It is from this reservoir of gas
that galaxies draw fuel for star formation.

Star formation and active galactic nuclei (AGN) are in turn believed
to have a significant effect on the IGM.  Winds generated by
supernovae in starburst events are observed to eject the interstellar
medium (ISM) with velocities of hundreds of \kms\ with respect to the
host galaxy \citep{veilleux_galactic_2005} and are believed to be able
to influence the IGM more than 100~kpc away from the galaxy
(e.g. Wilman et al., 2005). Jets from AGN have been observed to extend
hundreds of kpc from their host galaxy. There is considerable evidence
for winds being commonplace in star-forming galaxies at $z\sim3$. The
spectra of Lyman break galaxies (LBGs, so-called because they are
selected by their drop in flux at rest-frame 912~\AA) show a
systematic velocity offset between their \lya\ emission features,
absorption features (such as \CIV, \SiIV), and their nebular emission
features (such as \Ha).  Absorption features that are associated with
the galaxy's ISM are blue-shifted by 200-300~\kms\ with respect to the
true position of the galaxy, assumed to be given by the nebular
emission features \citep{pettini_rest-frame_2001}. This blueshift is
interpreted as an outflow velocity: the metal-enriched ISM gas is
being pushed out of the galaxy towards the surrounding IGM. Metal
enriched gas has also been observed up to $\sim 80$~kpc around $z
\approx 2.3$ LBGs, consistent with models for accelerating outflows
\citep{Steidel_structure_10}.

This ejection of matter and energy into the IGM is also an important
requirement for simulations of galaxy formation, where it is needed to
regulate star formation. Feedback from supernovae in starbursts is
required in semi-analytic models to reproduce the faint end of the
present day galaxy luminosity function
\citep[e.g.][]{baugh_canfaint_2005}. Hydrodynamical simulations have
also shown that such winds can enrich the IGM with metals to levels
required by observations \citep[e.g.][]{theuns_galactic_2002,
  oppenheimer_cosmological_2006}.

LBGs close to background quasar sightlines allow us to measure the gas
properties of the IGM close to galaxies, and search for any direct
evidence of feedback and winds at
$z\sim3$. \citet{adelberger_galaxies_2003} and
\citet{adelberger_connection_2005} (hereafter A03 and A05) pioneered
the first of these analyses.  By measuring the cross-correlation
between \HI\ \lya\ absorption with nearby LBGs, they showed there is
more \lya\ absorption within 5~\hMpc\ of LBGs compared to the mean
absorption level. This was interpreted as clustering of \HI\ gas
around the galaxies, consistent with LBGs being found in overdense
regions. For a significant fraction of LBGs within 1~\hMpc\ of a
quasar sightline, however, the observed absorption {\it decreased}
substantially. This was interpreted as the background quasar sightline
intercepting a bubble of ionized gas around some LBGs, possibly due to
star formation feedback from these galaxies heating their surrounding
IGM.

Winds were not expected to have such a large effect on the neutral
hydrogen surrounding LBGs. Using a smoothed particle hydrodynamical
(SPH) simulation \citet{theuns_galactic_2002} found winds had little
effect on nearby \HI\ absorption because they tended to deposit their
energy into low density regions around the galaxy, leaving much of the
\HI\ gas undisturbed in filamentary structures.  Subsequent
theoretical SPH models \citep[e.g.][]{Bruscoli03, Kollmeier03,
  Kollmeier06, kawata_galactic_2007} were also unable to reproduce the
distribution of \HI\ absorption close to LBGs measured by A03 and A05
without invoking exotic scenarios.  Semi-analytic models
\citep[e.g.][]{Desjacques04, bertone_do_2006} were able to reproduce
the distribution, but these assumed spherical symmetry, and so
side-stepped the geometrical considerations above.

Uncertainties in the galaxy redshift or \lya\ absorption redshift can
have a large influence on their cross-correlation at small scales.
Such uncertainties are caused by redshift measurement errors and any
intrinsic velocity dispersion between the galaxies and nearby
absorbing \HI\ gas. If there is a narrow feature in the real-space
cross-correlation, it will be suppressed by velocity dispersions when
it is measured in redshift-space. It is important to include these
effects when attempting to reconstruct the real-space correlation
function from that measured in redshift-space.

We have undertaken a program to observe LBGs with the Visible Imaging
\& Multi-Object Spectrograph (VIMOS) on the Very Large Telescope
(VLT).  This program will assemble a spectroscopic sample of $z \sim
3$ LBGs over nine fields, each chosen to be centred on a bright
($R\sim18$) background quasar with emission redshift $\gtrsim 3$, most
of which have archived echelle spectra available. The total area
covered by the nine fields is $\sim 3.17$~deg$^2$, corresponding to 45
VIMOS pointings. In addition to the central bright background quasar,
we have assembled a further spectroscopic sample of $R\sim 19-20$,
$z\sim 3$ quasars in each field. With these data we intend to measure
the galaxy-galaxy and galaxy-IGM clustering properties at redshifts
$\sim 2.5 - 3$. We shall both extend the A03, A05 data samples to
larger LBG-LBG and \lya-LBG transverse separations, and increase the
number of known small separation LBG-quasar sightline pairs at $z \sim
3$.

In this paper we present spectroscopy of the quasars inside and around
our LBG fields, describe the selection of quasar candidates and list
the new quasars we have identified. Our analysis focuses on the small
scale correlations between LBGs and the \lya\ forest at separations $<
10$~\hMpc. We compare our results to simple models and the results of
A03 and A05. The paper is structured as follows: in
Sections~\ref{sec:galaxy-sampl} and \ref{sec:quas-sampl} we describe
the galaxy and quasar samples used in our analysis.  In
Section~\ref{sec:quas-spectra} we describe the quasar
spectra. Sections~\ref{sec:civ} and \ref{sec:lya_autocorr} present
measurements of the \CIV-LBG cross-correlation and
\lya\ auto-correlation. Section~\ref{sec:lyblya_xcorr} presents the
main result of our paper, the cross-correlation between
\lya\ transmissivity and LBGs. Section~\ref{sec:summary} summarises
the main findings of the paper.

We assume a cosmology with $\Omega_{\rm m}=0.3$,
$\Omega_{\Lambda}=0.7$ and $H_0 \equiv 100$\,$h$~\kmsmpc, where
$\Omega_{\rm m}$ and $\Omega_{\Lambda}$ are the ratios of the matter
density and cosmological constant energy density to the critical
density. Unless stated otherwise all distances are comoving, and
magnitudes use the AB system, or asinh system for Sloan Digital Sky
Survey (SDSS) magnitudes.

\section{Galaxy Sample}
\label{sec:galaxy-sampl}

We obtained galaxy spectra for this project using the VIMOS
multi-object spectrograph on the VLT. A detailed description of the
LBG selection and sample properties is given by \citet{Bielby10}.  In
short, galaxies were selected using the Lyman break technique
\citep[e.g.][]{Steidel96}, yielding a sample with $2.2 < z <
3.5$. Deep $UBR$ or $UBVI$ imaging was used to select LBG candidates,
and these candidates were observed with the VIMOS multi-object
spectrograph at a resolution of 180.  In this paper we use the initial
set of LBG data presented by Bielby et al. It contains $1020$ LBGs
with spectroscopic redshifts $z>2$ spread across 19 VIMOS pointings in
a total area of 1.44~deg$^2$, over five of the nine fields that make
up the complete survey area.

For our present analysis we are most concerned with uncertainties on
the measured LBG redshifts. There are several contributions to the
redshift uncertainty; Bielby et al. quote $\sim 150$~\kms\ due to the
wavelength calibration, $\sim 450$~\kms\ from centroiding the
\lya\ emission lines, and $\sim 200$~\kms\ uncertainty in transforming
from the emission and absorption redshifts to the intrinsic galaxy
redshift. They estimate a total error of $\sim 500$~\kms\ in their
measured intrinsic LBG redshifts, corresponding to $\Delta z = 0.007$
at $z=3$.

\section{Quasar Sample}
\label{sec:quas-sampl}

Our quasar sample consists of $R < 22$ quasars with emission redshifts
$2 < z < 4$ in and around five LBG fields where we have reduced galaxy
spectra, and in four further fields that have as yet unreduced LBG
observations. These consist of:
\begin{itemize}

\item[(1)] Bright quasars at the centre of each LBG field. The LBG
  fields were chosen to be centred around bright quasars with emission
  redshifts $3 < z_{\rm em} < 4$ and over a wide RA range to enable
  observations to be made throughout the year. The central quasars in
  the five LBG fields analysed in this paper are Q0042$-$2627,
  J0124+0044, HE0940$-$1050, PKS2126$-$158 and J1201+0116. Archived
  echelle spectra taken using the Ultraviolet Echelle Spectrograph
  (UVES) on the VLT or the HIRES spectrograph on the Keck Telescope
  exist for most of these quasars.  The echelle spectra have
  resolution $> 30,000$ and hence resolve the linewidths of \HI\ lines
  in the \lya\ forest. There are four further fields where we will
  soon obtain LBG redshifts; around the central bright quasars
  Q0301$-$0035, Q2231$-$0015, Q2348$-$011 and Q2359+0653. Details for all
  nine central bright quasars are given in
  Table~\ref{tab:brightqsos}. \\
\item[(2)] Previously-known quasars in and around each field.  In
  addition to the central bright quasars, we searched for any other
  known quasars with the appropriate redshift and magnitude in either
  the NASA Extragalactic
  Database\footnote{http://nedwww.ipac.caltech.edu/} or the
  survey by \citet{worseck_slitless_2008}. \\
\item[(3)] New quasars discovered in each field. We obtained spectra
  for further quasar candidates in and around the LBG fields, selected
  from a variety of imaging sources.
\end{itemize}

We conducted a spectroscopic quasar survey targeting previously-known
quasars and quasar candidates using the AAOmega spectrograph on the
Anglo-Australian Telescope. AAOmega is a fibre-fed, multi-object
spectrograph \citep{saunders_aaomega:scientific_2004,
  smith_aaomega:multipurpose_2004, sharp_performance_2006} with a
resolution of $1300$ for the $385$R and $580$V gratings we used during
most of our observations.  In a single pointing, up to 400 fibres can
be targeted on objects over a circular field of view with radius of
one degree.

\begin{table*}
\centering
\begin{tabular}{lccccccc}
\hline
   &  &  &   &    & \multicolumn{3}{c}{Archived spectra}\\
{Name} &  {R.A. (J2000)} &  {Dec (J2000)} & {$z$}  &{Mag.}  & {Instrument} & {PI} & {ID} \\
\hline
Q2359+0653    & 00:01:40.6  & $+$07:09:54   & 3.23  & $V=18.5$    & - & - & -\\
Q0042$-$2627  & 00:44:33.95 & $-$26:11:19.9 & 3.289 & $B_j=18.47$ & HIRES & Chaffee   & K01H\\
J0124+0044    & 01:24:03.78 & $+$00:44:32.7 & 3.83  & $g=19.2$    & UVES  & Bouch\'e  & 073.A-0653\\
Q0301$-$0035  & 03:03:41.05 & $-$00:23:21.0 & 3.23  & $g=17.6$    & HIRES & Prochaska & U11H\\
HE0940$-$1050 & 09:42:53.50 & $-$11:04:25.9 & 3.06  & $B=17.2$    & UVES  & Bergeron  & 166.A-0106\\
J1201+0116    & 12:01:44.37 & $+$01:16:11.7 & 3.233 & $g=17.7$    & HIRES & Prochaska & U012Hb\\
PKS2126$-$158 & 21:29:12.15 & $-$15:38:40.9 & 3.268 & $V=17.3$    & UVES  & Bergeron  & 166.A-0106\\
Q2231$-$0015  & 22:34:08.99 & $+$00:00:01.7 & 3.02  & $r=17.29$   & UVES  & D'Odorico & 65.O-0296\\
Q2348$-$011   & 23:50:57.9  & $-$00:52:10   & 3.0235& $r=18.68$   & UVES  & Ubachs    & 079.A-0404\\
\hline
\end{tabular}
\caption{The nine central bright quasars around which LBG fields were
  targeted. The fourth and fifth columns give the emission redshift and
  a rough estimate of the quasar magnitude. The last three columns
  give the instrument, principle investigator and unique ID number for
  the archived observations where they are available.}
\label{tab:brightqsos}
\end{table*}

\subsection{Quasar selection for our AAOmega survey}
\label{sec:quasar-selection}

To cross-correlate LBG positions with quasar absorption, we need
background quasars with an emission redshift such that the
\lya\ forest overlaps the redshift range of our LBG sample and a
bright enough magnitude to obtain the signal to noise ratio (S/N)
required to measure \lya\ forest absorption. An emission range of $2.5
< z_{\rm em} < 4$ satisfies the first constraint -- at lower redshift
only a small portion of the forest remains above the atmospheric
cutoff; at higher redshift the higher order Lyman transitions and
Lyman limit absorption from redshifts $> 4$ make it difficult to
identify \lya\ absorption at the LBG redshifts. We chose a magnitude
limit for candidates of $R=22$; this was motivated by the S/N
achievable over the \lya\ forest in several hours of exposure using
AAOmega. Wolf et al. (2003) estimate a sky density for quasars with
$z>2.2$ and $R_{\rm Vega}<22$ of $\sim 40 \deg^{-2}$, thus we
anticipated there would be up to $\sim10$ such quasars inside one of
our typically $0.5 \times 0.5$~deg$^2$ LBG fields.

To select candidates we used the theoretical quasar tracks in $ugr$
colour space from \citet[][see his fig.~13]{fan_simulation_1999} as a
guide. Our criteria for candidates were that they were (1) point-like,
(2) outliers in $ugr$ colour space from the stellar locus, with the
expected colours for $2.5 < z < 4.0$ quasars, and (3) had $r<22$.  The
first two criteria are known to select quasars with redshifts $>3.0$
with a relatively high completeness and efficiency for targets with $i
< 20.8$ selected using SDSS imaging
[\citet{richards_spectroscopic_2002}, but see also
  \citet{Worseck10}]. For $2.5<z<3.0$, the expected position of
quasars in $ugr$ colour space overlaps with A, F and horizontal giant
branch stars, making efficient selection difficult.  In an attempt to
find a significant number of the available quasars in this redshift
range, we included objects as close to the stellar locus as was
possible without introducing an unacceptably large level of
contamination.  However, the efficiency of our quasar selection in
this range is poor.

When possible, we checked that any known quasars in our fields were
recovered by our selection process. This sometimes led us to adjust
our colour cuts to ensure that known quasars in the desired redshift
range were included using our selection criteria. We also adjusted
colour cuts to provide a sufficiently high sky density of targets
($\sim600$ over the AAOmega field of view) that allowed
\textsc{configure}\footnote{http://www.aao.gov.au/AAO/2df/aaomega/aaomega\_manuals.html},
the software used to assign objects to the AAOmega fibres, to maximise
the number of fibres used. Due to restrictions on fibre placement not
all possible candidates could be observed. In general we prioritised
brighter quasar candidates with $r<21.5$, those with photometric
redshift estimates, and those close to areas where LBG redshifts were
to be measured.

Finally, for repeat observations of the same field we performed an
initial identification of objects in the first set of observations.
Any targets that could not be identified as quasars in the required
redshift range were removed and replaced with new candidates for
subsequent observations. If this exhausted our candidate list in a
field, we adjusted the colour cuts closer to the stellar locus to
provide more candidates.
\begin{table*}
\centering
\begin{tabular}{lccccccc}
  \hline
  & \multicolumn{4}{c}{Central Imaging} &  \multicolumn{3}{c}{Surrounding Imaging}\\
  {Field} & {Source} & {Bands} & {Area} & {Depth} & {Source} & {Bands} & {Depth}  \\
  \hline
  Q2359+0653    & MOSAIC & $UBR$ & $32\arcmin\times 32\arcmin$ & $R=25$ & Schmidt  & $BR$ & $R=21$ \\
  Q0042$-$2627  & MOSAIC & $UBR$ & $32\arcmin\times 32\arcmin$ & $R=24.7$ & Schmidt & $BR$ & $R=21$ \\
  J0124+0044    & MOSAIC$^a$ &$UBVI$ & $32\arcmin\times 32\arcmin$ & $I=24.5$ & SDSS Stripe 82 & $ugriz$ & $r=24.7$ \\
  Q0301$-$0035  & MOSAIC & $UBR$ & $32\arcmin\times 32\arcmin$ & $R=25$ & SDSS Stripe 82 & $ugriz$ & $r=24.7$ \\
  HE0940$-$1050 & MegaCAM& $ugri$ & $1\Deg\times 1\Deg$        & $r=24.7$ & Schmidt & $UBR$ & $R=21$ \\
  J1201+0116    & MOSAIC & $UBR$ & $32\arcmin\times 32\arcmin$ & $R=25.5$ & SDSS  & $ugriz$ & $r=22.6$\\
  PKS2126$-$158 & MOSAIC & $UBR$ & $32\arcmin\times 32\arcmin$ & $R=24.7$ & Schmidt & $UBR$ & $R=21$\\
  Q2231$-$0015  & WFC    & $UBR$ & $32\arcmin\times 32\arcmin$ & $r=25$ & SDSS  & $ugriz$ & $r=22.6$ \\
  Q2348$-$011   & MegaCAM& $ugri$ & $1\Deg\times 1\Deg$        & $r=25$ & SDSS Stripe 82 & $ugriz$ & $r=24.7$ \\
  \hline
\end{tabular}
\caption{Imaging used to select quasar candidates inside the central
  LBG regions and surrounding the central regions. The different bands
  available, the area of the central imaging and depths are also
  shown. Depths give approximate 50\% completeness depth (MOSAIC,
  SDSS, Stripe 82, MegaCAM, INT) or the magnitude limit (Schmidt). The
  image source abbreviations are; MOSAIC: the imaging was taken with
  the MOSAIC imager at either CTIO or KPNO; Schmidt: catalogues
  generated from Schmidt plates were used; MegaCAM: archived CFHT
  MegaCAM imaging; WFC: archived INT Wide Field Camera observations;
  SDSS: single epoch SDSS imaging; SDSS stripe 82: multi-epoch SDSS
  imaging in Stripe 82.  $^a$~For the J0124+004 field, we selected
  targets in the central region from an object catalogue generated
  from MOSIAC imaging; see \citet{Bielby10} and
  \citet{bouch_clustering_2004} for details.}
\label{tab:imselect}
\end{table*}

\subsubsection*{Selection of quasars overlapping the LBG fields}

To select quasars overlapping the LBG fields we generally used the
same $UBR$ or $UBVI$ imaging that was used to select the LBG
candidates (see Table~\ref{tab:imselect}).  Selections in the central
fields around Q0042$-$2627, Q0301$-$0035, J0124+0044, J1201+0116,
PKS2126$-$158, Q2359+0653 were performed using the imaging data from
the MOSAIC imagers at Kitt Peak National Observatory (KPNO) and the
Cerro-Tololo Inter-American Observatory (CTIO) described by
\citet{Bielby10}. This imaging covers a $32\arcmin\times 32\arcmin$
region around the central quasar. We used archived imaging taken with
the Wide Field Camera on the Isaac Newton Telescope (INT) for the
Q2231$-$0015 field. SExtractor \citep{bertin_sextractor:_1996}
catalogues were generated from the images. For the HE0940$-$1050 and
Q2348$-$011 fields, the central field selection was performed using
archived $ugr$ imaging data from the MegaCAM instrument at the
Canada-France-Hawaii Telescope (CFHT) rather than the MOSAIC imaging
used to select LBGs. The MegaCAM data reaches a similar depth to the
MOSAIC data, but extends over a larger field of view,
$1\Deg\times1\Deg$.

\begin{figure*}
\centering
\includegraphics[width=\textwidth]{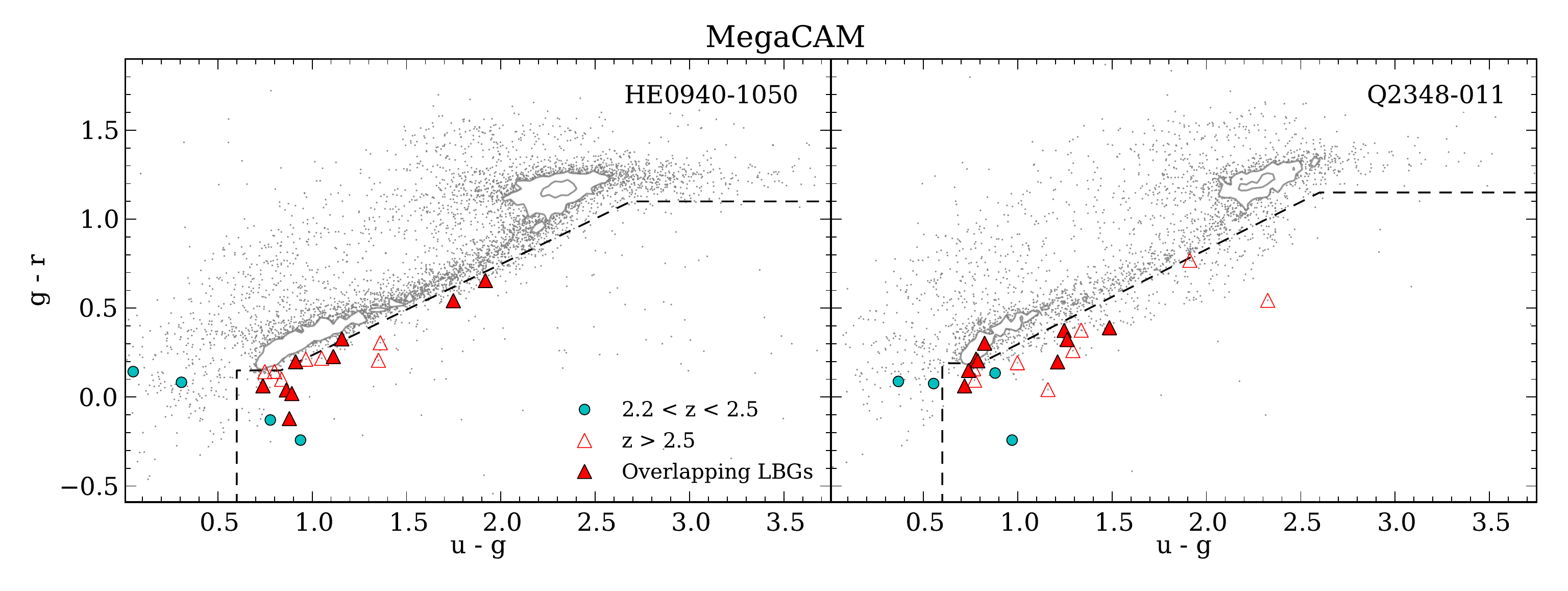}
\caption{Colour cuts used for $1\Deg \times 1\Deg$ CFHT MegaCam data
  in the HE0940$-$1050 and Q2348$-$011 fields.  Grey points and
  contours show all stellar objects (SExtractor CLASS\_STAR $> 0.85$)
  with $r<22$. Objects below the borders marked by the dashed line
  were selected as quasar candidates. Triangles show known quasars
  with $z \ge 2.5$, and circles show quasars with $2.2 \le z <
  2.5$. Open triangles are quasars outside the area where we have LBG
  observations; solid triangles are quasars overlapping the LBG fields
  on the sky.}
\label{fig:mega_select}
\end{figure*}

The colour cuts we used to select quasar candidates varied slightly
between fields, depending on the quality of the imaging, how well the
photometric zero points had been measured, and the filters used. As an
example, the cuts for the MegaCAM images in the HE0940$-$1050 field
were:
\begin{itemize}
\item $18 < r < 22$
\item $g - r < 1.1$
\item $g - r < 0.54 (u - g) - 0.35\,$, or $\,g - r < 0.15$
\item $u - g > 0.6$
\end{itemize}
They are shown in the left panel of Fig.~\ref{fig:mega_select} as
dashed lines, along with similar cuts for the other field with central
MegaCam imaging, Q2348$-$011. Candidates were required to be detected in
$g$ and $r$, but we included candidates undetected in $u$ if they
satisfied the above criteria. The precise selections used for the
MOSAIC data are given by Bielby et al. (2008, PhD thesis).

In total we obtained spectra of 50 $z>2.2$ quasars overlapping the LBG
fields in addition to the nine central bright quasars. Closed
triangles in Fig.~\ref{fig:mega_select} show such quasars in the cases
of the HE0940$-$1050 and Q2348$-$011 fields. Across all nine fields,
30 of these 50 were previously unknown quasars uncovered using the
selection process above; the remainder were previously known. The
total LBG area is 3.17~deg$^2$, over which we obtained a quasar sky
density of $\sim 18.6$~deg$^{-2}$.

In the five fields we use for the cross-correlation analysis there are
16 $z>2.2$ quasars. All of these were previously known. The Q0042$-$2627
field has been searched for quasars by
\citet{williger_large-scale_1996}, and the HE0940$-$1050 and PKS2126$-$158
fields by \citet{worseck_slitless_2008}, leaving few new quasars to be
found. However, our sky densities in these fields are also low
compared to other fields in the LBG area; the five fields cover
1.44~deg$^2$ giving a density of $11.1$~deg$^{-2}$, much lower than
the $18.6$~deg$^{-2}$ above. This lower density is in part due to the
absence of any new quasars in the J0124+0044 central area beyond the
central bright quasar. We are unsure of the reason for this. However,
similarly large areas with very few quasars are present in the
Q0301$-$0035 and Q2348$-$011 fields, which have deep Stripe 82 imaging
across the entire AAOmega field.  Clustering may be responsible for
the clumpy quasar distribution, and it may simply have been
unfortunate that a region largely empty of quasars occurs in the
J0124+0044 LBG area.

\subsubsection*{Selection of quasars around the LBG fields}

There were not enough quasar candidates overlapping each LBG area to
employ all the AAOmega fibres, so we searched for candidates outside
each LBG field over the full $3.1$~deg$^2$ AAOmega field of view.  Our
motivation for finding quasars with angular separations of tens of
arcminutes from LBGs was not to look for the effects of feedback --
the IGM probed is much further from the LBGs than the distances across
which winds are expected to have a significant effect.  However, with
a large number of $z\sim3$ quasars over a few square degrees we can
measure correlations in metal or forest absorption on scales of tens
of Mpc due to large scale structure \citep{williger_evidence_2000},
constrain the 3-d topology of the IGM \citep{pichon_inversion_2001}
and measure large scale anisotropies in the 2-d LBG-\lya\ correlation
function caused by velocity dispersion and infall. These projects are
beyond the scope of our current analysis, but our quasar sample
provides a valuable resource for future studies.

\begin{figure*}
\centering
\includegraphics[width=\textwidth]{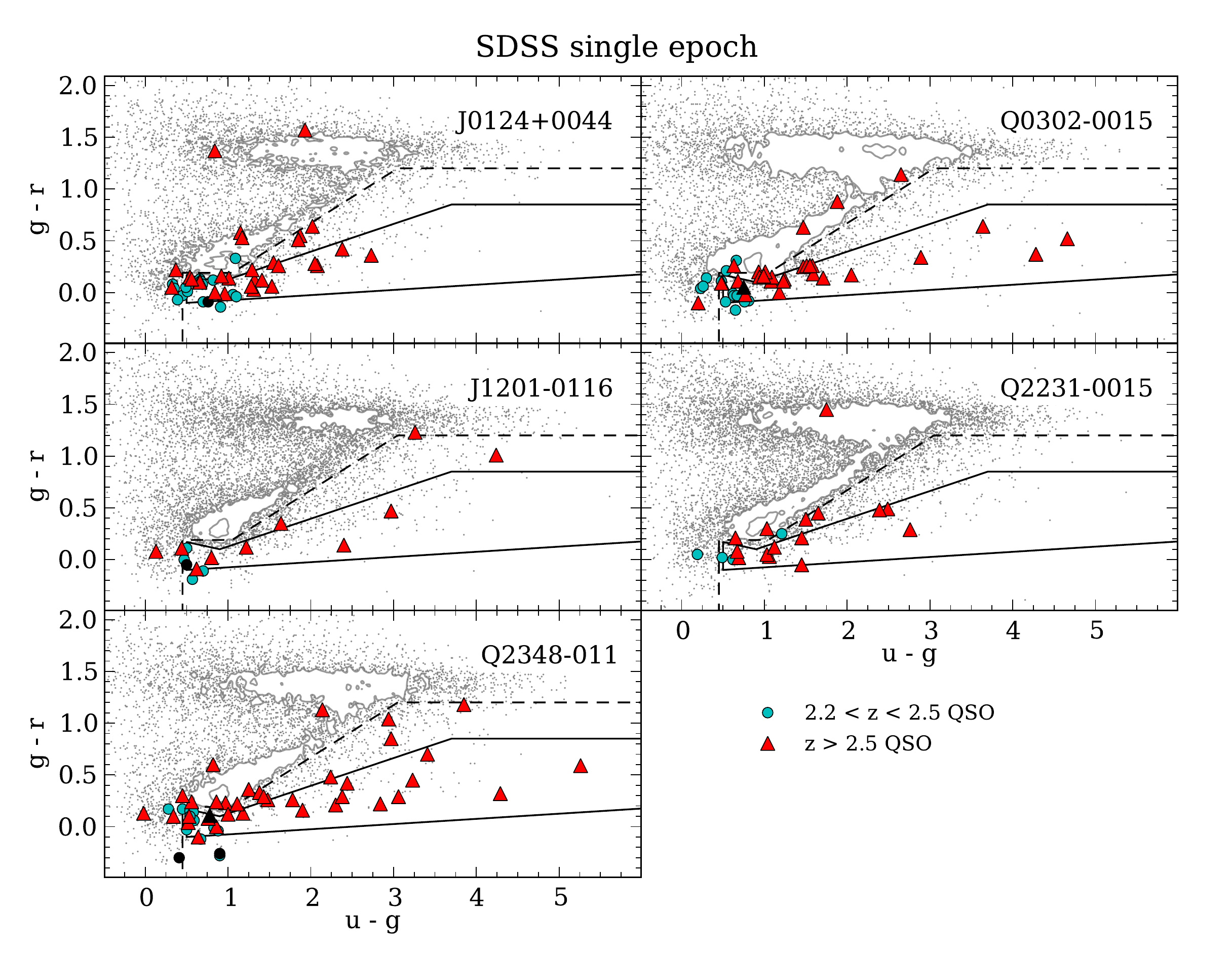}
\caption{Colour cuts for used for single-epoch SDSS data in the
  J0124+0044, Q0301$-$0035, J1201+0116, Q2231$-$0015, and Q2348$-$011
  fields. Objects classified by the SDSS pipeline as stellar with
  $r<22$ are shown by the grey points and contours.  Objects below the
  borders marked by the dashed line were selected as quasar
  candidates. The solid line shows the region used to select
  candidates for the efficiency calculations in
  Appendix~\ref{ap:effic}. Triangles show known quasars with $z >
  2.5$, and circles known quasars with $2.2 < z < 2.5$.  Several of
  these quasars were classified in SDSS single epoch imaging as
  non-stellar; they are shown by black circles and black
  triangles. They were included in our sample because of alternative
  selection criteria.}
\label{fig:sdss_select}
\end{figure*}

The deep imaging used to select LBGs does not extend across the full
AAOmega field of view, so we used different imaging sources to select
candidates outside the central LBG regions.  For the J0124+0044,
Q0301$-$0035, J1201+0116, Q2231$-$0015 and Q2348$-$011 fields we used
single epoch SDSS $ugr$ imaging catalogues. Quasar candidates were
selected in three ways. Firstly, we targeted any of the
photometrically-selected quasar candidates from
\citet{richards_efficient_2004} and \citet{richards_efficient_2009}
with appropriate photometric redshifts.  Secondly, we used the SDSS
pipeline classifications \citep{richards_spectroscopic_2002}: objects
were classified in the SDSS reduction pipeline as quasar candidates
based on their colours, stellar/non-stellar classification and radio
detection. Only candidates with $i<20.2$ were followed up for
spectroscopy by the SDSS, leaving many fainter candidates without
spectra.  Any of these with colours consistent with our desired
redshift range were added to our target list. Finally, we selected
additional candidates not already selected by the above two methods
using our own $ugr$ colour cuts.  Fig.~\ref{fig:sdss_select} shows
colour-colour plots for these five fields with single-epoch SDSS
imaging. The points and contours show SDSS stellar objects and the
$ugr$ selection cuts we used are shown by dashed lines. Quasars we
observed, both previously-known and newly discovered, inside and
outside the LBG areas, are shown as triangles and circles.

\begin{figure*}
\centering
\includegraphics[width=\textwidth]{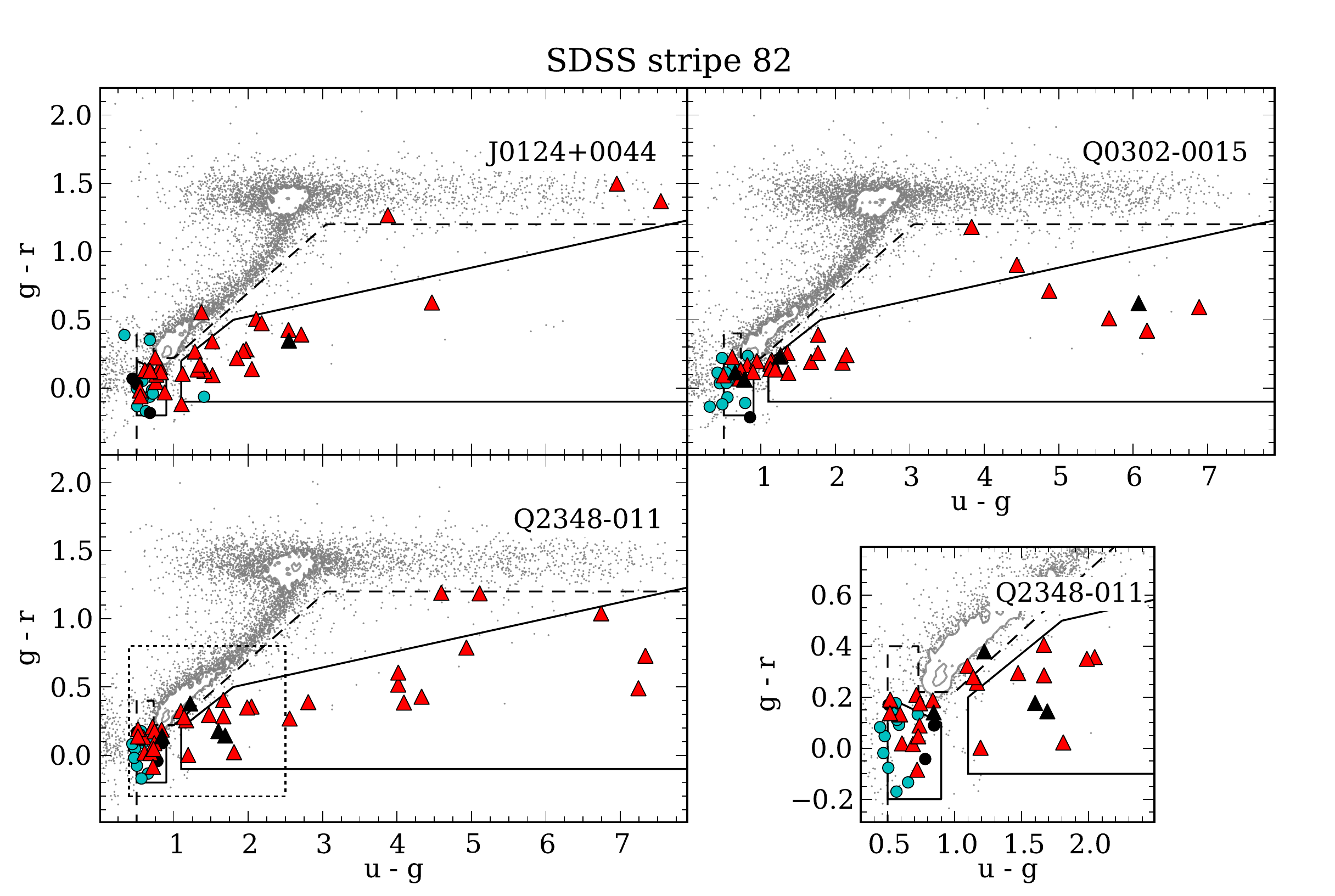}
\caption{Colour cuts for used for multi-epoch Stripe 82 SDSS data in
  the J0124+0044, Q0301$-$0035, and Q2348$-$011 fields. The lines and
  symbols are the same as in Fig.~\ref{fig:sdss_select}. The bottom
  right panel is a zoomed plot of the dotted region shown in the
  bottom left panel.}
\label{fig:s82_select}
\end{figure*}

Three of our equatorial fields (J0124+0044, Q0301$-$0035 and
Q2348$-$011) overlap the Stripe 82 region, where repeat SDSS
observations were taken for the Supernova Survey \citep{Frieman08}. In
these fields we selected candidates using catalogues generated by
combining the multi-epoch imaging. For J0124+0044 and Q2348$-$011 we
offset the AAOmega pointing centre from the central bright quasar to
maximise overlap with the Stripe 82 catalogue. The $ugr$ cuts used to
select candidates from the Stripe 82 catalogues are shown by dashed
lines in Fig.~\ref{fig:s82_select}. These cuts were modified slightly
from those used on other imaging catalogues to include a box with $0.5
< u - g < 0.75$ and $0.2 < g - r < 0.4$, based on the colours of
photometrically-selected targets from \citet{richards_efficient_2009}.

Where SDSS imaging was not available (Q0042$-$2627, HE0940$-$1050,
PKS2126$-$158 and Q2359+0653 fields), we used $B$ and $R$ catalogues
generated from Schmidt photographic plates processed by the automated
plate measuring machine. For the HE0940$-$1050 and PKS2126$-$158
fields we also had access to Schmidt $U$ catalogues. Candidates were
selected using similar criteria to the central areas where $U$ imaging
was available or using only $B-R$ cuts otherwise.

In total we obtained spectra for 193 $z>2.2$ quasars outside the LBG
areas. 134 of these are newly discovered: 31 photo-$z$ candidates from
single epoch SDSS imaging, 21 selected from deep LBG imaging that
extended beyond the LBG areas, 40 using $ugr$ cuts with Stripe 82
imaging, 28 selected using similar cuts with single epoch SDSS imaging
and the remaining 14 from Schmidt imaging.

\subsection{Catalogue of quasars found in the AAOmega survey and completeness}

\begin{figure}
\centering
\includegraphics[width=0.5\textwidth]{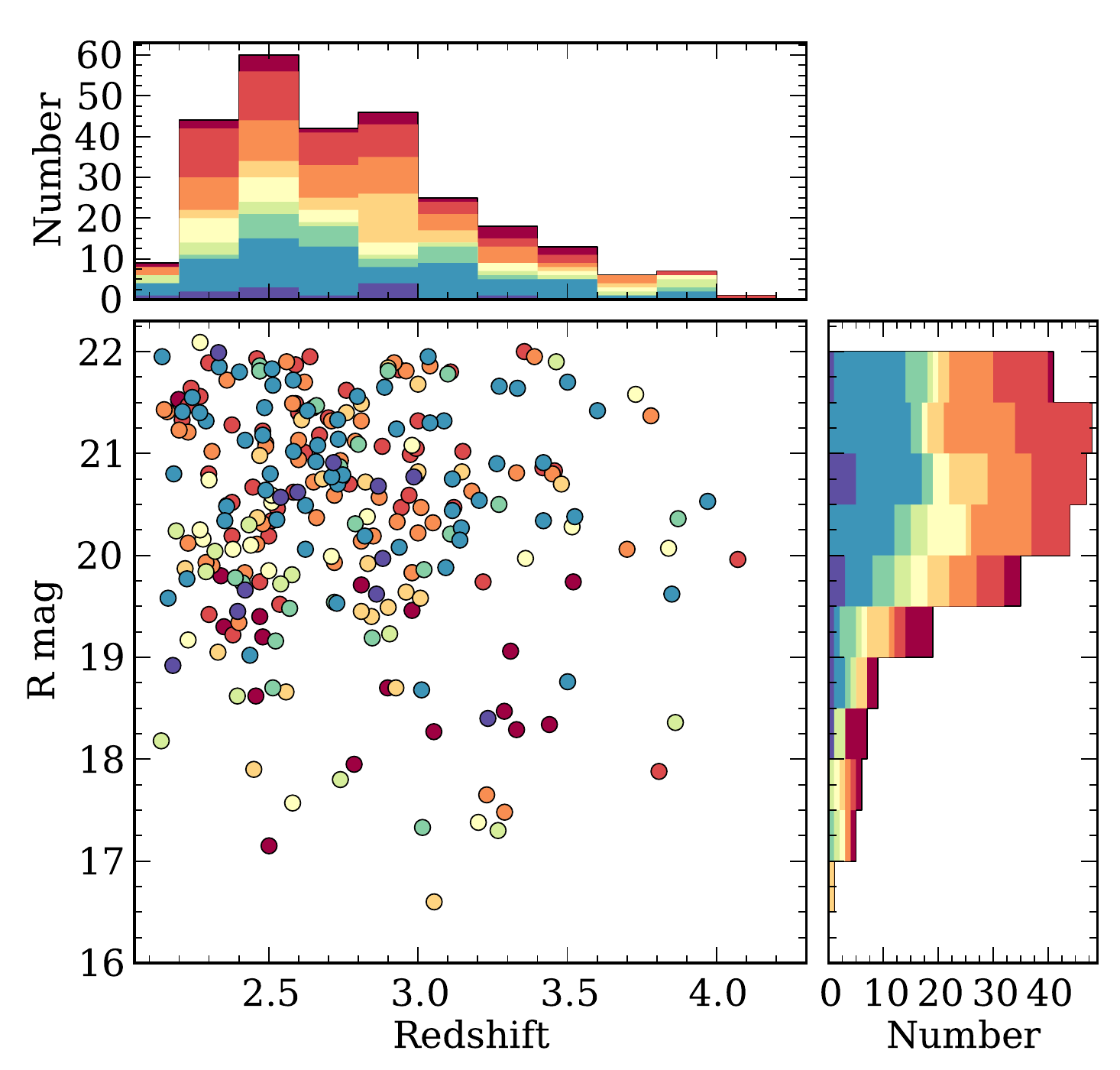}
\caption{The redshift and $R$ magnitude distribution for quasars with
  $R <22$ and $z>2.2$ in nine AAOmega fields. Each colour represents
  quasars from a different field. From the top of the histogram to the
  bottom, colours denote the Q0042$-$2627, J0124+0116, Q0301$-$0035,
  HE0940$-$1050, J1201+0116, PKS2126$-$158, Q2231$-$0015, Q2348$-$011 and
  Q2359+0653 fields.}
\label{fig:qsohist}
\end{figure}

Our AAOmega survey obtained spectra of 243 $z > 2.2$ quasars, of which
164 are newly discovered. The number of quasars found per field
and their selection source are given in Table~\ref{tab:qsonums}, and
their details are in Table~\ref{tab:qsos}. Their magnitude and
redshift distributions are shown in Fig.~\ref{fig:qsohist}. Some
quasars with $2.1 < z < 2.5$ were also recovered by our selection
criteria for $z > 2.5$ quasars. Even though little of their
\lya\ forest is observable above the atmospheric cutoff, they are
still useful for other purposes such as identifying \CIV\ absorption
near LBGs, or cross-correlating AGN with LBGs. There are also 10 faint $R \gtrsim 22$
quasars overlapping the LBG areas in Table~\ref{tab:qsos} that we
discovered in our VIMOS observations. We do not use these in the
present analysis; they are described further by \citet{Bielby10}.

\begin{figure}
  \centering
  \includegraphics[width=0.5\textwidth]{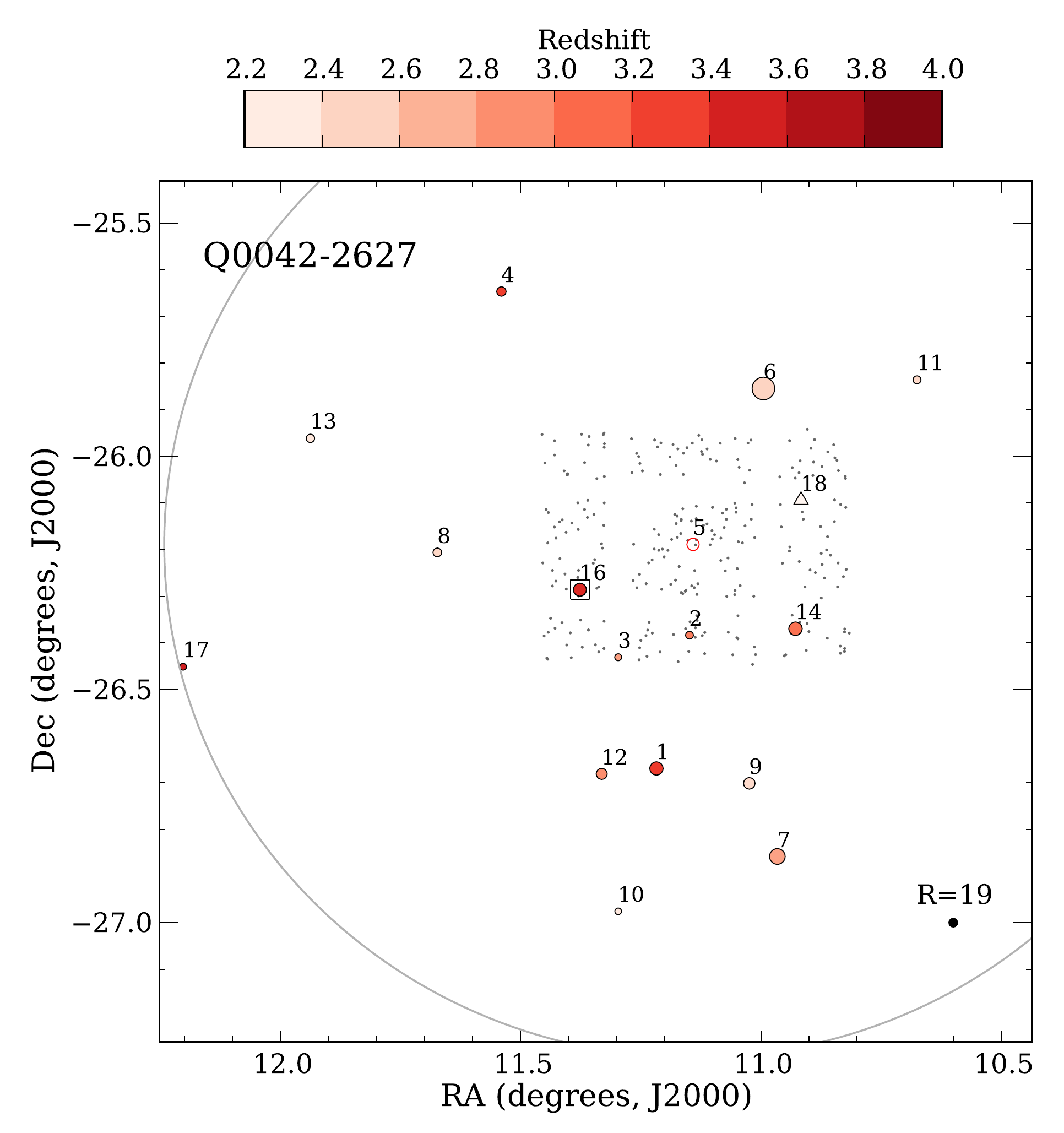}
  \caption{Quasar positions in the Q0042$-$2627 field.  Grey dots show
    LBGs with spectroscopically-confirmed redshifts. Each quasar is
    labelled with its number from Table~\ref{tab:qsos} and the open
    circle is the bright central quasar. Quasars surrounded by a
    square box (and the central quasar) have been observed at high
    resolution. The area of each circle is proportional to the R band
    quasar luminosity. The size of a quasar with $R=19$ is shown at
    the bottom right for comparison. Triangles show very faint quasars
    ($R>23$) that were discovered serendipitously in our VIMOS
    observations. North is up and East is to the left. The remaining
    eight fields are shown in Fig.~\ref{fig:radec0124} and
    \ref{fig:radec2126}.}
  \label{fig:radec0042}
\end{figure}

\begin{figure*}
  \centering
  \includegraphics[width=\textwidth]{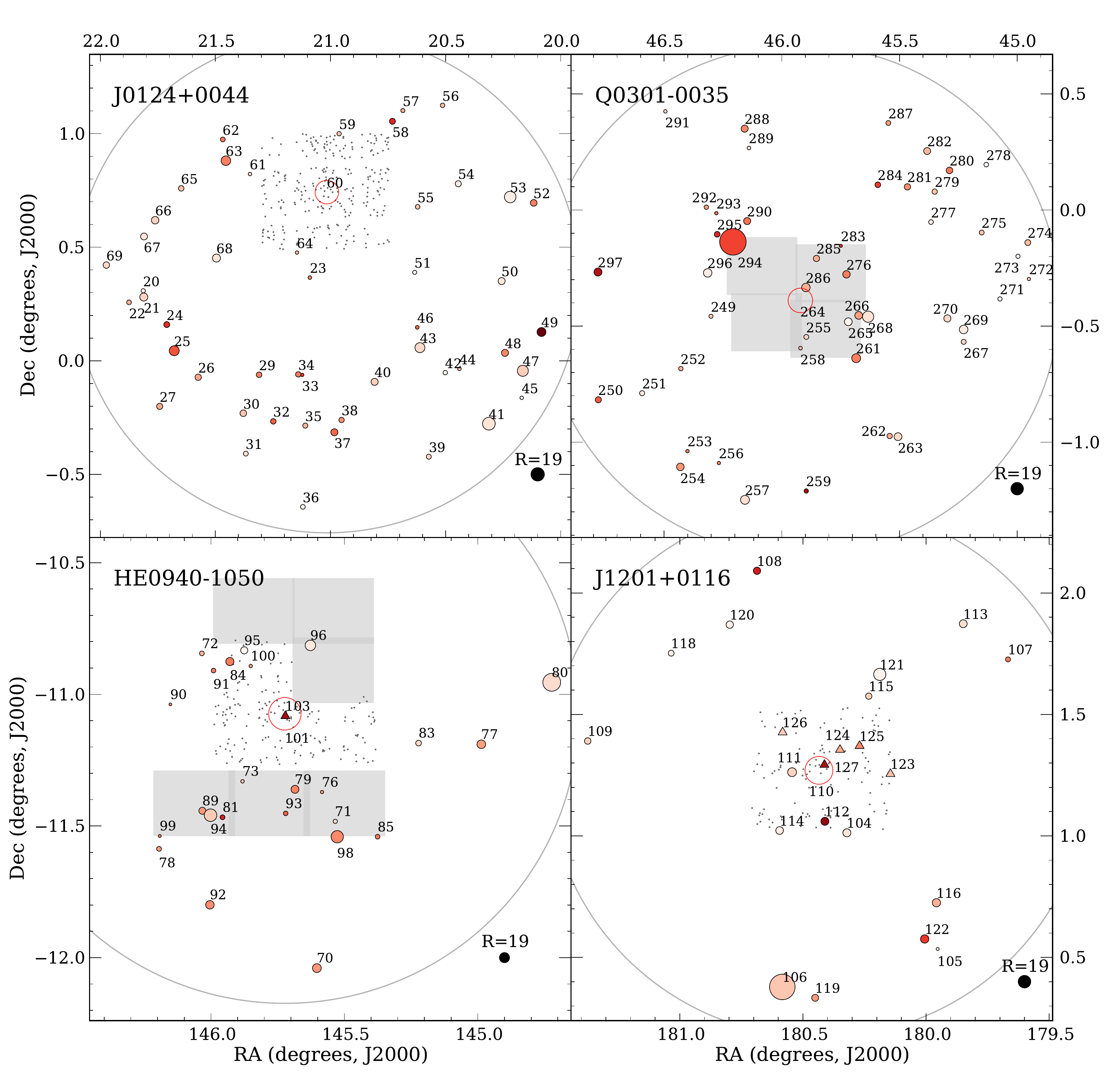}
  \caption{Quasar positions in the J0124+0044, Q0301$-$0035, HE0940$-$1050
    and J1201+0116 fields. Grey shaded regions show observed VIMOS
    fields where spectroscopic LBG redshifts have not yet been
    measured. Other symbols are the same as in
    Fig.~\ref{fig:radec0042}. The J0124+0044 pointing is offset from
    the central quasar to maximise overlap with Stripe 82 imaging. }
  \label{fig:radec0124}
\end{figure*}

\begin{figure*}
  \centering
  \includegraphics[width=\textwidth]{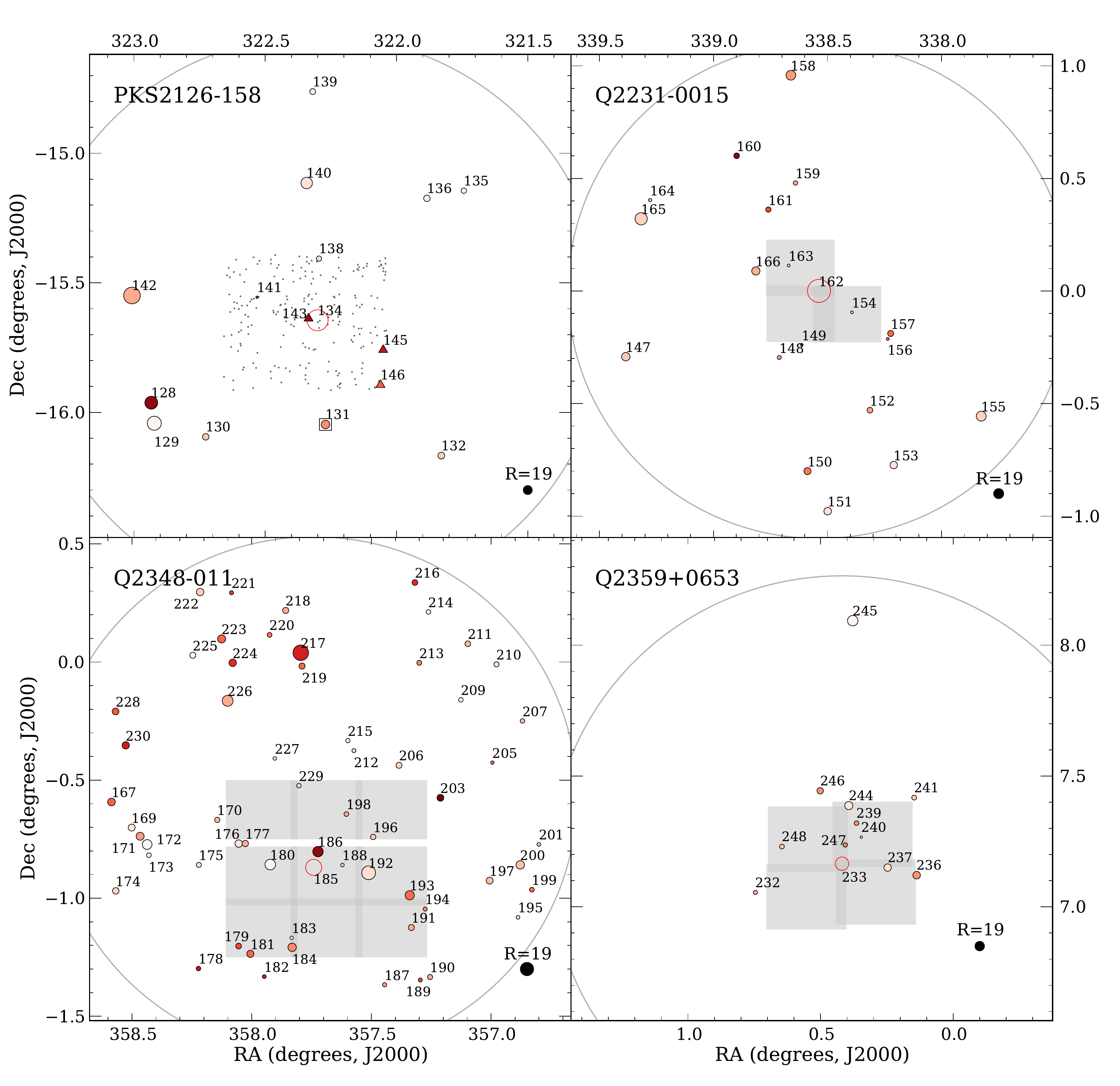}
  \caption{Quasar positions in the PKS2126$-$158, Q2231$-$0015, Q2348$-$011
    and Q2359+0653 fields. The Q2348$-$011 pointing is offset from the
    central quasar to maximise overlap with Stripe 82 imaging. Symbols
    are the same as in Fig.~\ref{fig:radec0042} and
    ~\ref{fig:radec0124}.}
  \label{fig:radec2126}
\end{figure*}

The right ascension and declination of quasars and LBGs in each field
are shown in Fig.~\ref{fig:radec0042} to \ref{fig:radec2126}. It is
apparent from these figures that our quasar sample is not uniformly
distributed on the sky. This is mostly due to the variable imaging
depths used in the different fields; few quasars were found outside
the LBG regions in the J1201+0116 and Q2359+0653 fields, where only
Schmidt imaging was available, but we achieved a much higher density
in the J0124+0044, Q0301$-$0035 and Q2348$-$011 fields where Stripe 82
imaging was available across the entire field. However, as we remarked
earlier, there are large regions free from quasars even in areas where
we have deep imaging. This may be due in part to large scale structure
and quasar clustering.

\begin{figure}
  \centering \includegraphics[width=0.5\textwidth]{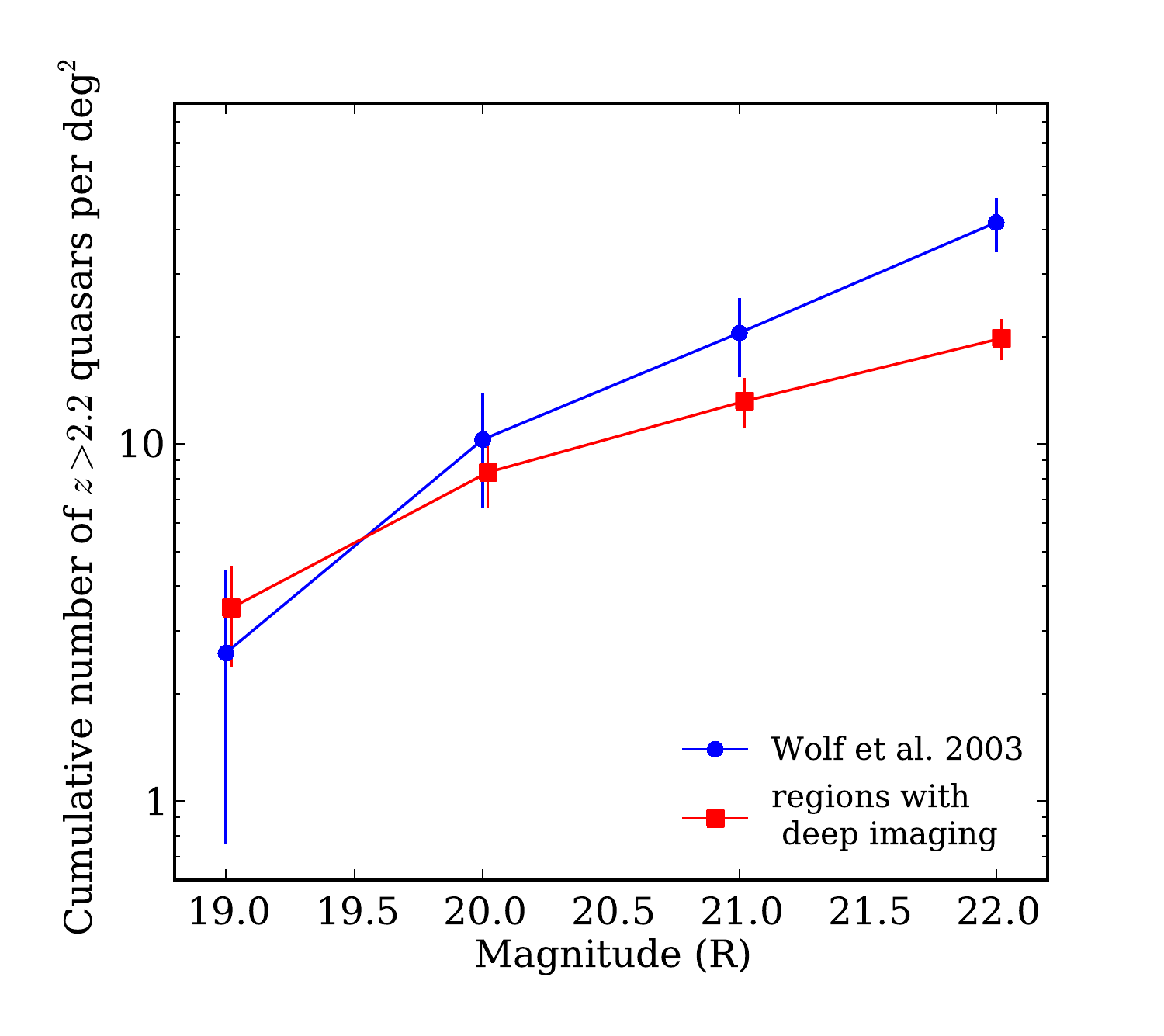}
  \caption{The cumulative sky densities of quasars with $z > 2.2$ as a
    function of magnitude from \citet{wolf_evolution_2003} compared to
    our quasar sample in regions where deep imaging (to $R\sim25$) was
    used to select quasars. Our completeness is high up to $R\sim21$,
    but drops to 50\% at $R=22$.}
  \label{fig:skydens}
\end{figure}

We can roughly estimate our sample's completeness -- the fraction of
the total number of quasars in our redshift range and to our magnitude
limits we have recovered -- by comparing our sky densities to those
for the COMBO-17 quasar survey \citep{wolf_evolution_2003}.  Wolf et
al. measured the number density of $R_{\rm Vega} <22$ quasars with $z>2.2$ to be
$\sim 40$ per square degree. In Fig.~\ref{fig:skydens} we show the
cumulative sky densities for quasars in the central region of each of
our nine fields, where deep imaging was used to select quasar
candidates, compared to the incompleteness-corrected sky densities
found by Wolf et al. (see their table~3). Up to $R=21$, our densities
are consistent, suggesting our completeness is high. At $R=22$, our
sky densities drop to 50\% of those found by Wolf et al., suggesting
we recover only 50\% of these fainter quasars. This low completeness
level is not surprising: for single-epoch SDSS candidates we
prioritised bright ($R < 21.5$) targets and did not observe many
fainter targets; a significant fraction of candidates overlaps with
the stellar locus; and for very faint targets even if a quasar was
observed in poor conditions, we may have failed to identify it. For
areas outside the central deep imaging, our completeness will be much
poorer. Appendix~\ref{ap:effic} describes the efficiency of our quasar
selection process and suggests ways to improve the selection
efficiency for $2.5 < z < 4.0$ quasars in future surveys.

\begin{table}
\centering
\begin{tabular}{lrccc}
  \hline
  {Field} & {Type}  & {Cand.} & {New} & {Known}  \\
  \hline
   Q0042$-$2627 & Central           &      &     &     \\
              & $32\arcmin\times 32\arcmin$ & 12 & 0 & \\
              & $ugr$             & 263  & 0   &     \\
              & Total             & 275  & 0   & 16  \\
  J0124+0044  & Central           &      &     &     \\
              & $32\arcmin\times 32\arcmin$ & 19 & 0 & \\
              & Photo-$z$         & 92   & 14  &     \\
              & Stripe 82         & 145  & 20  &     \\
              & SDSS $ugr$        & 377  & 2   &     \\
              & Total             & 633  & 36  & 13  \\
  Q0301$-$0035  & Central           &      &     &     \\
              & $32\arcmin\times 32\arcmin$ & 129 & 5 & \\
              & Photo-$z$         & 80   & 8   &     \\
              & Stripe 82         & 48   & 9   &     \\
              & SDSS $ugr$        & 351  & 8   &     \\
              & Total             & 608  & 30  & 16  \\
  HE0940$-$1050 & Central           &      &     &     \\
              & $1\Deg \times1 \Deg$ & 113  & 14 & \\
              & Schmidt $UBR$        & 382  & 8   &     \\
              & Total             & 479  & 22  & 2   \\
  J1201+0116  & Central           &      &     &     \\
              & $32\arcmin\times 32\arcmin$ & 0& 0   & \\
              & Photo-$z$         & 2    & 1   &     \\
              & SDSS $ugr$        & 322  & 4   &     \\
              & Total             & 324  & 5   & 13  \\
  PKS2126$-$158 & Central           &      &     &     \\
              & $32\arcmin\times 32\arcmin$ & 346 & 2 & \\
              & Schmidt $UBR$         & 355  & 6   &     \\
              & Total             & 701  & 8   & 3   \\
  Q2231$-$0015  & Central           &      &     &     \\
              & $32\arcmin\times 32\arcmin$ & 77 & 4 & \\
              & Photo-$z$         & 2    & 0   &     \\
              & SDSS $ugr$        & 83   & 5   &     \\
              & Total             & 162  & 9   & 11  \\
  Q2348$-$011   & Central           &      &     &     \\
              & $1\Deg \times1 \Deg$ & 219  & 16  & \\
              & Photo-$z$         & 49   & 8   &     \\
              & Stripe 82         & 68   & 11  &     \\
              & SDSS $ugr$        & 633  & 9   &     \\
              & Total             & 969  & 44  & 13  \\
  Q2359+0653  & Central           &      &     &     \\
              & $32\arcmin\times 32\arcmin$ & 251& 10& \\
              & Schmidt $BR$          & 360  & 0   &     \\
              & Total             & 611  & 10  & 1   \\
              &                   &      &     &     \\
  All         &                   & 4762 & 164 & 88  \\
  \hline
\end{tabular}
\caption{Quasar candidates observed with AAOmega, and the number of
  new quasars found. Columns show the field names, the candidate
  source, the observed candidates for that source, and the number of
  identified quasars with $z > 2.2$.  The total number of observed
  candidates, new quasars and the number of quasars previously known
  in each field is also shown.}
\label{tab:qsonums}
\end{table}

\section{Quasar spectra}
\label{sec:quas-spectra}
\subsection{High resolution quasar spectra}
\label{sec:hires_qso}

UVES archived spectra are available for the bright central quasars
J0124+0044, HE0940$-$1050 and PKS2126$-$158 and Keck/HIRES
archive spectra are available for Q0042$-$2627 and J1201+0116.  These
spectra have resolution full widths at half maximum (FWHM) of
$6-8$~\kms. The UVES spectra were reduced using the UVES pipeline, and
individual exposures were combined with \textsc{uves popler}
software\footnote{http://astronomy.swin.edu.au/$\sim$mmurphy/UVES\_popler}. The
Keck spectra were reduced using the \textsc{makee}
package\footnote{http://spider.ipac.caltech.edu/staff/tab/makee/}. In
addition we have obtained HIRES spectra for bright quasars in the
Q0042$-$2627 and PKS2126$-$158 fields; $[$WHO91$]$ 0043$-$265, with emission redshift
$z=3.45$, and Q212904.90$-$160249.0, with $z=2.94$. Archived
high-resolution spectra are also available for the central bright
quasars in the four fields without LBG data; these will be presented
in a future paper.

The observations of $[$WHO91$]$ 0043$-$265 and Q212904.90$-$160249.0 were taken on
the night of the 22nd of August 2007 using Keck/HIRES with the red
cross-disperser and C1 dekker, giving a slit width of 0.861 arcsec and
resolution of 6.7~\kms. Exposures were extracted and wavelength
calibrated using the \textsc{makee} package.


Fig.~\ref{fig:hires_spec} shows the final reduced echelle spectra.
In some of the spectra there are breaks in the wavelength
coverage. These are due to either separate wavelength settings that
did not overlap, gaps between the CCD detectors, or regions where
echelle orders were too wide to be completely recorded by the
detector.

\subsection{Lower resolution quasar spectra}

\label{sec:lower-resol-qsos}
G. Williger kindly provided us with electronic versions of the spectra
that overlap the Q0042$-$2627 LBG field
\citep{williger_large-scale_1996}. These spectra had been extracted,
wavelength calibrated and flux calibrated. They have a typical S/N of
$\sim20$ per $1$~\AA\ pixel, and resolution FWHM of $\sim
2$~\AA. Quasars for which these spectra are available are marked by
`Wil' in the comment field of Table~\ref{tab:qsos}.

The remaining low resolution spectra were obtained with AAOmega.  Each
night of the AAOmega observations, arcs and flat fields were taken for
every wavelength setting used. The central coordinates, exposure times
per grating for each field and observation dates are shown in
Table~\ref{tab:obsexptime}.  For the initial observations we used the
blue 1500V grating (resolution 3700) and red 1000R grating (resolution
3500) covering wavelengths 4230--6860~\AA.  However, for faint quasars
the S/N at wavelengths covering the \lya\ forest, 4230--5700~\AA, was
poorer than anticipated.  Therefore for the majority of our
observations we used the lower resolution blue (580V) and red (385R)
gratings.  Both have a resolution of 1300 and together provide a
wavelength range of 3750--8900~\AA. In addition to enabling better S/N
at wavlengths corresponding to the \lya\ forest, the larger wavelength
range for these gratings allowed us to identify more emission
features, and thus make more secure quasar identifications.

\begin{table*}
  \begin{center}
\begin{tabular}{rcccccccl}
  \hline
& & &  \multicolumn{5}{c}{Exposure time per grating (hours)} &  \\
 &              &            &  580V & 385R  &1500V  &1500V  &1000R & \\
  Field  & {R.A.}   & {Dec.} & c4800 & c7300 & c4625 & c5365 & c6280 & {Dates observed} \\
  \hline
  Q2359+0653    & 00:01:45.87  & $+$07:11:45.3 &  3.5  &  3.5  &   0   &   0   &   0  & 3 Jul 2008, 26 Oct 2008  \\
  Q0042$-$2627  & 00:44:34.00  & $-$26:11:33.0 &   0   &   0   &  2.0  &  2.5  &  4.5 & 10-12 Jul 2007  \\
  J0124+0044    & 01:24:03.77  & $+$00:20:32.7 &  6.0  &  6.0  &   0   &   0   &   0  & 24-25 Oct 2008 \\
  Q0301$-$0035  & 03:03:41.05  & $-$00:23:21.8 &  6.0  &  6.0  &   0   &   0   &   0  & 24-26 Oct 2008  \\
  HE0940$-$1050 & 09:42:52.77  & $-$11:04:19.9 &  1.5  &  3.5  &  5.0  &   0   &  3.0 & 18 Mar 2007, 5 Feb 2008  \\
  J1201+0116    & 12:01:43.90  & $+$01:16:00.1 &  1.5  &  2.5  &  1.5  &  1.5  &  3.0 & 10-12 Jul 2007, 5-6 Feb 2008 \\
  PKS2126$-$158 & 21:29:12.40  & $-$15:38:46.1 &  2.5  &  2.5  &  3.0  &  2.5  &  5.5 & 10-11 Jul 2007, 29 Jun 2008 \\
  Q2231$-$0015  & 22:34:09.16  & $+$00:00:05.0 &  3.5  &  3.5  &   0   &   0   &   0  & 30 Jun 2008, 3 Jul 2008  \\
  Q2348$-$011   & 23:50:57.90  & $-$00:34:10.0 &  6.5  &  6.5  &   0   &   0   &   0  & 2 Jul 2008, 24-26 Oct 2008  \\
  \hline
\end{tabular}
\caption{ Exposure times and central coordinates
  (J2000) for the AAOmega pointings.  Note the central pointings do
  not always coincide with the positions of the bright central quasars
  in Table~\ref{tab:brightqsos}. The total exposure times for each
  combination of grating and central wavelength are shown. The final
  column gives the dates when each field was observed.}
\label{tab:obsexptime}
\end{center}
\end{table*}

\subsection{Reduction of AAOmega spectra}
\label{sec:reduct-aaom-spectra}

AAOmega spectra were reduced using the \textsc{2dfdr}
program\footnote{http://www.aao.gov.au/AAO/2df/aaomega/aaomega\_manuals.html}. Each
set of AAOmega observations consists of science, arc and flat field
exposures for the blue and red gratings.  \textsc{2dfdr} processes
each science image by subtracting a combined bias image, dividing by a
combined flat field, tracing and rectifying the spectrum of each
object and generating a wavelength solution using an arc
exposure. Finally it extracts each spectrum, producing a 1-d spectrum
for each fibre. The RMS for the wavelength solution for an AAOmega
spectrum was typically 0.2 pixels, corresponding to $\sim
15$~\kms. Once the 1-d spectra were extracted, we used \textsc{2dfdr}
to combine spectra taken on the same night using the same grating and
wavelength setting into a single spectrum.

The final reduction steps removed the instrumental response from the
spectra and combined the multiple wavelength settings.  For our
analysis we are interested in the absorption properties along each
sightline, which do not require an accurate flux calibration. However,
we still performed an approximate correction for the instrumental
response to guide our object identification and continuum fitting. For
most of the AAOmega pointings we targeted a bright quasar with an
existing flux-calibrated spectrum in the literature.  We obtained an
instrumental response curve for this quasar by dividing the AAOmega
spectrum by the flux-calibrated spectrum, and applied this curve to
the rest of the AAOmega spectra in that pointing.  For pointings
without a flux-calibrated target, we used a response curve from a
similar observation taken during the same night.

For fields where we obtained spectra using both 1500V/1000R gratings
and 580V/385R gratings, we only used spectra from the lower resolution
gratings. The typical S/N per \AA\ in our combined spectra was $\sim
10$ at $R=19$, and $\sim 3$ at $R=21$.

\subsection{Measurement of quasar redshifts}
\label{sec:meas-quas-redsh}
We identified quasars in the required redshift range by their \lya,
\CIV, \SiIV\ and \CIII\ emission features and forest absorption. The
identifications were performed by eye. We measured quasar emission
redshifts by fitting a Gaussian profile to the \CIV\ emission line
where it could be measured, or \CIII\ when \CIV\ was not usable. Care
was taken to account for broad absorption that can shift the apparent
position of emission peaks for broad absorption line (BAL) quasars; in
these cases we fitted only the red wing of the emission line when
measuring the emission line position.


\subsection{Quasar continuum fitting}
\label{sec:quas-cont}

In order to perform the cross-correlation analysis, we require the
transmissivity in the \lya\ forest for each of the quasars. This is
defined as
\begin{equation}
T=\frac{f}{f_c},
\end{equation}
where $f$ is the measured flux and $f_c$ is the flux level of the
continuum (the intrinsic unabsorbed quasar spectrum) in the \lya\
forest. We therefore require an estimate of $f_c$ from the forest
profile. To find this we perform a continuum fitting method based on
that of \citet{young_high-resolution_1979} and
\citet{carswell_observations_1982}.

First the quasar spectrum is split into wavelength intervals and the
mean and standard deviation are calculated within each
interval. Pixels that fall below the mean by more than an arbitrary
factor $n$ times the standard deviation are rejected, and the mean and
standard deviation are re-calculated using the remaining pixels. This
process is repeated iteratively until the remaining pixel fluxes show
an approximately Gaussian distribution, with standard deviation equal
to the expected $1\sigma$ flux errors. With the continuum level
determined in these discrete intervals, a cubic spline was then used
to interpolate across the whole of the spectrum. Finally, this
continuum was adjusted by hand in regions where the fit still appeared
poor, generally over damped \lya\ absorption systems and emission
lines. $n$ was determined by trial and error; a typical value was
$1.2$, but the best value varied with the signal to noise ratio and
resolution of the spectrum, and inside and outside the
\lya\ forest. The widths of the wavelength intervals were similarly
chosen by trial and error.  Narrow intervals were required over
emission features and wider intervals were appropriate for the
\lya\ forest.

\begin{figure}
  \begin{center}
\includegraphics[width=0.5\textwidth]{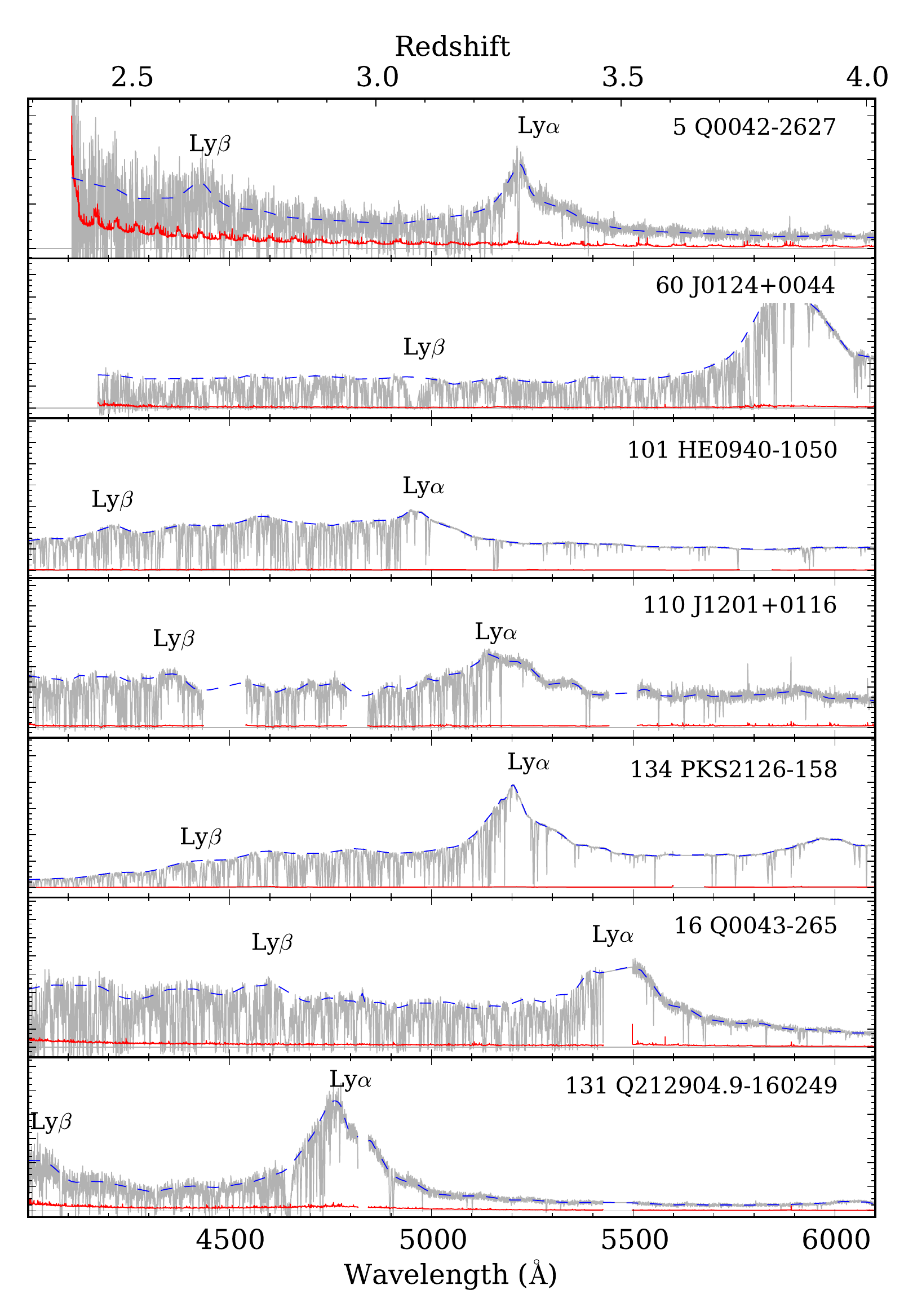}
\caption{\label{fig:hires_spec} High resolution spectra of quasars
  overlapping or near our LBG fields. The top five panels show
  archived spectra of the central bright quasars in our five fields
  with reduced LBG data.  The lower two panels show spectra of two
  non-central quasars we observed at high resolution: $[$WHO91$]$ 0043$-$265, in
  the Q0042$-$2627 field, and Q212904.90$-$160249.0, in the PKS2126$-$158
  field.  The quasar \lya\ and \lyb\ emission are labelled, and the
  number of each quasar from Table~\ref{tab:qsos} is listed before the
  quasar name. Continua (dashed blue curves) and the $1\sigma$ error
  per $\sim 0.05$\AA\ pixel are also shown.}
\end{center}
\end{figure}

\begin{figure}
  \centering
  \includegraphics[width=0.5\textwidth]{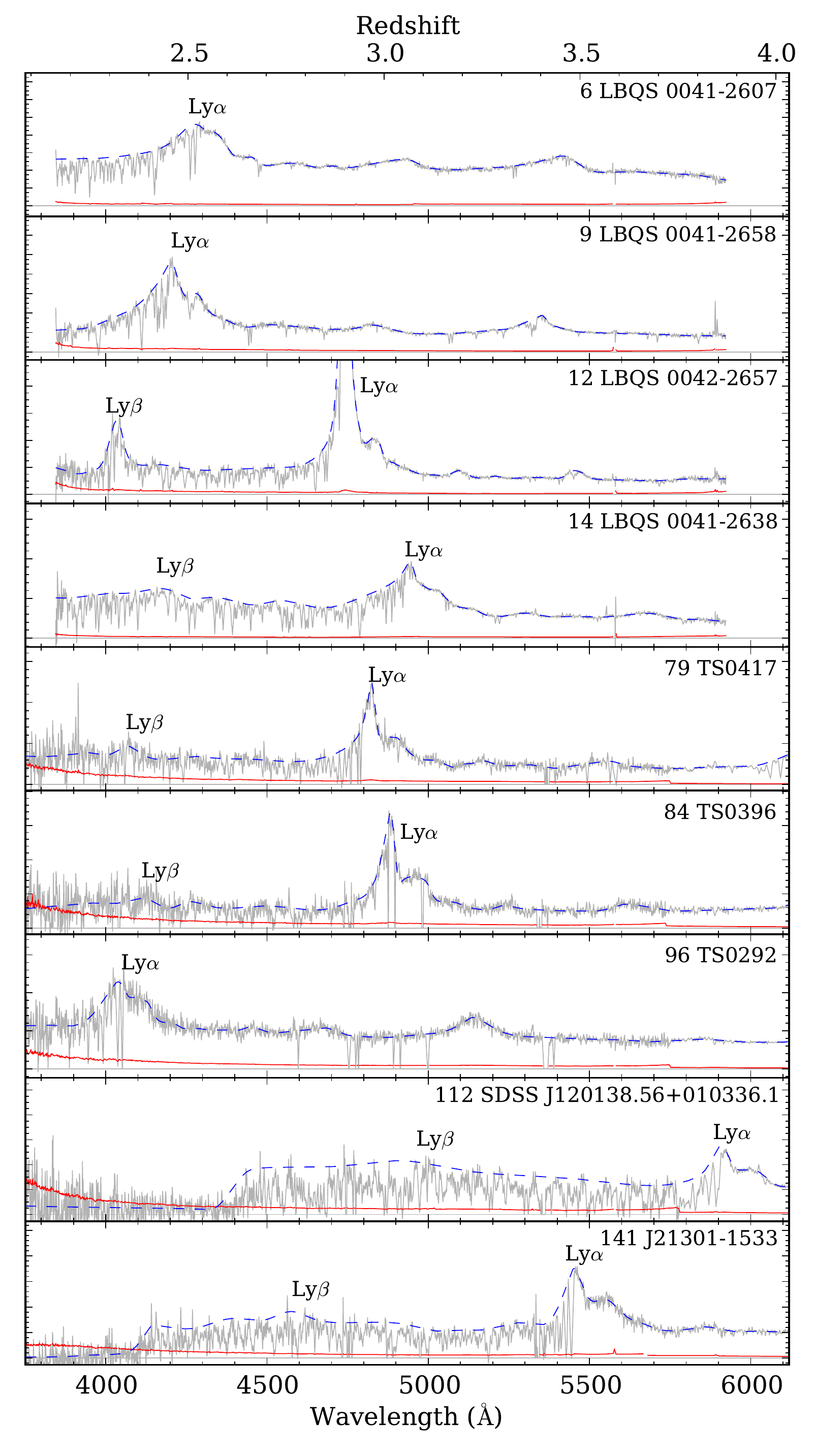}
  \caption{Low resolution spectra of quasars we used to measure the
    \lya\ transmissivity as a function of distance from LBG. Only
    quasars that contribute to the \lya-LBG cross-correlation on
    scales $< 20$~\hMpc\ are shown. The top four panels are spectra
    from Williger et al. (2000) in the Q0042$-$2627 field. The next
    three panels show our AAOmega spectra for three quasars in the
    HE0940$-$1050 field.  The final two spectra are for quasars in the
    J1201+0116 and PKS2126$-$158 fields. The quasar \lya\ and \lyb\
    emission are labelled, and the number of each quasar from
    Table~\ref{tab:qsos} is listed before the quasar name. Continua
    (dashed blue curves) and the $1\sigma$ error per $\sim1$~\AA\
    pixel are also shown.}
  \label{fig:lores_spec}
\end{figure}

The results of this fitting process for each of the central bright
quasars and two non-central bright quasars observed at high resolution
are shown in Fig.~\ref{fig:hires_spec}. Typical continua fitted to
the fainter AAOmega survey quasars are shown in
Fig.~\ref{fig:lores_spec}.

\subsection{Damped \lya\ and Lyman limit systems in the sample}

During our analysis of spectra in the first five LBG fields we
identified several candidate damped \lya\ systems (DLAs) and
sub-damped \lya\ systems [also known as super Lyman limit systems
  (SLLS)]. Regions of the \lya\ forest affected by strong DLA damping
wings were removed for the correlation analysis. Such systems are of
interest for potential follow-up studies with higher resolution
spectroscopy and we list them in Table~\ref{tab:dlas}. Candidate
systems were identified as strong \lya\ absorption features with two
or more associated metal features. We caution that the systems were
identified by eye in spectra of varying quality and S/N, and thus are
subject to selection biases.

\subsection{Quasar sub-sample used in the LBG-\lya\ cross-correlation}
\label{sec:xcorr_sample}

We constructed a sub-sample of quasars with a \lya\ forest suitable
for cross correlation with LBGs in the following way.  We began with
all quasars within $20\arcmin$ of a spectroscopically-confirmed
LBG. From these we removed any quasars that were clearly broad
absorption line quasars, showing very strong absorption in the blue
wing of their \CIV\ emission line.


In the remaining quasars we defined the \lya\ forest region used to
cross-correlate with the LBGs.  We used only the quasar spectral range
between the quasar \lyb\ and \lya\ emission lines. By discarding the
region below \lyb\ emission, we avoid any contamination of the
\lya\ lines by Lyman series lines from higher redshift forest
absorbers. We also excluded the range within 3000~\kms\ of the
quasar's \lya\ emission both to avoid absorbers affected by the
ionising radiation from the background quasar, and to minimise the
number of absorbers in our sample that may be ejected from the
background quasar.  Additionally, we removed any damped \lya\ systems
present in the spectra from the analysis.

We excluded any regions in either the low-resolution spectra where the
S/N at the continuum is very poor ($<3$ per pixel), or there were
clearly problems with the reduction, such as poor subtraction of sky
lines. For such very low S/N regions the reliability of the continuum
fit is likely to be low, and systematics associated with the data
reduction process to be significant. For some quasars the entire
\lya\ region was removed due to poor S/N.

Any remaining quasars after applying the criteria above formed the
sample we used to cross-correlate with LBG positions. This final
sample was split into two further sub-samples; seven quasars with high
resolution ($<10$~\kms) spectra and nine with low resolution
($>100$~\kms) spectra. As the systematic errors affecting the low
resolution and high resolution samples are different, we calculate the
cross-correlation for each sample separately.  The quasars in each
sample are given in Tables~\ref{tab:qsoscorrhires} and
\ref{tab:qsoscorrlores} and their spectra are shown in
Fig.~\ref{fig:hires_spec} and \ref{fig:lores_spec}. For five quasars
in the low resolution sample we use spectra taken in our AAOmega
survey and for the remaining four we use spectra from
\citet{williger_large-scale_1996}.

\begin{table*}
\begin{center}
\begin{tabular}{clcccccc}
\hline
{Num.} & {Name} & {Field} & {R.A. (J2000)} & {Dec (J2000)} & {$z$} & {Resolution} & {S/N}\\
 & & & & & & (\kms) & (per pixel)\\
\hline
  5&          Q0042$-$2627 & Q0042$-$2627  & 00:44:33.95 & $-$26:11:19.9 & 3.29 & 6.7 & 8 \\
 16&$[$WHO91$]$ 0043$-$265 & Q0042$-$2627  & 00:45:30.47 & $-$26:17:09.2 & 3.44 & 6.7 & 18 \\
 60&            J0124+0044 & J0124+0044    & 01:24:03.78 & $+$00:44:32.7 & 3.81 & 7.5 & 38 \\
101&         HE0940$-$1050 & HE0940$-$1050 & 09:42:53.50 & $-$11:04:25.9 & 3.05 & 7.5 & 90 \\
110&            J1201+0116 & J1201+0116    & 12:01:44.37 & $+$01:16:11.7 & 3.20 & 6.7 & 24 \\
131& Q212904.90$-$160249.0 & PKS2126$-$158 & 21:29:04.90 & $-$16:02:49.0 & 2.90 & 6.7 & 9 \\
134&         PKS2126$-$158 & PKS2126$-$158 & 21:29:12.15 & $-$15:38:40.9 & 3.27 & 7.5 & 100 \\
\hline
\end{tabular}
\caption{Quasars with high resolution spectra used to measure the
  cross-correlation between LBGs and the \lya\ forest. These comprise
  the high resolution sample used for the correlation function in
  Fig.~\ref{fig:lyalbg_hilo}.  Their spectra are shown in
  Fig.~\ref{fig:hires_spec}. Five of these (Q0042$-$2627, J0124+0044,
  HE0940$-$1050, J1201+0116 and PKS2126$-$158) are quasars at the
  centre of an LBG field. The last column gives the approximate S/N
  per pixel over the \lya\ forest.}
\label{tab:qsoscorrhires}
\end{center}
\end{table*}

\begin{table*}
\begin{center}
\begin{tabular}{clccccc}
\hline
{Num.}   &{Name} &{Field}  & {R.A. (J2000)}  & {Dec (J2000)}  & {$z$} & {Resolution}\\
 & & & & &  & (\kms)\\
\hline
  6 &         LBQS 0041$-$2607   & Q0042$-$2627  & 00:43:58.80 & $-$25:51:15.7  & 2.50 & 120 \\
  9 &         LBQS 0041$-$2658   & Q0042$-$2627  & 00:44:05.85 & $-$26:42:04.4  & 2.46 & 120 \\
 12 &         LBQS 0042$-$2657   & Q0042$-$2627  & 00:45:19.57 & $-$26:40:50.9  & 2.90 & 120 \\
 14 &         LBQS 0041$-$2638   & Q0042$-$2627  & 00:43:42.79 & $-$26:22:10.2  & 3.05 & 120 \\
 79 &      Q094244.40$-$112138.7 & HE0940$-$1050 & 09:42:44.40 & $-$11:21:38.7  & 2.96 & 230 \\
 84 &      Q094342.99$-$105231.6 & HE0940$-$1050 & 09:43:42.99 & $-$10:52:31.6  & 3.01 & 230 \\
 96 &      Q094230.55$-$104850.8 & HE0940$-$1050 & 09:42:30.55 & $-$10:48:50.8  & 2.33 & 230 \\
112 & SDSS J120138.56+010336.1   & J1201+0116    & 12:01:38.564& $+$01:03:36.22 & 3.84 & 230 \\
141 &      Q213007.46$-$153320.9 & PKS2126$-$158 & 21:30:07.460& $-$15:33:20.90 & 3.46 & 230 \\
  \hline
\end{tabular}
\caption{Quasars with low resolution spectra in and around the LBG
  fields. Listed are all quasars that contribute to the \lya-LBG
  cross-correlation for the low resolution sample shown in
  Fig.~\ref{fig:lyalbg_hilo}. Their spectra are shown in
  Fig.~\ref{fig:lores_spec}. }
\label{tab:qsoscorrlores}
\end{center}
\end{table*}

\section{\CIV\ -- LBG cross-correlation}
\label{sec:civ}

Supernovae-driven winds are one of the processes believed to be able
to enrich the IGM with metals.  If they are a dominant mechanism for
enriching the IGM, we might expect to see metal-rich gas surrounding
LBGs, which are known to be undergoing significant star formation. The
cross-correlation between \CIV\ absorption systems and LBGs allows us
to examine the clustering of metal enriched gas around known
star-forming galaxies. 

By analysing the projected transverse correlation function A03 found
that on scales $< 5$~\hMpc, the clustering strength between LBGs and
\CIV\ systems was comparable to the LBG-LBG clustering strength for
log~$\NCIV \sim 13$, smaller for smaller log~\NCIV\ and larger for
larger log~\NCIV.  One explanation for this is that many of the
strongest \CIV\ systems arise in gas directly associated with LBGs
seen in the same data sample, perhaps a larger scale extension of the
winds inferred from \CIV\ absorption in the LBG spectra. The weaker
correlation at low \NCIV\ may be explained by \CIV\ arising in the
same large scale structures as LBGs. By measuring absorption from
foreground LBGs in the spectrum of a nearby ($<10$\arcsec) background
LBGs, A05 found that \CIV\ gas with log~$\NCIV> 12.5$ is found around
LBGs out to radius of 80~kpc. This analysis was recently extended to a
larger sample of higher resolution LBG spectra by
\citet{Steidel_structure_10}.  We defer a similar analysis of our LBG
spectra to a future paper. In this section we look for
\CIV\ absorption that may be associated with LBGs that are very close
to background quasar sightlines, and measure the \CIV-LBG
cross-correlation for our sample.

\subsection{Creating a \CIV\ absorption line catalogue}

We identified absorption due to the \CIV~$\lambda1548,1550$ doublet in
background quasar spectra in the following way. First we identified
significant absorption features in the quasar spectrum, then we
scanned each spectrum to identify possible features that were part of
a \CIV\ doublet. For the lower resolution AAOmega spectra we also
fitted Gaussian profiles to the absorption doublets. Finally we
measured the rest equivalent width of the \CIV~$\lambda1548$
transition for each \CIV\ system.

Significant absorption features were identified in a similar manner to
that described by \citet{schneider_hubble_1993}. The equivalent width
per pixel was calculated taking into account the instrumental
resolution, assumed to be a Gaussian profile. We identified all
features with a minimum significance level (the ratio of the
equivalent width to the equivalent width error) of four. We then
scanned each spectrum by eye, searching for possible \CIV\ systems
between the quasar \lya\ and \CIV\ emission lines. For the HIRES and
UVES spectra, once we identified \CIV\ systems, we measured the total
equivalent width of the \CIV~$\lambda1548$ transition. We considered
any \CIV\ absorption components that were separated by less than 500
\kms\ to be part of a single system. The lower resolution spectra
could not resolve individual \CIV\ components, so we fitted the
candidate systems identified by eye with Gaussian profiles. We checked
these fits were consistent with the relative oscillator strengths of
the transitions, taking into account possible line saturation. In
these lower resolution spectra we used the deblended profile for the
\CIV~$\lambda1548$ transition to measure a system's equivalent
width. Table~\ref{tab:civ} gives the redshifts and observed equivalent
widths for \CIV\ systems we identified towards the quasars in our
sample.

\subsection{\CIV\ close to LBGs}

The three LBGs in our sample closest to background sightlines have
proper impact parameters of 100, 140 and 150~$h^{-1}$~kpc. The two closest of
these do not show any \CIV\ absorption within 1000~\kms\ of the LBG
nebular redshift to 3$\sigma$ column density detection limits of
$10^{13}$~\cmm\ (100~kpc) and $10^{12}$~\cmm\ (140~kpc). For the
furthest there is a probable Lyman limit system with both high (\CIV,
\SiIV, \OVI) and low (\HI, \OI, \CII, \SiII, \SiIII) ionization
transitions $\sim500$~\kms\ away from the LBG redshift. The two
non-detections suggest that if the \CIV\ enveloping LBGs extends
beyond $\sim$100~kpc, its covering factor must be less than unity or
its column density lower than $10^{13}$~\cmm. The Lyman limit system
could be associated with the nearby LBG, but it could also be
associated with a fainter, closer galaxy that does not appear in our
survey. We intend to use the large number of transitions to explore
the physical conditions of this system in a future analysis \citep[see
  also][]{simcoe_observations_2006}.

\subsection{\CIV-LBG cross-correlation}

 To measure the 3-d comoving separation between an LBG and
 \CIV\ system, $\Delta s$, we first find the comoving distance $r$ to
 each object using
\begin{equation}
\label{eq:dc}
r = \frac{c}{H_0} \int^z_0 {dz' \over \sqrt{\Omega_m (1+z')^3 + \Omega_\Lambda}}\, ,
\end{equation}
where $c$ is the speed of light and $z$ is the redshift of the LBG or
absorber. The separation is then given by
\begin{equation}
\label{eq:ds}
  \Delta s = \sqrt{r_1^2 + r_2^2 - 2r_1r_2 \,cos\,\theta},
\end{equation}
where $r_1$ and $r_2$ are the comoving distances to the LBG and
absorber, and $\theta$ is their angular separation. We calculate the
cross-correlation using the ratio of the number of \CIV-galaxy pairs
in the real data to the number for a random distribution of
\CIV\ absorbers for different separation bins.  We generated a random
\CIV\ absorber catalogue in the following way: for each sightline
where we measured \CIV\ absorption, we generate $1000 \times N$ random
absorbers with redshifts drawn at random between the maximum and
minimum \CIV\ redshifts able to be detected along that sight-line,
where $N$ is the number of real absorbers found along that sightline.
We generated 1000 times more random absorbers to ensure the Poisson
noise introduced by the number of random pairs had a negligible
contribution to the final error estimate. This method assumes that the
detection limits do not change significantly along a single sightline,
which is a reasonable approximation for our spectra.

\begin{figure}
\begin{center}
\includegraphics[width=0.5\textwidth]{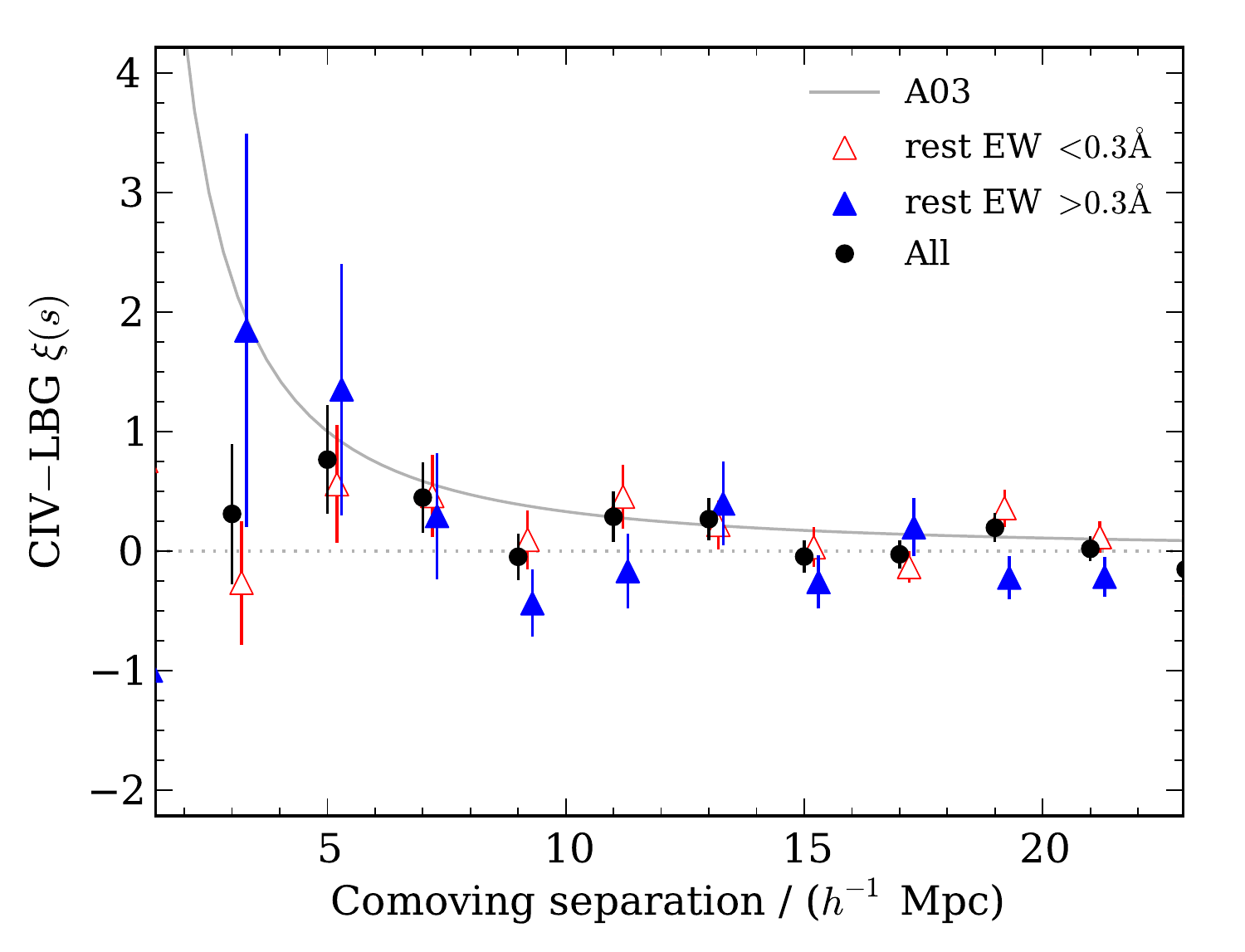}
\caption{ The cross-correlation of \CIV\ systems with LBGs as a
  function of 3-d distance between the \CIV\ absorbers and LBGs. The
  cross-correlation for the subsets of \CIV\ systems with rest
  equivalent widths $< 0.3$~\AA\ and $>0.3$~\AA\ are shown by open and
  filled triangles, each slightly offset for clarity. The model used
  in A05 to fit the \CIV-LBG cross-correlation at scales $< 5$~\hMpc,
  $\xi(r) = (r/5.0)^{-1.6}$, is shown by the solid line.}
\label{fig:CIVlbg}
\end{center}
\end{figure}

Fig.~\ref{fig:CIVlbg} shows the \CIV-LBG correlation function as
filled circles.  Our \CIV~$\lambda 1548$ rest equivalent width
distribution ranges from 0.005 to 2~\AA, with a median of 0.31~\AA, or
$\NCIV = 10^{13.9}$\cmm\ assuming unsaturated absorption.  At
separations $< 5$~\hMpc\ A05 fitted their \CIV-LBG correlation
function with a function of the form $\xi(r) = (r/r_0)^{-1.6}$. They
measured a clustering strength $r_0 \sim 5$~\hMpc\ between LBGs and
absorbers with $N \approx 10^{13.9}$\cmm, slightly higher than both
their and our LBG-LBG $r_0$ values \citep{Bielby10}. This
relation is shown in Fig.~\ref{fig:CIVlbg} as a solid curve. In an
attempt to increase statistical power, we measured the
cross-correlation in a single bin in the range $5 - 15$~\hMpc. This
yielded a correlation of $0.20 \pm 0.16$, where we assume a $1\sigma$
Poisson error from the number of absorbers contributing to this bin.
Thus the A05 relation is consistent with our measurement, but the
strength is too low for us to detect the clustering signal with our
sample size. We also split our sample into high and low equivalent
width sub-samples to measure the clustering strength as a function of
column density (solid and open triangles in
Fig.~\ref{fig:CIVlbg}). However, the results were inconclusive due to
the small number of galaxy-\CIV\ pairs. We are in the process of
assembling a larger sample of \CIV\ absorbers near LBGs using the
X-shooter spectrograph on the VLT.

\section{\lya\ auto-correlation}
\label{sec:lya_autocorr}

Before embarking on the \lya-LBG cross-correlation we measure the
\lya\ flux auto-correlation along quasar sightlines in our sample,
with the aim of measuring the velocity dispersion of \HI\ gas giving
rise to \lya\ absorption at the redshift of our sample. As will be
shown in the next section, any velocity smoothing in the redshift
direction has a large effect on our ability to detect a peak in the
\lya\ transmissivity around LBGs. If \HI\ gas does not share the
intrinsic velocity dispersion of nearby LBGs, then its own dispersion
will contribute to this velocity smoothing in the \lya-LBG signal.

Simulations show that at $z\sim 2.5$ and at scales where linear theory
holds, the \lya\ forest flux auto-correlation function is given by the
dark matter correlation function scaled by a constant and largely
scale-independent factor \citep[e.g.][]{croft_towardprecise_2002,
  slosar_acoustic_2009}. By comparing the measured flux correlation
function to the dark matter correlation, we can estimate the magnitude
of the \HI\ gas velocity dispersion by the size of the departure from
the expected correlation function on non-linear scales.

We measure the correlation function in the following way. For each
pixel in the \lya\ forest region of each quasar we calculate the
quantity
\begin{equation}
\delta = T/\Tbar - 1,
\end{equation}
where $T$ is the measured transmissivity and $\Tbar$ is the mean
transmissivity at the pixel redshift. To calculate the separation in
\hMpc\ between two pixels, we convert the redshift of each pixel to a
comoving distance using equation~\ref{eq:dc}. The correlation along
the sightline is then given by
\begin{equation}
\xi(\Delta r) = \langle \delta(r) \delta(r + \Delta r) \rangle,
\end{equation}
where $\langle \, \rangle$ denotes a sum over all pixels with a comoving separation
$\Delta r$. In practice we select some finite range of separations
around $\Delta r$ and include all pixels with separations that fall
inside that range.

\begin{figure}
\begin{center}
  \includegraphics[width=0.5\textwidth]{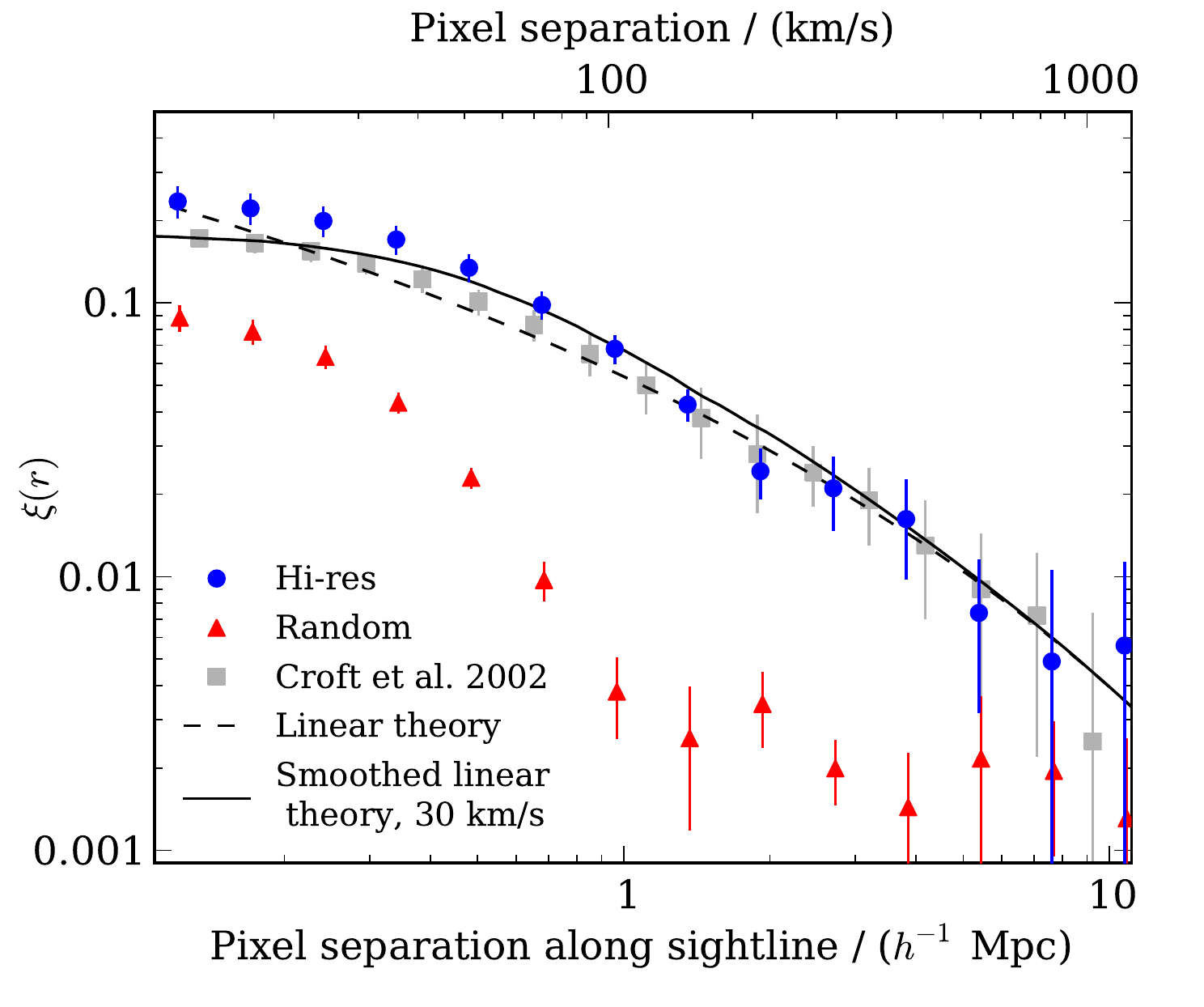}
  \caption{The auto-correlation of \lya\ forest pixels along quasar
    sightlines for our high resolution sample. Circles show our measured
    auto-correlation as a function of pixel separation along the
    quasar sightline. Separations are given in \hMpc\ (bottom) and
    \kms\ (top, using our assumed cosmology and $z=3$).  The
    correlation measured by \citet{croft_towardprecise_2002} at $z
    \sim 3$ are shown as pale grey squares. Triangles show the
    correlation measured using a set of mock spectra with thermal and
    instrumental broadening, but otherwise random, uncorrelated
    \lya\ forest absorption. Also shown is the linear theory matter
    correlation function, convolved in the redshift direction with a
    Gaussian velocity distribution with FWHM 30~\kms\ and scaled to
    match the observed correlation function. The velocity dispersion
    of \HI\ gas is small compared to velocity errors on the LBG
    positions.}
\label{fig:lya_xiz}
\end{center}
\end{figure}

Fig.~\ref{fig:lya_xiz} shows the \lya\ forest flux auto-correlation
from our high resolution quasar sample (resolution FWHM $\sim
7$~\kms). We use this rather than the low resolution sample because it
makes the largest contribution to the LBG-\lya\ correlation at small
scales, and probes the \lya\ $\xi(\Delta r)$ down to $\sim 100$~kpc
scales. We masked any DLAs or regions with poor sky subtraction in the
spectra.  Error bars were estimated using a jackknife technique; we
calculate $\xi(\Delta r)$ seven times, each time removing a different
quasar from the sample, and the error is then given by the standard
deviation of these around the value from the full sample, times $(7-1)
= 6$. We compare our results to those of
\citet{croft_towardprecise_2002}, who measure the auto-correlation
using a sample of 30 high resolution ($\sim 8$~\kms), high S/N
spectra.  Our result is slightly higher than that of Croft et al.;
this is likely due to the different redshift ranges of our samples
($z\sim3$ for the Croft et al. sample and $z\sim 3.3$ for ours).

To explore the effect of intrinsic line broadening, instrumental
resolution, and incomplete wavelength coverage due to the removal of
parts of spectra affect by DLAs or sky lines, we generate synthetic
spectra and measure their correlation. Each synthetic spectrum has a
\lya\ forest generated by adding absorption lines with a redshift, $b$
parameter and column density drawn at random from the distributions
over the redshift range corresponding to an observed spectrum (see
Appendix~\ref{ap:random}). This sample of synthetic spectra has
\lya\ forest lines placed at random such that they reproduce the mean
flux and $b$ parameter distributions, which are well known from large
samples of high-resolution spectra. However, the synthetic spectra do
not show any correlation in absorption other than that caused by line
broadening and instrumental effects. The auto-correlation for these
synthetic spectra is shown in Fig.~\ref{fig:lya_xiz} by
triangles. There is significant correlation at small separations,
mostly due to intrinsic broadening, but past separations of
70~\kms\ ($\sim0.7$~\hMpc) it is many times smaller than the signal
from the real \lya\ forest. Therefore, we believe that there are no
systematic effects in our sample that introduce a spurious
auto-correlation signal.

The black line in Fig.~\ref{fig:lya_xiz} shows the expected dark
matter correlation function from linear theory using $\sigma_8=0.8$,
multiplied by a factor of 0.08 and convolved with a Gaussian velocity
distribution of width $30$~\kms\ (note we did not include any
contribution from gravitational infall).  Much larger dispersions,
$\sim 100$~\kms, flatten $\xi(\Delta r)$ at small separations, and are
not consistent with the data. We conclude that the velocity dispersion
is $\sim 30$~\kms. This is much smaller than the measured velocity
dispersion of the LBGs, and so we do not consider it any further in
our analysis of the LBG-\lya\ cross-correlation.

\section{LBG -- \lya\ cross-correlation}
\label{sec:lyblya_xcorr}
To measure the LBG-\lya\ cross-correlation we must compare the
measured \lya\ transmissivity close to LBGs to the mean
\lya\ transmissivity. The mean transmissivity in the \lya\ forest
decreases with increasing redshift due to evolution of the UV ionising
background, structure formation, and the expansion of the
Universe. This is a significant effect; from $z=2.5$ to $z=3.5$ the
mean transmissivity drops from $\sim0.8$ to $\sim0.6$. We used two
methods for estimating the mean transmissivity as a function of
redshift. The first uses the measured mean transmissivity from
\citet{mcdonald_observed_2000},
\begin{equation}
  \Tbar = 0.676 - 0.220 (z - 3).
\end{equation}
This is measured from the mean flux along eight quasar sightlines and
was used by \citet{adelberger_galaxies_2003}. Over our redshift range
it is very similar to the more recent result by
\citet{faucher-gigure_direct_2008}.  Our second approach was to
measure $\Tbar$ in the \lya\ forest region for each individual quasar
sightline. For the high resolution and low resolution samples, we used
the $\Tbar$ estimate that gives the smallest quasar to quasar scatter
in the \lya-LBG cross correlation function. For the high resolution
sample, this is the McDonald relation, and for the low resolution
sample, the measured mean.  For the low resolution sample, we believe
the measured mean gives less scatter by compensating for errors in the
inferred continuum level over the \lya\ forest.  At the AAOmega
spectral resolution \lya\ forest lines are not resolved, and line
blending means that no part of a spectrum returns to the intrinsic
continuum level. Thus our continuum fitting process will likely be
offset from the true continuum level by $5-10$\%, and using the
measured mean flux minimises the effect of such an offset.

\subsection{Systematic effects}

There are two issues that could affect the \lya-LBG cross-correlation
measurement. In addition to \lya\ absorption, the forest region
contains absorption from metal transitions. Thus there will be a small
contribution to the measured transmissivity by metal absorption lines,
decreasing the transmissivity below the expected mean value. However,
we do not expect metal absorption at redshifts significantly different
from a galaxy's redshift to correlate with the galaxy
position. Therefore, we do not expect metal absorption to bias any
correlation signal, instead it will tend to reduce the strength of any
measured correlation.

We also cannot completely rule out a systematic offset in our LBG
redshifts. As we can only measure the redshifts for the ISM absorption
lines and \lya\ emission lines, which are affected by winds and
\HI\ absorption, respectively, we must infer the intrinsic redshift of
the galaxies using the relation from Adelberger et al. (2005). This
could introduce a systematic offset between our inferred LBG redshifts
and the true redshifts, and thus an offset between the LBG positions
and \lya\ absorption. A more recent relation between the \lya, ISM and
intrinsic LBG redshifts is given by \citet{Steidel_structure_10}.
However, this was calibrated using an LBG sample with $2 < z < 2.6$,
and our LBG distribution extends to $z \sim 3.5$. Thus for our
analysis we choose to use the A05 relation that was calibrated using a
$2 < z < 3.5$ LBG sample.  The best way to quantify any systematic
redshift offsets in our sample is to obtain NIR emission lines for
LBGs close to the quasar sightlines; we are pursuing such observations
for LBGs where these lines are observable.

\subsection{Measuring the cross-correlation}
\label{sec:meas-cross-corr}

We performed the cross-correlation using the normalised quasar
transmissivity profiles, $T'=T/\Tbar$.  We calculated the \lya-LBG
cross-correlation function in 0.5~\hMpc\ bins by measuring the mean
normalised transmissivity in all regions of the \lya\ forest enclosed
in spherical shell around each LBG with inner and outer radii given by
the bin edges. The separations, $s$, between each \lya\ pixel and an
LBG were calculated using equations~\ref{eq:dc} and \ref{eq:ds}, and
the bin size was chosen to match that used by A05.

The mean transmissivity in each bin was taken to be the mean of the
individual transmissivity values for each LBG contributing to that
bin.  The errors on each bin value were taken to be the standard error
in the mean of the LBG to LBG transmissivity values. We note that for
points where few LBGs contribute to a bin, this will probably
underestimate the error. Finally, we scaled the mean transmissivity
for the \lya\ forest of each quasar to 0.76 to enable a comparison
with the results of A05.

\subsection{Results}

\begin{figure}
\begin{center}
\includegraphics[width=0.5\textwidth]{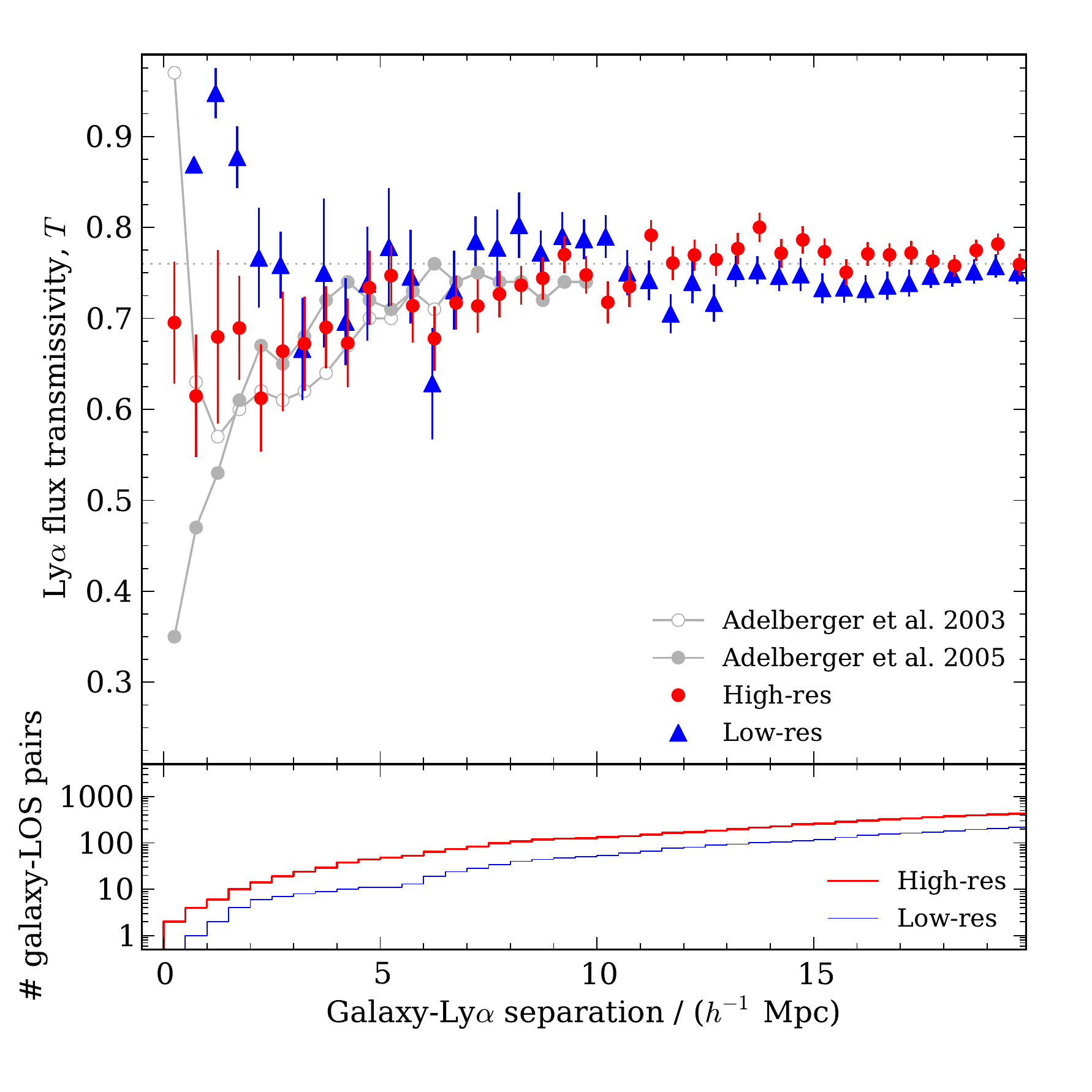}
\caption{A comparison between the high resolution ($\sim 40,000$) and
  low resolution ($\sim 1300$) quasar samples. The mean \lya\ forest
  transmissivity in quasar spectra plotted as a function of distance
  from the nearest LBG. Errors are the standard error on the mean of
  each transmissivity measurement from a single LBG-\lya\ forest
  region pair in a particular bin (see the discussion in the text).
  The results of Adelberger et al. 2003 and Adelberger et al. 2005 are
  also shown. The bottom panel shows the number of LBGs contributing
  to each bin in each sample.}
\label{fig:lyalbg_hilo}
\end{center}
\end{figure}

The VLT LBG-\lya\ cross-correlation function is shown separately for
the high resolution and low resolution quasar samples in
Fig.~\ref{fig:lyalbg_hilo}.  On scales $3-11$~\hMpc\ the two samples
agree.  There appears to be an offset between the low resolution and
high resolution samples at large separations, likely due to residual
continuum-fitting errors in the low-resolution spectra.

For transmissivities in the three bins with separations $<2$~\hMpc, the
low resolution sample increases above the mean, apparently becoming
inconsistent with the high resolution sample. However, we do not believe this
inconsistency is real; rather it is a result of small number
statistics.  In the three smallest separation bins, only four LBGs
contribute to the measured transmissivities.

The LBG-\lya\ cross-correlation function for the combined low and
high resolution samples is shown in Fig.~\ref{fig:lyalbg_combined}. At scales
$>2$~\hMpc\ the combined VLT result is consistent with both A03 and
A05 and the relationship seems reasonably well described by a
power-law, $T=0.77-(s/0.3$~\hMpc$)^{-1}$, which also describes the A05
results (see Fig.~\ref{fig:lyalbg_adelberger}) and the GIMIC
simulation results of Tummuangpak et al. (in preparation, see also
\citealt{Crain09}), even at smaller scales. The VLT data aims for good
statistics at large scales, so the errors are generally larger than
the A03 or A05 data at smaller scales. In the case of the first point
at $s=0.25$~\hMpc\ the VLT error may be underestimated by the simple
LBG-LBG error shown. In experiments where the two LBG redshifts were
perturbed randomly by a velocity error drawn from a Gaussian with
width 390~\kms, a larger error was obtained for this point by a factor
of $\approx3$. However, the transmissivity value is robust to changes
in the estimation procedure. For example, weighting by numbers of
\lya\ pixels contributing to a bin rather than the number of LBGs did
not significantly change this result. At the smallest scales,
$s<2$~\hMpc, our result lies between these two previous results, but
taking the errors into account there is no significant disagreement
with either.  We also note that our LBG sample has a mean redshift of
3, closer to A03 than the A05 mean redshift of 2.5. Thus it is
possible that evolution in the LBG-IGM relation from $z\approx3$ to
$z\approx2.5$ explains the difference between A03 and A05 as well as
the slightly better agreement of the VLT result with A03. But we
suggest that statistical fluctuations remain a more likely source of
any differences observed between these 3 datasets than evolution. In
the next section we examine the role that LBG velocity dispersion and
redshift errors play in the comparison and interpretation of these
results.

\subsection{Interpretation}

The distance between LBGs and \lya\ pixels is measured assuming we can
convert velocity differences into distances, not taking into account
any velocity dispersion or redshift errors. Intrinsic velocity
dispersion of the LBGs and the \HI\ gas, outflows, and LBG velocity
measurement errors will thus smear out any correlation between the two
along the redshift direction.  Bielby et al. analysed the LBG-LBG
correlation function and found it can be modelled with a real-space
correlation function of $\xi(r) = (r/r_0)^{-\gamma}$ with
$r_0=3.98$~\hMpc\ and $\gamma =1.9$, if convolved with a pairwise
velocity dispersion of $<w_z^2>^{1/2}=720$~\kms. In
Section~\ref{sec:lya_autocorr} we showed that the velocity dispersion
of the \HI\ gas is likely to be low, therefore we assume that the only
contribution to the LBG-\lya\ intrinsic velocity dispersion comes from
the LBGs. For a single LBG the velocity dispersion measured by Bielby
et al. is $720/\sqrt 2 = 510$~\kms, comprising 200~\kms\ for the
\lya\ emission line outflow error and 450~\kms\ for the VLT VIMOS
velocity measurement error, leaving 140~\kms\ for the intrinsic
velocity dispersion.

\begin{figure}
\begin{center}
\includegraphics[width=0.5\textwidth]{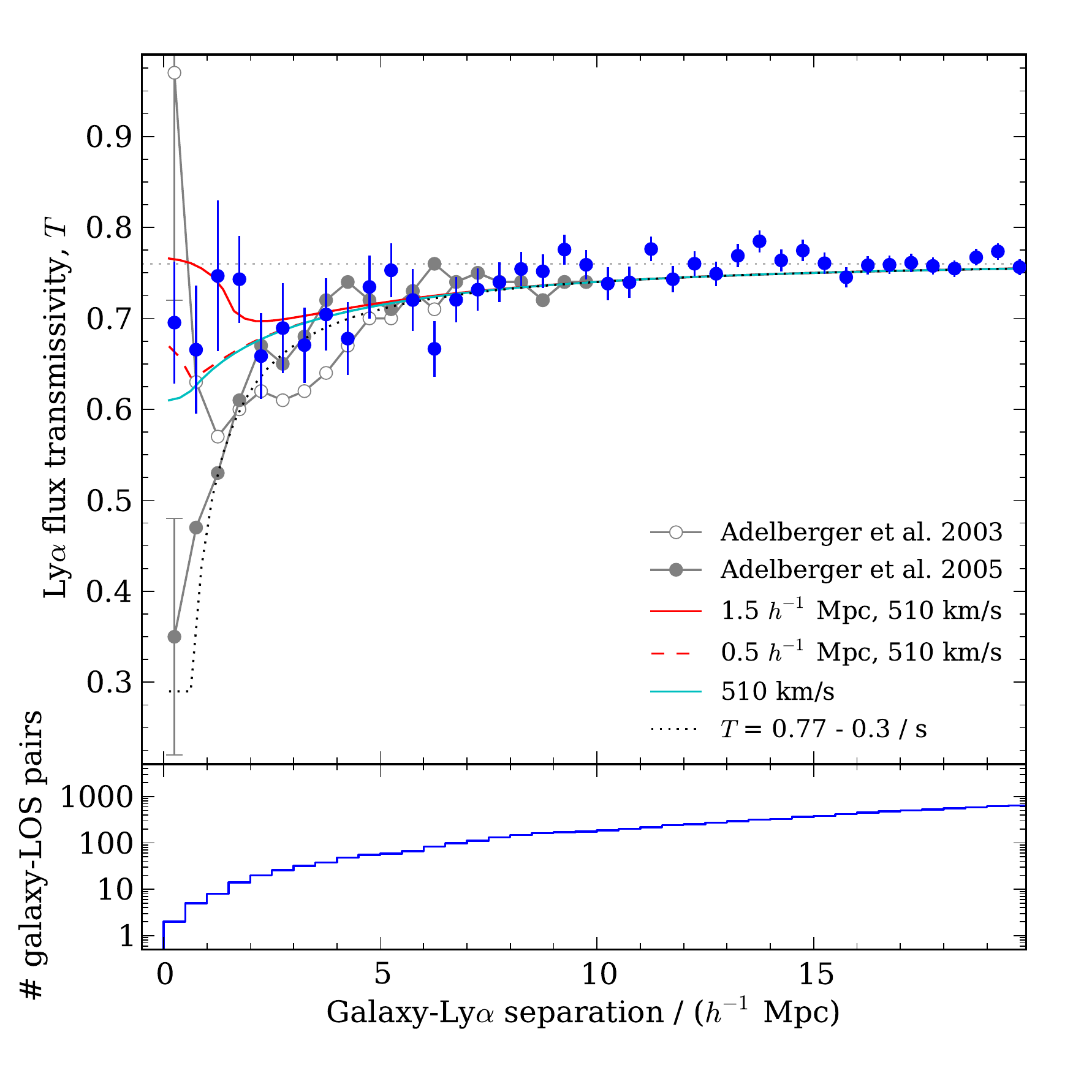}
\caption{The mean \lya\ forest transmissivity in quasar spectra
  plotted as a function of distance from the nearest LBG for the
  combined (high and low resolution) samples. We also show power law
  model correlation functions as described in the text. The dotted
  line shows the real-space power law described in the text without
  the effects of velocity dispersion. The cyan solid line shows this
  model convolved in the redshift direction with a velocity dispersion
  of $510$~/kms, the typical uncertainty on a LBG redshift for our
  sample. The red lines show the power law model with a real-space
  $T=1$ transmission spike at separations $<1.5$~Mpc (solid) and
  $<0.5$~Mpc (dashed), also convolved with a $510$~/kms/
  dispersion. Error bars for the values from A03 and A05 are only
  shown for the smallest separation point.}

\label{fig:lyalbg_combined}
\end{center}
\end{figure}

To model the effect of this dispersion, we have taken the real-space
LBG-\lya\ cross-correlation function that approximately fits the A05
results and GIMIC galaxy-\lya\ simulations (Tummuangpak et al., in
preparation), with
\begin{equation}
T(s) = \Tbar - (s/s_0)^{-1},
\end{equation}
where $s_0 = 0.3$~\hMpc. We convolved this in the redshift direction
only using a velocity dispersion of $510$~\kms. We do not include any
contribution from infall velocities. These are not negligible
\citep[e.g.][]{Padilla02}, but their contribution is likely to be
small compared to the redshift uncertainties and intrinsic velocity
dispersion. The result is shown in Fig.~\ref{fig:lyalbg_combined} by
the cyan solid line. The smoothing is considerable; this is expected
given that 510~\kms\ is $\sim 5$~\hMpc\ at $z=3$. We have also added a
central transmissivity spike to the above power-law model for $T(s)$
such that $T=1$ for $s<1.5$~\hMpc, simulating the case where high
transmissivity caused by putative galactic winds are found at small
scales (red solid line in Fig.~\ref{fig:lyalbg_combined}). Both
convolved results in Fig.~\ref{fig:lyalbg_combined} match the VLT
observations.  A spike of width 1.5~\hMpc\ is wider than the
sub-500~kpc spike suggested by A03's results. Thus we also consider a
narrower spike of width 0.5~\hMpc\ (red dashed line in
Fig.~\ref{fig:lyalbg_combined}). Such a spike has very little effect
on the correlation function for the uncertainties on our LBG
redshifts.

\begin{figure}
\begin{center}
\includegraphics[width=0.5\textwidth]{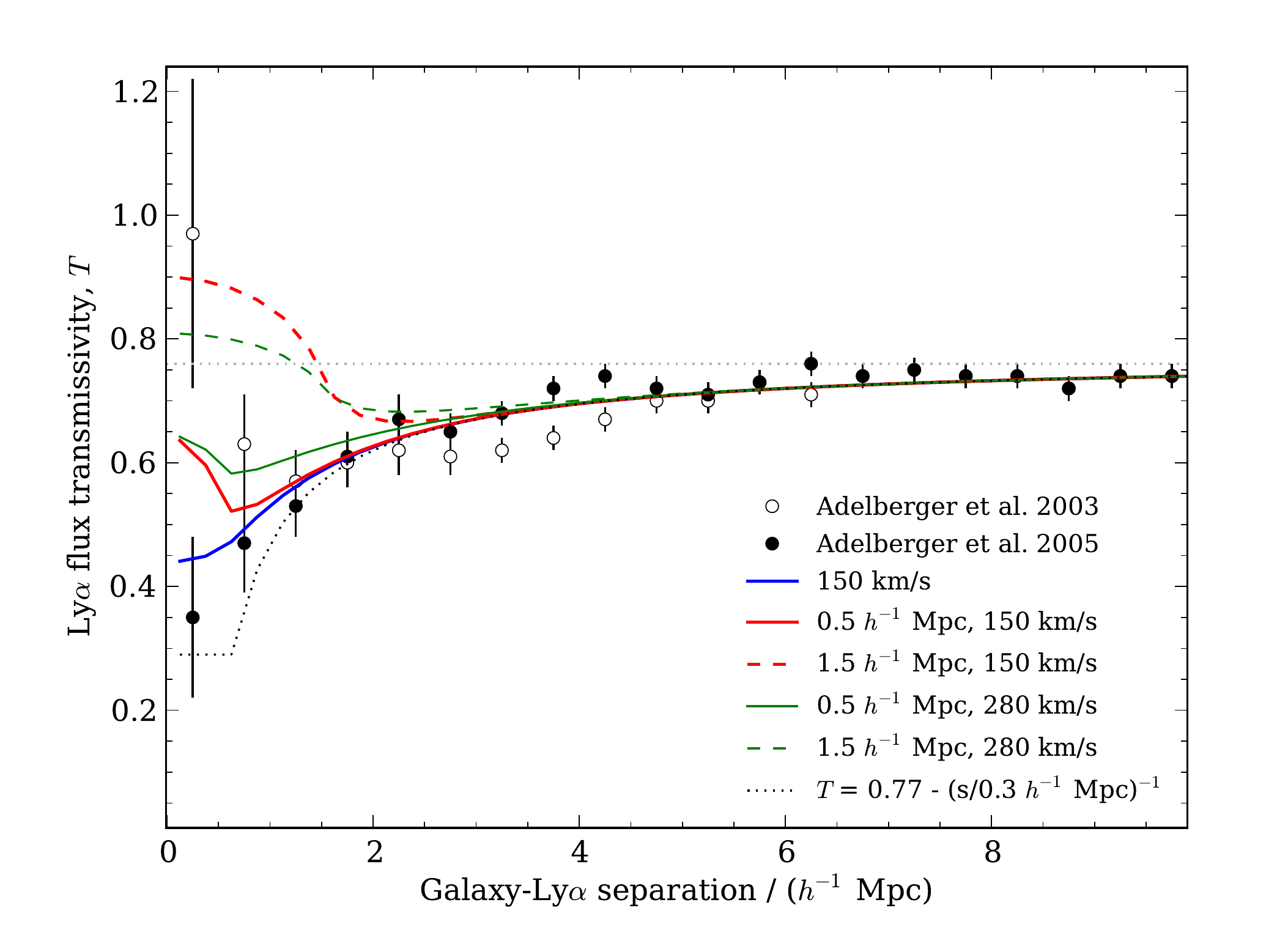}
\caption{The mean \lya\ forest transmissivity in quasar spectra
  plotted as a function of distance from the nearest LBG from
  Adelberger et al. (2003 \& 2005). Also plotted are power law model
  correlation functions as described in the text. The dotted line
  shows the unconvolved power law model, and the blue line shows this
  model convolved with the expected uncertainty for LBGs in the
  NIRSPEC sample of A05. The green and red lines show models with a
  transmission spike of width 1.5 or 0.5~Mpc, convolved with two
  different velocity dispersions, as described in the text.}
\label{fig:lyalbg_adelberger}
\end{center}
\end{figure}

The original results from A03 are also affected by velocity dispersion
and velocity errors, and we now estimate the correlation function that
would have been measured by A03 and A05 given the models above and the
redshift uncertainties of the Keck galaxy spectra. \citet{daAngela08}
fitted a pairwise velocity dispersion of 400~\kms\ to the Keck LBG-LBG
$z$-space correlation function. This converts to a dispersion of
280~\kms\ for a single LBG.  If we assume 200~\kms\ for outflow error
and 140~\kms\ for intrinsic velocity dispersion, this leaves
140~\kms\ velocity measurement error.  280~\kms\ translates to
2.8~\hMpc\ and so it is hard to see how a narrow spike width of
0.5~\hMpc\ seen by A03 in redshift-space could be physical. In
Fig.~\ref{fig:lyalbg_adelberger} we show how such a narrow spike,
shown by a green solid line, is almost smoothed away by this velocity
dispersion. Even the NIRSPEC $H_\alpha$ based LBG redshifts of A05,
which have 60~\kms\ velocity error and 140~\kms\ intrinsic velocity
dispersion will have 150~\kms\ or 1.5~\hMpc\ smoothing of the real
LBG-\lya\ $T(s)$. In Fig.~\ref{fig:lyalbg_adelberger} we show the
effect that such a velocity dispersion has on our real-space model
with (red solid line) and without (blue solid line) the narrow
spike. We also show a model with a broader 1.5~\hMpc\ spike as dashed
lines for each velocity dispersion.

We conclude that at $s<2$~\hMpc\ the VLT data could be consistent with
either the existence of the $T\approx1$ transmission spike as found by
A03 or the $T\approx0.3$ absorption found by A05. However, any $T=1$
spike is unlikely to be as narrow as the spike originally suggested by
the results of A03. The smoothing effect of even a 150~\kms\ velocity
dispersion on a spike of width $\approx0.5$~\hMpc\ is likely to be
significant.

The velocity dispersions we are using for the Keck data could be a
lower limit, since Bielby et al. find that a pairwise velocity
dispersion of 700~\kms\ best fits the LBG-LBG redshift space,
$\xi(\sigma, \pi)$, correlation function from the combined Keck LRIS
and VLT data. The best estimate from the Keck LRIS data alone is
similar. This would imply a much larger intrinsic pairwise LBG
velocity dispersion of $\approx 600$~\kms\ and an individual LBG
velocity dispersion of $\approx 500$~\kms. This would make the models
shown in Fig.~\ref{fig:lyalbg_adelberger} to be just as appropriate
for the Keck LRIS data as the VLT data, given the increasing dominance
of the intrinsic dispersion. This would reinforce the conclusion that
the apparent spike seen by A03 is unphysically narrow. It would also
suggest the absorption feature seen by A05 has the same problem in
being somewhat too narrow given the likely effect of the velocity
dispersion.

Alternatively, if the absorbing \HI\ gas has no net peculiar velocity
with respect to nearby LBGs, then the contribution of the intrinsic
velocity dispersion may be overestimated in the models we have
presented. However, for the VLT data the measurement error is as large
as any plausible intrinsic velocity dispersion, and even for the Keck
data with NIRSPEC redshifts the smoothing from measurement errors
alone is considerable. Thus even if LBGs and nearby \HI\ gas share the
same velocities our modelled real-space cross-correlation will not
change significantly.

\begin{figure}
\begin{center}
\includegraphics[width=0.5\textwidth]{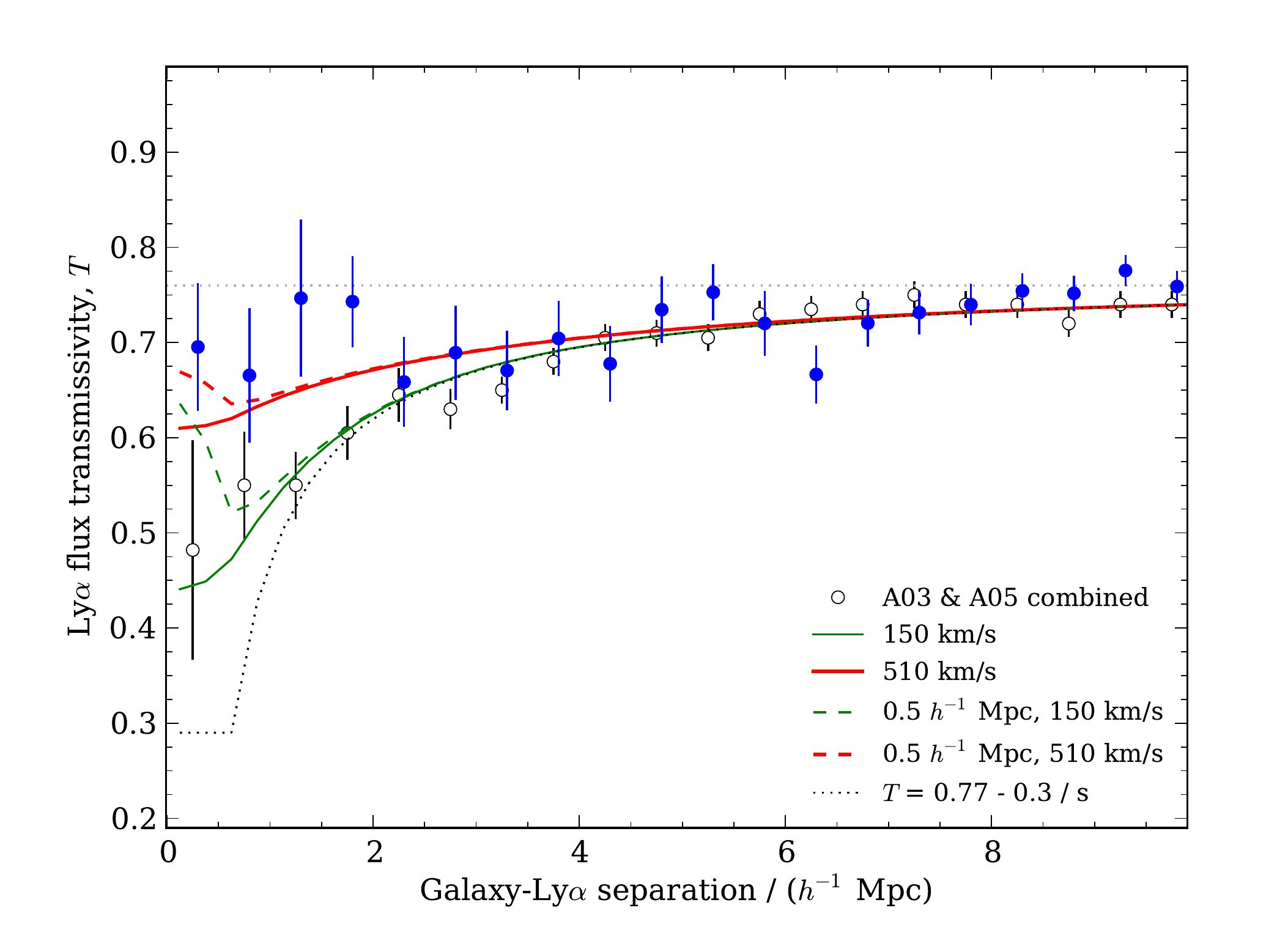}
\caption{The mean \lya\ forest transmissivity for the combined
  Adelberger (2005) sample, including LBGs with and without NIRSPEC
  redshifts, compared to our combined sample. Also plotted are power
  law model correlation functions as described in
  Fig. ~\ref{fig:lyalbg_combined} \&
  ~\ref{fig:lyalbg_adelberger}. Taking LBG velocity dispersion into
  account, the combined A05 results and our results are consistent
  with both a power law relation with no transmissivity spike, and the
  same relation with a narrow 0.5~Mpc transmissivity spike.}
\label{fig:lyalbg_adelberger_all}
\end{center}
\end{figure}

Despite these warnings about the effect of inter-comparing different
LBG-\lya\ samples with different velocity errors, we finally compare
our result to the combined A05 results, constructed by taking a
weighted mean of the \lya-LBG transmissivity results for LBGs with and
without NIRSPEC redshifts in A05.  We compare this to the combined VLT
result in Fig.~\ref{fig:lyalbg_adelberger_all}.

Although there is excellent agreement on scales $s>2$~\hMpc, the VLT
data remains slightly higher than the Keck data on scales
$s<2$~\hMpc. This could mean that there is still a hint of feedback in
the VLT result compared to the combined Keck data. However, if the
individual galaxy velocity dispersions are as low as 150~\kms\ for the
Keck (NIRSPEC) data and as high as 510~\kms\ for the VLT data, then
both datasets may be consistent with the $(s/0.3$~\hMpc$)^{-1}$ power
law transmissivity decline (dotted line) when it is convolved with the
respective velocity dispersions in the $z$ direction (red and green
solid lines). Since a 1.5~\hMpc\ wide, velocity convolved,
transmission spike is ruled out by the Keck data alone (see
Fig.~\ref{fig:lyalbg_adelberger}) the only other models consistent
with the data are the 0.5~\hMpc\ wide transmission spike models,
convolved with the appropriate velocity dispersions (green and red
dashed lines in Fig.~\ref{fig:lyalbg_adelberger_all}). In both models,
the transmission spike is suppressed by the velocity convolution such
that they are not ruled out by either the Keck or VLT data.

We conclude that the VLT data confirms the distribution of neutral
hydrogen around an LBG as indicated by the A05 results. Despite the
fact that the original A03 redshift-space spike is probably
unphysically narrow given the velocity effect and hence a likely
statistical fluke, a narrow, real-space transmission spike is
consistent with the data. Its existence would demand the presence of a
high absorption region in the immediate surroundings ($s<5$~\hMpc)
which would disguise the spike when both are convolved with the
velocity dispersion. Such an absorption region appears to be supported
by both the Keck and VLT data.  Models without such a
$<0.5$~\hMpc\ spike are equally consistent with both sets of data; in
this case the higher VLT result in
Fig.~\ref{fig:lyalbg_adelberger_all} can be explained by a higher LBG
velocity dispersion for the VLT data. However, a model with a
1.5~\hMpc\ $T=1$ spike seems to be rejected, particularly by the Keck
data.

Evidence for outflows from LBGs has also been seen in the systematic
offsets between the \lya\ and interstellar line redshifts (A03, A05,
Bielby et al.) and so it remains of interest to improve the
LBG-\lya\ data to check further for the transmission spike as evidence
of star-formation feedback on the IGM. In a future paper we therefore
intend to combine the full VLT and Keck LBG-\lya\ datasets to study
the LBG-\lya\ $\xi(\sigma, \pi)$ correlation function. We can then
test again for the presence of feedback by checking if any high or low
spike at small scales is more clearly seen in the angular direction,
where it will be less affected by velocity dispersion and redshift
errors.

Spectra with better resolution and S/N for those quasars with nearby
LBGs will improve the S/N of these results. We also need nebular line
redshifts for LBGs near quasar lines of sight to determine better the
width of any transmission spike in the VLT results at small
scales. More S/N will result from the forthcoming doubling in size of
our sample of VLT VIMOS LBGs in quasar survey fields. With this larger
sample it will be possible to divide the LBG-\lya\ cross-correlation
function into components in the redshift and angular directions, and
so search for signs of infall or outflow using anisotropies between
the redshift and angular correlation functions.

\section{Summary}
\label{sec:summary}

We have presented redshifts for 252 quasars across nine fields where
we are surveying LBG spectroscopic redshifts with VIMOS, 164 of which
are newly discovered. Using an initial sample of 1020 LBGs in five
fields and 16 of our background quasars with small impact parameters
from the LBGs, we have presented a measurement of the
LBG-\lya\ transmissivity and LBG-\CIV\ cross-correlation
functions. Using the 7 quasars in our sample with high-resolution
spectra we have also measured the \lya-\lya\ auto-correlation
function. The main results of our paper are:
\begin{itemize}

\item We have presented colour cuts that can be used with $ugr$
  imaging to select quasars in the redshift range $2.5 < z < 4$. With
  imaging similar to that available for Stripe 82 they yield an
  efficiency of 40\% at a sky density of $11.2$~deg$^{-2}$ for targets to
  $r = 22$.  With single epoch SDSS imaging, they yield an efficiency
  of 17\% at a sky density of $4.2$~deg$^{-2}$ to $r=21$.

\item We have identified \CIV\ absorption systems towards quasars
  inside the five fields where we have measured LBG redshifts. Using
  these systems we have measured the cross-correlation between
  \CIV\ absorbers and LBGs for $5-15$~\hMpc, and find it to be
  consistent with the power law relation measured by Adelberger et
  al. (2003) at smaller separations.

\item The \lya\ auto-correlation function can be qualitatively
  reproduced on non-linear scales by smoothing the expected dark
  matter correlation function with a Gaussian velocity dispersion
  distribution. The width of the velocity distribution is $\sim
  30$~\kms, much smaller than the typical velocity errors on our LBG
  positions, and thus they do not have a significant effect on the
  measured \lya-LBG cross-correlation.

\item We have measured the \lya-LBG cross-correlation for a sample of
  16 background quasars with small impact parameters from foreground
  LBGs.  The cross-correlation is consistent with the results of
  Adelberger et al. (2005). In particular, we also see a decrease in
  \HI\ transmission at \lya-LBG separations of $2-7$~\hMpc. This
  decrease can be described by a power law with index $\gamma = -1$.

\item Uncertainties in the LBG redshifts have a significant effect on
  the detectability of any \lya\ transmission spike around the
  galaxies.  We have examined the effect of velocity dispersion and
  redshift errors on the LBG-\lya\ cross correlation using simple
  models of a transmission spike, and found that with the measured LBG
  velocity dispersions a narrow transmission spike, even if present,
  is difficult to measure against the underlying power law. The A03
  and A05 results rule out a broad (1.5~\hMpc) transmission spike, but
  both their and our data are consistent with both a narrow
  (0.5~\hMpc) spike or no spike, once the velocity dispersion is taken
  into account.

\end{itemize}

In a future paper we intend to exploit our survey's wide field to
analyse the \lya-LBG cross-correlation function on $\sim 20$~Mpc
scales, in both radial (redshift) and transverse directions, using a
sample of $\approx2000$ LBGs.

\section{Acknowledgments}

NC thanks the Science and Technology Facilities Council for the
rolling grant that funds his position.

We thank the referee for their comments, G. Williger for providing us
with electronic versions of the spectra in his paper, G. Richards for
sharing photometrically selected quasar candidates from the SDSS with
us and Mike Read for scanning Schmidt $U$ plates for us. We also thank
Paul Dobbie and Rob Sharp for their help during AAOmega observations,
and advice on the use of \textsc{2dfdr}. This study used data from the
Keck Observatory archive and ESO archive.

This work was partially supported by the Consejo Nacional de
Investigaciones Cient\'ificas y T\'ecnicas and Secretaria de Ciencia y
T\'ecnica de la Universidad Nacional de C\'ordoba, and the European
Union Alfa II Programme, through LENAC, the Latin American European
Network for Astrophysics and Cosmology. DM, NP and LI are supported by
FONDAP CFA 15010003, and BASAL CATA PFB-06.

\bibliographystyle{./mn2e}

\clearpage

\appendix

\section[]{Efficiency of Quasar Selection}
\label{ap:effic}

How can the efficiency be improved for future surveys for $2.5<z<4.0$
quasars? Using our quasar sample we can identify optimised colour cuts
that attempt to recover a significant fraction of the available
quasars while maintaining a relatively high efficiency. We use the
three Stripe 82 fields for this purpose, where we have both single
epoch SDSS imaging and deep Stripe 82 imaging covering the entire
AAOmega fields, and we have a relatively high completeness level for
quasars with $21 < r < 22$. This will allow us to determine the effect
of using two different imaging depths (SDSS and Stripe 82) to select
quasars.

Optimised colour cuts for SDSS imaging are shown in
Fig.~\ref{fig:sdss_select} as solid black lines. These cuts were made
to recover as many of the observed $2.5 < z < 4.0$ quasars in our
sample, while maintaining a high efficiency. All stellar targets with
$r < 22$, and $u-g$, $g-r$ colours satisfying the optimised cuts are
selected as candidates.

For Stripe 82 imaging we devised similar cuts shown by black solid
lines in Fig.~\ref{fig:s82_select}. The region between the two Stripe
82 selection polygons is not included as it is populated by A and F
stars at brighter magnitudes ($r \lesssim 21$). At $r > 21$, we expect
to probe past even the most distant A and F dwarfs [assuming
  M$_R\sim2.1$ from \citet{Covey07}, and a maximum distance of 50~kpc;
  \citet{jiang_spectroscopic_2006} also mention this point]. Thus we
could in principle include this region in our selection for targets
with $21 < r < 22$, but we have not done so here for simplicity.

Table~\ref{tab:eff} shows the efficiencies for the two different
imaging catalogues (single epoch SDSS and Stripe 82) using the
optimised cuts, and for the photo-$z$ candidates selected from
single-epoch SDSS imaging for comparison. For each selection method we
define the efficiency as follows: given the total number of candidates
in an AAOmega field, $N_{\rm cand}$, the number of those we observed,
$N_{\rm obs}$, the number of candidates that were previously known to
be quasars with $2.5 < z < 4.0$, $N_{\rm known}$, and the number of
candidates that turned out to be new quasars in the same redshift
range, $N_{\rm new}$, then
\begin{equation*}
  {\rm efficiency} = \left( N_{\rm new} +  \frac{N_{\rm obs}}{N_{\rm cand}} N_{\rm known} \right) / N_{\rm obs}\, .
\end{equation*}
The second term inside the parentheses is required as we have observed
only a subset of all the candidates, and this subset is biased by
including known quasars.

Efficiencies to $R=22$ for the Stripe 82 catalogue are around 40\%,
and for the single epoch SDSS imaging 13\%, or 15\% for photo-$z$
selected candidates.  If we extrapolate the measured efficiencies to
candidates that were not observed, we expect these selection methods
to yield sky densities of 11.2 quasars per deg$^2$ for Stripe 82, 4.0
for photo-$z$ targets, and 9.1 for single epoch SDSS targets. This
extrapolation is reasonable for the Stripe 82 and photo-$z$ targets,
but is more uncertain for the SDSS targets, as our observed candidates
are highly skewed towards targets with $r < 21$. If the efficiency at
$r > 21$ is much lower than that at $r < 21$ (due to larger errors in
the $ugr$ photometry, for example), then the true efficiencies and sky
densities will be lower those calculated here. At $r < 21$ we have
observed a large fraction of the available candidates using all the
selection methods and so the extrapolation of the efficiencies is more
secure. In this case the single epoch SDSS selection efficiency is
17\%, yielding a sky density of 4.2 quasars per square degree.

We conclude that it is possible to recover a sky density of $\sim
11$~deg$^{-2}$ for quasars with $2.5 < z < 4$ and $R<22$ using simple
colour cuts and star/galaxy separation on imaging of a similar depth
to the combined Stripe 82 catalogues, at an efficiency of $\sim
40$\%. Jiang et al. performed a similar survey for quasars using
Stripe 82 imaging and they report an efficiency of 43\% in the
redshift range $2 < z < 3$ for targets with $g < 22.5$.  This is
consistent with our efficiency estimates, though over a lower redshift
range and to a slightly different magnitude limit. We note that it is
possible to recover higher densities: we achieved densities of $\sim
14.2$~deg$^{-2}$ in the LBG areas for this redshift range, but at a
much reduced efficiency of $\sim 5$\%.

\begin{table*}
  \centering

\begin{tabular}{crccccc}
  \hline
  {Source} & {Field} & {Total} & {Obs.} & {Known} & {New}  & {Eff.}  \\
                  &                & {Cand.} & {Cand.} & {QSOs} & {QSOs} &   \\
  \hline
  Stripe 82, & J0124+0044  & 94  & 69  & 4  & 21 & 0.35 \\
    $r<22$   & Q0301$-$0035  & 92  & 41  & 8  & 13 & 0.40 \\
             & Q2348$-$011   & 79  & 54  & 8  & 17 & 0.42 \\
             & Total       &     &     &    &    & 0.39 \\
             &             &     &     &    &    &      \\
  SDSS $ugr$,& J0124+0044  & 233 & 125 & 2  & 14 & 0.12 \\
     $r<22$  & Q0301$-$0035  & 308 & 126 & 8  & 10 & 0.11 \\
             & Q2348$-$011   & 151 & 84  & 5  & 12 & 0.18 \\
             & Total       &     &     &    &    & 0.13 \\
             &             &     &     &    &    &      \\
SDSS photo-z,& J0124+0044  & 85  & 72  & 3  & 8  & 0.15 \\
    $r<22$   & Q0301$-$0035  & 100 & 63  & 7  & 3  & 0.12 \\
             & Q2348$-$011   & 71  & 54  & 9  & 4  & 0.20 \\
             & Total       &     &     &    &    & 0.15 \\
             &             &     &     &    &    &      \\
  Stripe 82, & J0124+0044  & 40  & 31  & 4  & 10 & 0.42 \\
   $r<21$    & Q0301$-$0035  & 44  & 30  & 8  & 9  & 0.48 \\
             & Q2348$-$011   & 34  & 26  & 8  & 9  & 0.58 \\
             & Total       &     &     &    &    & 0.49 \\
             &             &     &     &    &    &      \\
SDSS $ugr$,  & J0124+0044  & 79  & 65  & 2  & 8  & 0.15 \\
    $r<21$   & Q0301$-$0035  & 115 & 90  & 7  & 7  & 0.14 \\
             & Q2348$-$011   & 44  & 41  & 5  & 6  & 0.26 \\
             & Total       &     &     &    &    & 0.17 \\
             &             &     &     &    &    &      \\
SDSS photo-z,& J0124+0044  & 37  & 33  & 3  & 5  & 0.23 \\
    $r<21$   & Q0301$-$0035  & 49  & 45  & 6  & 2  & 0.17 \\
             & Q2348$-$011   & 39  & 36  & 9  & 4  & 0.34 \\
             & Total       &     &     &    &    & 0.24 \\
             &             &     &     &    &    &      \\
SDSS $ugr$,  & J0124+0044  & 57  & 48  & 2  & 8  & 0.20 \\
  $20<r<21$  & Q0301$-$0035  & 84  & 69  & 4  & 6  & 0.13 \\
             & Q2348$-$011   & 31  & 29  & 3  & 6  & 0.30 \\
             & Total       &     &     &    &    & 0.19 \\
             &             &     &     &    &    &      \\
  \hline
\end{tabular}
\caption{ Efficiencies for selecting $2.5 < z < 4.0$ quasars for each
  candidate source.  For all but the photo-$z$ candidates, these are
  not the efficiencies we achieved in our observations, rather they
  are achievable efficiencies calculated using selection criteria
  applied after the observations. The columns show the total number of
  quasar candidates using the selection criteria described in
  Appendix~\ref{ap:effic}, the number of those candidates we observed,
  the number of candidates previously known to be quasars with $2.5 <
  z < 4.0$, the number of such quasars newly discovered in the
  observed candidates, and the efficiency as defined in
  Appendix~\ref{ap:effic}.  The total efficiency for each selection
  method is also shown.}
\label{tab:eff}
\end{table*}

\section[]{Generating synthetic quasar spectra}

One synthetic spectrum covering the \lya\ forest region was created to
match each observed spectrum. For each synthetic spectrum, \lya\
forest absorption was simulated by drawing many absorption lines with
redshifts, $b$ parameters and column densities drawn from the
distributions observed in high resolution spectra of the \lya\ forest.
In generating forest lines, we follow closely the procedure of
\citep{DallAglio08}. The distributions we use are as follows: for
line redshifts $dn / dz \propto (1+z)^{\gamma}$, where $\gamma =
2.37$; for column densities $f(\NHI) \propto \NHI^{-\beta}$, where
$\beta = 1.5$; and for Doppler $b$ parameters $dn/db \propto
b^{-5}{\rm exp}\left[{-{b_{\sigma}^4}/{b^4}}\right]$, where $b_{\sigma}=
24$~\kms. Values for the parameters $\gamma$, $\beta$ and
$b_{\sigma}$, are taken from \citet{Kim01}, who measure them from
the distribution of lines fitted to a sample of high-resolution quasar
spectra.

We add forest lines to the spectrum until the mean flux in the forest
region matches the measured value from McDonald et al. (2000) at the
mean redshift of the forest. This flux array is then convolved with
the instrumental spread function of the real spectrum. The forest
absorption is multiplied by a quasar continuum generated using
components derived from a principle component analysis of the
continuum of low-redshift quasars \citep{Suzuki05}. Finally we
added Gaussian noise to the spectrum, varying as a function of
wavelength in the same way as the real spectra.

Thus we expect the simulated spectra to reproduce the same S/N,
resolution, and wavelength ranges as the observed spectra, but with
uncorrelated absorption (apart from the correlations introduced by
intrinsic line broadening and the instrumental broadening). We did not
introduce metal lines or absorption from high column density systems
(Lyman limit and damped \lya\ systems) to the simulated spectra.

\section[]{Tables with Quasar, DLA/LLS candidate and \CIV\ information}
\label{ap:random}
\clearpage

\onecolumn

\begin{center}
\begin{longtable}{clccccccr}

\multicolumn{9}{c}{\sc Quasar Sample}\\
\\[-2ex]
\hline
\\[-2ex]
\multicolumn{1}{c}{{\#}} &
\multicolumn{1}{l}{{Name}} &
\multicolumn{1}{c}{{Field}} &
\multicolumn{1}{c}{{R.A.}} &
\multicolumn{1}{c}{{Dec.}} &
\multicolumn{1}{c}{{$z$}} &
\multicolumn{1}{c}{{Mag.}} &
\multicolumn{1}{c}{{O/lap?}} &
\multicolumn{1}{r}{{Comments}} \\
\\[-2ex]
\hline
\\[-2ex]
\endfirsthead
\multicolumn{9}{c}{{\tablename} \thetable{} -- Continued} \\
\\[-2ex]
\hline
\\[-2ex]
\multicolumn{1}{c}{{\#}} &
\multicolumn{1}{l}{{Name}} &
\multicolumn{1}{c}{{Field}} &
\multicolumn{1}{c}{{R.A.}} &
\multicolumn{1}{c}{{Dec.}} &
\multicolumn{1}{c}{{$z$}} &
\multicolumn{1}{c}{{Mag.}} &
\multicolumn{1}{c}{{O/lap?}} &
\multicolumn{1}{c}{{Comments}} \\
\\[-2ex]
\hline
\\[-2ex]
\endhead

\\
  \multicolumn{9}{l}{{Continued on Next Page\ldots}} \\
\endfoot

\\[-2ex]
\hline \\
\caption{Quasars with $z>2.1$ for which we have obtained AAOmega
  spectra. Coordinates are in J2000, and magnitudes are $R$ band. The
  second last column indicates whether the quasar overlaps the LBG
  survey fields. The final comments field shows the manner in which
  the quasar was selected. The abbreviations in the comments field
  are: Wil -- A spectrum for this quasar was available from Williger
  et al. (1996); NED -- selected from the NED database; VIMOS
  serendip. -- discovered in our VIMOS LBG observations; Stripe 82 --
  selected using Stripe 82 imaging; photo-$z$ -- this quasar was a
  candidate with a photometric redshift from Richards et al. (2004,
  2009); SDSS -- selected using single epoch SDSS imaging; Schmidt
  -- selected using Schmidt plate imaging; Central Imaging -- selected
  using the central imaging described in Table~\ref{tab:imselect}; WW
  -- taken from \citep{worseck_slitless_2008}; }
\label{tab:qsos}
\endlastfoot

  1 & $[$WHO91$]$ 0042$-$269	      & Q0042$-$2627 & 00:44:52.24& $-$26:40:09.3& 3.33 & 18.3 &n &  Wil NED $z=3.33$ \\
  2 & $[$WHO91$]$ 0042$-$266	      & Q0042$-$2627 & 00:44:35.72& $-$26:23:00.2& 2.98 & 19.5 &y &  Wil NED $z=2.98$ \\
  3 & $[$WHO91$]$ 0042$-$267	      & Q0042$-$2627 & 00:45:11.33& $-$26:25:50.7& 2.81 & 19.7 &y &    NED $z=2.81$   \\
  4 & $[$WHO91$]$ 0043$-$259	      & Q0042$-$2627 & 00:46:09.67& $-$25:38:47.2& 3.31 & 19.1 &n &    NED $z=3.31$   \\
  5 & Q0042$-$2627		      & Q0042$-$2627 & 00:44:33.95& $-$26:11:19.9& 3.29 & 18.5 &y &  Wil NED $z=3.289$ \\
  6 & LBQS 0041$-$2607	              & Q0042$-$2627 & 00:43:58.80& $-$25:51:15.7& 2.50 & 17.1 &n &  Wil NED $z=2.501$ \\
  7 & LBQS 0041$-$2707	              & Q0042$-$2627 & 00:43:51.84& $-$26:51:28.6& 2.79 & 17.9 &n &  Wil NED $z=2.786$ \\
  8 & $[$D87$]$ UJ3682P$-$038	      & Q0042$-$2627 & 00:46:41.65& $-$26:12:21.7& 2.48 & 19.2 &n &    NED $z=2.48$   \\
  9 & LBQS 0041$-$2658	              & Q0042$-$2627 & 00:44:05.85& $-$26:42:04.4& 2.46 & 18.6 &n &  Wil NED $z=2.457$ \\
 10 & $[$D87$]$ UJ3682P$-$028	      & Q0042$-$2627 & 00:45:11.33& $-$26:58:32.1& 2.34 & 19.8 &n &  Wil NED $z=2.34$ \\
 11 & $[$VCV96$]$ Q 0040$-$2606       & Q0042$-$2627 & 00:42:42.10& $-$25:50:09.0& 2.47 & 19.4 &n &    NED $z=2.47$   \\
 12 & LBQS 0042$-$2657	              & Q0042$-$2627 & 00:45:19.57& $-$26:40:50.9& 2.90 & 18.7 &n &  Wil NED $z=2.898$ \\
 13 & $[$VCV96$]$ Q 0045$-$2614       & Q0042$-$2627 & 00:47:45.07& $-$25:57:40.7& 2.35 & 19.3 &n &    NED $z=2.35$   \\
 14 & LBQS 0041$-$2638	              & Q0042$-$2627 & 00:43:42.79& $-$26:22:10.2& 3.05 & 18.3 &y &  Wil NED $z=3.053$ \\
 16 & $[$WHO91$]$ 0043$-$265	      & Q0042$-$2627 & 00:45:30.47& $-$26:17:09.2& 3.44 & 18.3 &y &  Wil NED $z=3.44$ \\
 17 & $[$WHO91$]$ 0046$-$267	      & Q0042$-$2627 & 00:48:48.65& $-$26:27:04.1& 3.52 & 19.7 &n &    NED $z=3.52$   \\
 18 & Q004340.02$-$260538.1	      & Q0042$-$2627 & 00:43:40.02& $-$26:05:38.1& 2.20 & 21.5 &y & VIMOS serendip.  \\
 20 & Q012715.19$+$001828.9	      & J0124+0044   & 01:27:15.19& $+$00:18:28.9& 2.27 & 21.6 &n &   Stripe 82 \\
 21 & Q012714.65$+$001650.3	      & J0124+0044   & 01:27:14.65& $+$00:16:50.3& 2.50 & 20.2 &n &     photo-$z=2.60$ \\
 22 & Q012730.08$+$001525.7	      & J0124+0044   & 01:27:30.08& $+$00:15:25.7& 2.70 & 21.4 &n &   Stripe 82 \\
 23 & Q012421.49$+$002158.3	      & J0124+0044   & 01:24:21.49& $+$00:21:58.3& 2.94 & 21.8 &n &   Stripe 82 \\
 24 & SDSS J012650.71+000933.3        & J0124+0044   & 01:26:50.71& $+$00:09:33.4& 3.42 & 20.9 &n &   NED $z=3.43$ \\
 25 & SDSS J012642.91+000239.0        & J0124+0044   & 01:26:42.91& $+$00:02:39.0& 3.22 & 19.7 &n &   NED $z=3.23$ \\
 26 & Q012617.91$-$000421.4	      & J0124+0044   & 01:26:17.91& $-$00:04:21.4& 2.77 & 20.7 &n &     photo-$z=2.44$ \\
 27 & SDSS J012658.10$-$001202.4      & J0124+0044   & 01:26:58.11& $-$00:12:02.5& 2.75 & 20.8 &n &   NED $z=2.76$ \\
 29 & Q012514.44$-$000342.2	      & J0124+0044   & 01:25:14.44& $-$00:03:42.2& 2.97 & 21.0 &n &   Stripe 82 \\
 30 & Q012530.89$-$001351.8	      & J0124+0044   & 01:25:30.89& $-$00:13:51.8& 2.58 & 20.6 &n &     SDSS \\
 31 & Q012528.25$-$002431.9	      & J0124+0044   & 01:25:28.25& $-$00:24:31.9& 2.38 & 21.3 &n &   Stripe 82 \\
 32 & Q012459.61$-$001600.7	      & J0124+0044   & 01:24:59.61& $-$00:16:00.7& 3.15 & 21.0 &n &   Stripe 82 \\
 33 & Q012429.48$-$000344.6	      & J0124+0044   & 01:24:29.48& $-$00:03:44.6& 3.36 & 22.0 &n &   Stripe 82 \\
 34 & Q012433.58$-$000335.6	      & J0124+0044   & 01:24:33.58& $-$00:03:35.6& 2.99 & 21.1 &n &     photo-$z=2.90$ \\
 35 & Q012426.25$-$001708.1	      & J0124+0044   & 01:24:26.25& $-$00:17:08.1& 2.67 & 21.2 &n &     photo-$z=2.90$ \\
 36 & Q012428.81$-$003835.5	      & J0124+0044   & 01:24:28.81& $-$00:38:35.5& 2.21 & 21.3 &n &     photo-$z=2.69$ \\
 37 & Q012355.95$-$001853.4	      & J0124+0044   & 01:23:55.95& $-$00:18:53.4& 3.12 & 20.5 &n &   Stripe 82 \\
 38 & Q012348.47$-$001538.8	      & J0124+0044   & 01:23:48.47& $-$00:15:38.8& 2.88 & 21.1 &n &     photo-$z=2.90$ \\
 39 & Q012217.48$-$002520.5	      & J0124+0044   & 01:22:17.48& $-$00:25:20.5& 2.48 & 21.2 &n &   Stripe 82 \\
 40 & Q012314.00$-$000534.4	      & J0124+0044   & 01:23:14.00& $-$00:05:34.4& 2.53 & 20.5 &n &     photo-$z=2.44$ \\
 41 & SDSS J012114.86$-$001637.3      & J0124+0044   & 01:21:14.86& $-$00:16:37.4& 2.38 & 19.2 &n &   NED $z=2.39$ \\
 42 & Q012200.31$-$000308.5	      & J0124+0044   & 01:22:00.31& $-$00:03:08.5& 2.23 & 21.5 &n &   Stripe 82 \\
 43 & SDSS J012226.76+000327.5        & J0124+0044   & 01:22:26.77& $+$00:03:27.5& 2.47 & 19.7 &n &   NED $z=2.48$ \\
 44 & Q012145.53$-$000208.6	      & J0124+0044   & 01:21:45.53& $-$00:02:08.6& 2.59 & 21.9 &n &   Stripe 82 \\
 45 & Q012040.64$-$000947.5	      & J0124+0044   & 01:20:40.64& $-$00:09:47.5& 2.30 & 21.9 &n &   Stripe 82 \\
 46 & Q012229.58$+$000849.1	      & J0124+0044   & 01:22:29.58& $+$00:08:49.1& 3.11 & 21.8 &n &   Stripe 82 \\
 47 & SDSS J012039.47$-$000239.4      & J0124+0044   & 01:20:39.47& $-$00:02:39.4& 2.54 & 19.5 &n &   NED $z=2.51$ \\
 48 & SDSS J012058.06+000205.0        & J0124+0044   & 01:20:58.07& $+$00:02:05.0& 2.94 & 20.5 &n &   NED $z=2.96$ \\
 49 & SDSS J012019.99+000735.5        & J0124+0044   & 01:20:20.00& $+$00:07:35.6& 4.07 & 20.0 &n &   NED $z=4.10$ \\
 50 & Q012101.58$+$002102.6	      & J0124+0044   & 01:21:01.58& $+$00:21:02.6& 2.36 & 20.5 &n &   Stripe 82 \\
 51 & Q012232.22$+$002321.2	      & J0124+0044   & 01:22:32.22& $+$00:23:21.2& 2.24 & 21.6 &n &   Stripe 82 \\
 52 & Q012028.15$+$004141.9	      & J0124+0044   & 01:20:28.15& $+$00:41:41.9& 2.97 & 20.6 &n &     photo-$z=2.90$ \\
 53 & SDSS J012052.64+004315.5        & J0124+0044   & 01:20:52.64& $+$00:43:15.6& 2.30 & 19.4 &n &   NED $z=2.30$ \\
 54 & Q012146.78$+$004645.2	      & J0124+0044   & 01:21:46.78& $+$00:46:45.2& 2.30 & 20.8 &n &     photo-$z=2.54$ \\
 55 & Q012229.16$+$004039.7	      & J0124+0044   & 01:22:29.16& $+$00:40:39.7& 2.60 & 21.4 &n &     SDSS \\
 56 & Q012203.15$+$010728.1	      & J0124+0044   & 01:22:03.15& $+$01:07:28.1& 2.65 & 21.4 &n &   Stripe 82 \\
 57 & Q012244.55$+$010604.4	      & J0124+0044   & 01:22:44.55& $+$01:06:04.4& 2.76 & 21.6 &n &   Stripe 82 \\
 58 & SDSS J012255.42+010315.3        & J0124+0044   & 01:22:55.42& $+$01:03:15.4& 3.46 & 20.8 &n &   NED $z=3.51$ \\
 59 & Q012351.00$+$005958.6	      & J0124+0044   & 01:23:51.00& $+$00:59:58.6& 2.59 & 21.5 &n &     photo-$z=2.79$ \\
 60 & J0124+0044		      & J0124+0044   & 01:24:03.78& $+$00:44:32.7& 3.81 & 17.9 &y &   NED $z=3.84$ \\
 61 & Q012523.94$+$004918.6	      & J0124+0044   & 01:25:23.94& $+$00:49:18.6& 2.46 & 21.9 &n &   Stripe 82 \\
 62 & Q012552.28$+$005827.6	      & J0124+0044   & 01:25:52.28& $+$00:58:27.6& 3.00 & 21.3 &n &     photo-$z=3.06$ \\
 63 & Q012549.03$+$005250.8	      & J0124+0044   & 01:25:49.03& $+$00:52:50.8& 2.98 & 19.8 &n &     photo-$z=2.90$ \\
 64 & Q012434.92$+$002834.5	      & J0124+0044   & 01:24:34.92& $+$00:28:34.5& 2.64 & 21.9 &n &   Stripe 82 \\
 65 & Q012635.64$+$004532.0	      & J0124+0044   & 01:26:35.64& $+$00:45:32.0& 2.62 & 21.0 &n &   Stripe 82 \\
 66 & Q012702.83$+$003707.4	      & J0124+0044   & 01:27:02.83& $+$00:37:07.4& 2.51 & 20.3 &n &     photo-$z=2.40$ \\
 67 & SDSS J012714.39+003249.6        & J0124+0044   & 01:27:14.39& $+$00:32:49.6& 2.38 & 20.5 &n &   NED $z=2.39$ \\
 68 & Q012558.89$+$002707.6	      & J0124+0044   & 01:25:58.89& $+$00:27:07.6& 2.38 & 20.2 &n &     photo-$z=2.40$ \\
 69 & SDSS J012753.69+002516.4        & J0124+0044   & 01:27:53.70& $+$00:25:16.4& 2.45 & 20.7 &n &   NED $z=2.46$ \\
 70 & Q094224.73$-$120222.9	      & HE0940$-$1050& 09:42:24.73& $-$12:02:22.9& 2.84 & 19.4 &n &    Schmidt \\
 71 & Q094208.20$-$112856.7	      & HE0940$-$1050& 09:42:08.20& $-$11:28:56.7& 2.47 & 21.0 &y &    Central Imaging      \\
 72 & Q094408.14$-$105040.0	      & HE0940$-$1050& 09:44:08.14& $-$10:50:40.0& 2.68 & 20.8 &n &    Central Imaging      \\
 73 & Q094331.59$-$111949.3	      & HE0940$-$1050& 09:43:31.59& $-$11:19:49.3& 2.61 & 21.3 &y &    Central Imaging      \\
 76 & Q094220.07$-$112215.9	      & HE0940$-$1050& 09:42:20.07& $-$11:22:15.9& 2.81 & 21.5 &y &    Central Imaging      \\
 77 & Q093956.82$-$111122.6	      & HE0940$-$1050& 09:39:56.82& $-$11:11:22.6& 2.81 & 19.4 &n &    Schmidt \\
 78 & Q094446.69$-$113512.8	      & HE0940$-$1050& 09:44:46.69& $-$11:35:12.8& 2.82 & 20.7 &n &     Central Imaging     \\
 79 & Q094244.40$-$112138.7	      & HE0940$-$1050& 09:42:44.40& $-$11:21:38.7& 2.96 & 19.6 &y &     WW $z=2.96$ \\
 80 & J0938535$-$105715	              & HE0940$-$1050& 09:38:53.50& $-$10:57:15.8& 2.45 & 17.9 &n &      NED $z=2.455$ \\
 81 & Q094349.59$-$112800.8	      & HE0940$-$1050& 09:43:49.59& $-$11:28:00.8& 3.48 & 20.7 &y &     Central Imaging     \\
 83 & Q094053.27$-$111107.2	      & HE0940$-$1050& 09:40:53.27& $-$11:11:07.2& 2.46 & 20.4 &n &     Central Imaging     \\
 84 & Q094342.99$-$105231.6	      & HE0940$-$1050& 09:43:42.99& $-$10:52:31.6& 3.01 & 19.6 &y &     WW $z=3.02$ \\
 85 & Q094130.12$-$113226.7	      & HE0940$-$1050& 09:41:30.12& $-$11:32:26.7& 3.00 & 20.8 &n &     Central Imaging     \\
 89 & Q094407.71$-$112632.2	      & HE0940$-$1050& 09:44:07.71& $-$11:26:32.2& 2.83 & 19.9 &y &     Central Imaging     \\
 90 & Q094436.51$-$110217.6	      & HE0940$-$1050& 09:44:36.51& $-$11:02:17.6& 2.90 & 21.8 &n &    Central Imaging      \\
 91 & Q094357.66$-$105435.1	      & HE0940$-$1050& 09:43:57.66& $-$10:54:35.1& 3.00 & 20.8 &n &    Central Imaging      \\
 92 & Q094400.94$-$114757.5	      & HE0940$-$1050& 09:44:00.94& $-$11:47:57.5& 2.90 & 19.5 &n &     Schmidt   \\
 93 & Q094252.79$-$112707.6	      & HE0940$-$1050& 09:42:52.79& $-$11:27:07.6& 3.15 & 20.8 &y &     Central Imaging     \\
 94 & Q094400.38$-$112732.7	      & HE0940$-$1050& 09:44:00.38& $-$11:27:32.7& 2.56 & 18.7 &y &     Central Imaging     \\
 95 & Q094330.04$-$104958.7	      & HE0940$-$1050& 09:43:30.04& $-$10:49:58.7& 2.22 & 19.9 &y &     WW $z=2.22$ \\
 96 & Q094230.55$-$104850.8	      & HE0940$-$1050& 09:42:30.55& $-$10:48:50.8& 2.33 & 19.1 &y &     WW $z=2.33$ \\
 98 & Q094206.40$-$113227.3	      & HE0940$-$1050& 09:42:06.40& $-$11:32:27.3& 2.93 & 18.7 &n &      Schmidt \\
 99 & Q094446.00$-$113216.9	      & HE0940$-$1050& 09:44:46.00& $-$11:32:16.9& 3.00 & 21.7 &n &      Central Imaging \\
100 & Q094324.23$-$105332.8	      & HE0940$-$1050& 09:43:24.23& $-$10:53:32.8& 2.76 & 21.4 &y &     WW $z=2.76$ \\
101 & HE0940$-$1050	              & HE0940$-$1050& 09:42:53.50& $-$11:04:25.9& 3.05 & 16.6 &y &      NED $z=3.05$ \\
103 & Q094253.07$-$110456.4	      & HE0940$-$1050& 09:42:53.07& $-$11:04:56.4& 3.79 & 24.7 &y &  VIMOS serendip.\\
104 & 2QZ J120117.1+010045	      & J1201+0116 & 12:01:17.11& $+$01:00:46.0& 2.38 & 20.1 &n &    NED $z=2.38$ \\
105 & 2QZ J115948.5+003203	      & J1201+0116 & 11:59:48.61& $+$00:32:04.3& 2.27 & 22.1 &n &    NED $z=2.27$ \\
106 & Q120220.06$+$002242.1	      & J1201+0116 & 12:02:20.06& $+$00:22:42.1& 2.58 & 17.6 &n &          SDSS \\
107 & Q115840.06$+$014335.2	      & J1201+0116 & 11:58:40.06& $+$01:43:35.2& 2.98 & 21.1 &n &          SDSS \\
108 & Q120244.72$+$020528.5	      & J1201+0116 & 12:02:44.72& $+$02:05:28.5& 3.52 & 20.3 &n &          SDSS \\
109 & 2QZ J120529.7+012326	      & J1201+0116 & 12:05:29.72& $+$01:23:26.3& 2.51 & 20.5 &n &    NED $z=2.51$ \\
110 & J1201+0116		      & J1201+0116 & 12:01:44.37& $+$01:16:11.7& 3.20 & 17.4 &y &    NED $z=3.23$ \\
111 & 2QZ J120210.5+011543	      & J1201+0116 & 12:02:10.55& $+$01:15:44.3& 2.50 & 19.9 &y &    NED $z=2.50$ \\
112 & SDSS J120138.56+010336.1        & J1201+0116 & 12:01:38.56& $+$01:03:36.2& 3.84 & 20.1 &y &    NED $z=3.88$ \\
113 & SDSS J115923.69+015224.2        & J1201+0116 & 11:59:23.70& $+$01:52:24.0& 2.44 & 20.1 &n &    NED $z=2.44$ \\
114 & 2QZ J120222.6+010119	      & J1201+0116 & 12:02:22.68& $+$01:01:20.1& 2.28 & 20.2 &n &    NED $z=2.28$ \\
115 & 2QZ J120055.7+013430	      & J1201+0116 & 12:00:55.77& $+$01:34:30.8& 2.51 & 20.6 &n &    NED $z=2.51$ \\
116 & 2QZ J115949.8+004329	      & J1201+0116 & 11:59:49.84& $+$00:43:29.6& 2.71 & 20.0 &n &    NED $z=2.71$ \\
118 & Q120408.38$+$014507.5	      & J1201+0116 & 12:04:08.38& $+$01:45:07.5& 2.30 & 20.7 &n &    Photo-$z=2.325$ \\
119 & 2QZ J120148.0+002000	      & J1201+0116 & 12:01:48.04& $+$00:20:00.8& 2.83 & 20.4 &n &    NED $z=2.83$ \\
120 & 2QZ J120311.2+015209	      & J1201+0116 & 12:03:11.30& $+$01:52:09.9& 2.27 & 20.2 &n &    NED $z=2.27$ \\
121 & SDSS J120045.05+013953.3        & J1201+0116 & 12:00:45.06& $+$01:39:53.2& 2.23 & 19.2 &n &    NED $z=2.23$ \\
122 & Q120001.29$+$003432.7	      & J1201+0116 & 12:00:01.29& $+$00:34:32.7& 3.36 & 20.0 &n &           SDSS \\
123 & Q120034.67$+$011518.4	      & J1201+0116 & 12:00:34.67& $+$01:15:18.4& 2.62 & 24.2 &y & VIMOS serendip. \\
124 & Q120123.83$+$012115.7	      & J1201+0116 & 12:01:23.83& $+$01:21:15.7& 2.73 & 24.3 &y & VIMOS serendip. \\
125 & Q120104.76$+$012213.8	      & J1201+0116 & 12:01:04.76& $+$01:22:13.8& 2.91 & 22.8 &y & VIMOS serendip. \\
126 & Q120219.73$+$012534.8	      & J1201+0116 & 12:02:19.73& $+$01:25:34.8& 2.53 & 23.0 &y & VIMOS serendip. \\
127 & Q120139.01$+$011733.9	      & J1201+0116 & 12:01:39.01& $+$01:17:33.9& 3.73 & 21.6 &y & VIMOS serendip. \\
128 & Q213144.16$-$155746.4	      & PKS2126$-$158& 21:31:44.16& $-$15:57:46.4& 3.86 & 18.4 &n &     Schmidt    \\
129 & Q213141.42$-$160231.5	      & PKS2126$-$158& 21:31:41.42& $-$16:02:31.5& 2.14 & 18.2 &n &     Schmidt \\
130 & Q213054.40$-$160540.4	      & PKS2126$-$158& 21:30:54.40& $-$16:05:40.4& 2.58 & 19.8 &n &     Schmidt  \\
131 & Q212904.90$-$160249.0	      & PKS2126$-$158& 21:29:04.90& $-$16:02:49.0& 2.90 & 19.2 &n &     Schmidt  \\
132 & Q212719.00$-$161001.1	      & PKS2126$-$158& 21:27:19.00& $-$16:10:01.1& 2.54 & 19.7 &n &     Schmidt  \\
134 & PKS2126$-$158		      & PKS2126$-$158& 21:29:12.15& $-$15:38:40.9& 3.27 & 17.3 &y &    NED $z=3.26$ \\
135 & Q212658.46$-$150839.8	      & PKS2126$-$158& 21:26:58.46& $-$15:08:39.8& 2.19 & 20.2 &n &     Schmidt \\
136 & Q212732.20$-$151026.6	      & PKS2126$-$158& 21:27:32.20& $-$15:10:26.6& 2.29 & 19.8 &n &     Schmidt \\
138 & Q212910.85$-$152423.7	      & PKS2126$-$158& 21:29:10.85& $-$15:24:23.7& 2.43 & 20.3 &y &  WW $z=2.480$ \\
139 & Q212916.60$-$144542.6	      & PKS2126$-$158& 21:29:16.60& $-$14:45:42.6& 2.32 & 20.0 &n &     Schmidt  \\
140 & Q212922.08$-$150653.2	      & PKS2126$-$158& 21:29:22.08& $-$15:06:53.2& 2.40 & 18.6 &n &      Schmidt   \\
141 & Q213007.46$-$153320.9	      & PKS2126$-$158& 21:30:07.46& $-$15:33:20.9& 3.46 & 21.9 &y &  WW $z=3.487$ \\
142 & Q213201.80$-$153256.4	      & PKS2126$-$158& 21:32:01.80& $-$15:32:56.4& 2.74 & 17.8 &n &      Schmidt \\
143 & Q212920.40$-$153816.1	      & PKS2126$-$158& 21:29:20.40& $-$15:38:16.1& 3.81 & 23.4 &y &   VIMOS serendip. \\
145 & Q212812.15$-$154533.0	      & PKS2126$-$158& 21:28:12.15& $-$15:45:33.0& 3.64 & 22.8 &y &   VIMOS serendip. \\
146 & Q212814.66$-$155338.6	      & PKS2126$-$158& 21:28:14.66& $-$15:53:38.6& 3.17 & 22.7 &y &   VIMOS serendip. \\
147 & Q223732.18$-$001730.6	      & Q2231$-$0015 & 22:37:32.18& $-$00:17:30.6& 2.57 & 19.5 &n &    SDSS  \\
148 & Q223450.81$-$001743.1	      & Q2231$-$0015 & 22:34:50.81& $-$00:17:43.1& 2.80 & 21.1 &n &     SDSS \\
149 & Q223427.19$-$001419.9	      & Q2231$-$0015 & 22:34:27.19& $-$00:14:19.9& 2.90 & 21.8 &n &         Central Imaging \\
150 & SDSS J223421.08$-$004800.5      & Q2231$-$0015 & 22:34:21.09& $-$00:48:00.5& 3.02 & 19.9 &n &   NED $z=3.02$ \\
151 & SDSS J223359.82$-$005841.0      & Q2231$-$0015 & 22:33:59.82& $-$00:58:41.0& 2.41 & 19.7 &n &   NED $z=2.41$ \\
152 & Q223315.39$-$003146.5	      & Q2231$-$0015 & 22:33:15.39& $-$00:31:46.5& 2.79 & 20.3 &n &    SDSS  \\
153 & SDSS J223250.37$-$004623.9      & Q2231$-$0015 & 22:32:50.38& $-$00:46:24.0& 2.39 & 19.8 &n &   NED $z=2.39$ \\
154 & Q223334.36$-$000541.2	      & Q2231$-$0015 & 22:33:34.36& $-$00:05:41.2& 2.47 & 21.9 &y &         Central Imaging \\
155 & SDSS J223118.34$-$003321.4      & Q2231$-$0015 & 22:31:18.35& $-$00:33:21.4& 2.52 & 19.2 &n &   NED $z=2.52$ \\
156 & Q223256.67$-$001246.4	      & Q2231$-$0015 & 22:32:56.67& $-$00:12:46.4& 3.10 & 21.8 &n &         Central Imaging \\
157 & SDSS J223253.56$-$001119.4      & Q2231$-$0015 & 22:32:53.57& $-$00:11:19.4& 3.11 & 20.2 &n &   NED $z=3.11$ \\
158 & SDSS J223438.52+005730.0        & Q2231$-$0015 & 22:34:38.52& $+$00:57:30.0& 2.85 & 19.2 &n &   NED $z=2.85$ \\
159 & SDSS J223433.70+002846.9        & Q2231$-$0015 & 22:34:33.71& $+$00:28:47.0& 2.74 & 20.9 &n &   NED $z=2.74$ \\
160 & SDSS J223535.59+003602.0        & Q2231$-$0015 & 22:35:35.59& $+$00:36:02.1& 3.87 & 20.4 &n &   NED $z=3.87$ \\
161 & SDSS J223502.19+002141.8        & Q2231$-$0015 & 22:35:02.20& $+$00:21:41.8& 3.27 & 20.5 &n &   NED $z=3.27$ \\
162 & Q2231$-$0015		      & Q2231$-$0015 & 22:34:09.00& $+$00:00:01.7& 3.02 & 17.3 &y &   NED $z=3.02$ \\
163 & Q223440.85$+$000647.7	      & Q2231$-$0015 & 22:34:40.85& $+$00:06:47.7& 2.47 & 21.8 &y &         Central Imaging \\
164 & Q223706.63$+$002413.4	      & Q2231$-$0015 & 22:37:06.63& $+$00:24:13.4& 2.66 & 21.5 &n &     SDSS \\
165 & SDSS J223716.03+001914.5        & Q2231$-$0015 & 22:37:16.03& $+$00:19:14.5& 2.51 & 18.7 &n &   NED $z=2.51$ \\
166 & Q223515.41$+$000519.7	      & Q2231$-$0015 & 22:35:15.41& $+$00:05:19.7& 2.72 & 19.5 &n &     SDSS \\
167 & SDSS J235420.60$-$003534.0      & Q2348$-$011  & 23:54:20.60& $-$00:35:34.0& 3.15 & 20.3 &n &   NED $z=3.14$ \\
169 & Q235400.37$-$004203.5	      & Q2348$-$011  & 23:54:00.37& $-$00:42:03.5& 2.36 & 20.5 &n &     photo-$z=2.46$ \\
170 & Q235234.52$-$004006.1	      & Q2348$-$011  & 23:52:34.52& $-$00:40:06.1& 2.66 & 21.1 &n &  SDSS \\
171 & Q235351.86$-$004416.1	      & Q2348$-$011  & 23:53:51.86& $-$00:44:16.1& 2.82 & 20.2 &n &     photo-$z=2.75$ \\
172 & SDSS J235344.85$-$004621.5      & Q2348$-$011  & 23:53:44.85& $-$00:46:21.5& 2.23 & 19.8 &n &   NED $z=2.23$ \\
173 & Q235343.09$-$004904.3	      & Q2348$-$011  & 23:53:43.09& $-$00:49:04.3& 2.29 & 21.3 &n &   Stripe 82 \\
174 & Q235416.34$-$005808.3	      & Q2348$-$011  & 23:54:16.34& $-$00:58:08.3& 2.49 & 20.6 &n &     SDSS \\
175 & Q235252.99$-$005132.7	      & Q2348$-$011  & 23:52:52.99& $-$00:51:32.7& 2.48 & 21.2 &n &  SDSS \\
176 & SDSS J235213.08$-$004607.7      & Q2348$-$011  & 23:52:13.08& $-$00:46:07.8& 2.35 & 20.3 &y &   NED $z=2.33$ \\
177 & SDSS J235206.44$-$004606.6      & Q2348$-$011  & 23:52:06.44& $-$00:46:06.6& 2.73 & 20.7 &y &   NED $z=2.76$ \\
178 & Q235253.41$-$011752.3	      & Q2348$-$011  & 23:52:53.41& $-$01:17:52.3& 3.60 & 21.4 &n &         Central Imaging \\
179 & Q235213.16$-$011209.7	      & Q2348$-$011  & 23:52:13.16& $-$01:12:09.7& 3.26 & 20.9 &n &         Central Imaging \\
180 & Q235141.42$-$005127.4	      & Q2348$-$011  & 23:51:41.42& $-$00:51:27.4& 2.16 & 19.6 &y &         Central Imaging \\
181 & Q235201.36$-$011408.2	      & Q2348$-$011  & 23:52:01.36& $-$01:14:08.2& 3.12 & 20.4 &y &  SDSS \\
182 & Q235147.40$-$011955.7	      & Q2348$-$011  & 23:51:47.40& $-$01:19:55.7& 3.50 & 21.7 &n &         Central Imaging \\
183 & Q235119.94$-$011004.4	      & Q2348$-$011  & 23:51:19.94& $-$01:10:04.4& 2.14 & 21.9 &y &   Stripe 82 \\
184 & Q235119.47$-$011229.2	      & Q2348$-$011  & 23:51:19.47& $-$01:12:29.2& 2.94 & 20.1 &y &   SDSS \\
185 & UM 184			      & Q2348$-$011  & 23:50:57.88& $-$00:52:09.9& 3.01 & 18.7 &y &   NED $z=3.02$ \\
186 & SDSS J235053.54$-$004810.2      & Q2348$-$011  & 23:50:53.55& $-$00:48:10.2& 3.85 & 19.6 &y &   NED $z=3.85$ \\
187 & Q234946.70$-$012159.7	      & Q2348$-$011  & 23:49:46.70& $-$01:21:59.7& 2.80 & 21.6 &n &         Central Imaging \\
188 & Q235029.07$-$005134.8	      & Q2348$-$011  & 23:50:29.07& $-$00:51:34.8& 2.51 & 21.8 &y &   Stripe 82 \\
189 & Q234910.97$-$012046.9	      & Q2348$-$011  & 23:49:10.97& $-$01:20:46.9& 3.27 & 21.7 &n &         Central Imaging \\
190 & Q234901.15$-$012002.0	      & Q2348$-$011  & 23:49:01.15& $-$01:20:02.0& 2.73 & 21.1 &n &         Central Imaging \\
191 & Q234919.94$-$010727.0	      & Q2348$-$011  & 23:49:19.94& $-$01:07:27.0& 2.75 & 20.8 &y &     SDSS \\
192 & SDSS J235002.78$-$005332.9      & Q2348$-$011  & 23:50:02.78& $-$00:53:33.0& 2.44 & 19.0 &y &   NED $z=2.41$ \\
193 & SDSS J234921.56$-$005915.1      & Q2348$-$011  & 23:49:21.56& $-$00:59:15.2& 3.09 & 19.9 &y &   NED $z=3.11$ \\
194 & Q234906.09$-$010245.8	      & Q2348$-$011  & 23:49:06.09& $-$01:02:45.8& 2.89 & 21.6 &y &         Central Imaging \\
195 & Q234732.98$-$010451.6	      & Q2348$-$011  & 23:47:32.98& $-$01:04:51.6& 2.40 & 21.8 &n &   Stripe 82 \\
196 & Q234958.23$-$004426.4	      & Q2348$-$011  & 23:49:58.23& $-$00:44:26.4& 2.58 & 21.0 &y &         Central Imaging \\
197 & Q234801.41$-$005532.4	      & Q2348$-$011  & 23:48:01.41& $-$00:55:32.4& 2.62 & 20.5 &n &     photo-$z=2.48$ \\
198 & Q235025.07$-$003838.1	      & Q2348$-$011  & 23:50:25.07& $-$00:38:38.1& 2.73 & 21.3 &y &         Central Imaging \\
199 & Q234719.09$-$005750.2	      & Q2348$-$011  & 23:47:19.09& $-$00:57:50.2& 3.09 & 21.3 &n &     SDSS \\
200 & Q234730.76$-$005131.5	      & Q2348$-$011  & 23:47:30.76& $-$00:51:31.5& 2.62 & 20.1 &n &     SDSS \\
201 & Q234712.10$-$004620.2	      & Q2348$-$011  & 23:47:12.10& $-$00:46:20.2& 2.58 & 21.7 &n &   Stripe 82 \\
203 & SDSS J234850.78$-$003429.4      & Q2348$-$011  & 23:48:50.79& $-$00:34:29.5& 3.97 & 20.5 &n &   NED $z=4.01$ \\
205 & Q234758.67$-$002533.2	      & Q2348$-$011  & 23:47:58.67& $-$00:25:33.2& 3.03 & 21.9 &n &   Stripe 82 \\
206 & SDSS J234932.30$-$002614.3      & Q2348$-$011  & 23:49:32.31& $-$00:26:14.4& 2.50 & 20.8 &n &   NED $z=2.17$ \\
207 & Q234728.48$-$001458.1	      & Q2348$-$011  & 23:47:28.48& $-$00:14:58.1& 2.63 & 21.4 &n &     SDSS \\
209 & Q234830.27$-$000935.5	      & Q2348$-$011  & 23:48:30.27& $-$00:09:35.5& 2.21 & 21.4 &n &   Stripe 82 \\
210 & Q234754.57$-$000036.0	      & Q2348$-$011  & 23:47:54.57& $-$00:00:36.0& 2.42 & 21.1 &n &   Stripe 82 \\
211 & Q234823.34$+$000437.8	      & Q2348$-$011  & 23:48:23.34& $+$00:04:37.8& 2.66 & 20.9 &n &     photo-$z=2.44$ \\
212 & Q235017.53$-$002230.6	      & Q2348$-$011  & 23:50:17.53& $-$00:22:30.6& 2.51 & 21.7 &n &         Central Imaging \\
213 & Q234911.97$-$000012.5	      & Q2348$-$011  & 23:49:11.97& $-$00:00:12.5& 2.93 & 21.2 &n &     photo-$z=2.40$ \\
214 & Q234902.85$+$001243.1	      & Q2348$-$011  & 23:49:02.85& $+$00:12:43.1& 2.27 & 21.4 &n &   Stripe 82 \\
215 & Q235023.60$-$001958.6	      & Q2348$-$011  & 23:50:23.60& $-$00:19:58.6& 2.24 & 21.6 &n &   Stripe 82 \\
216 & Q234916.38$+$002011.2	      & Q2348$-$011  & 23:49:16.38& $+$00:20:11.2& 3.42 & 20.9 &n &     SDSS \\
217 & Q2348$-$011		      & Q2348$-$011  & 23:51:10.82& $+$00:02:21.2& 3.50 & 18.8 &n &   NED $z=3.53$ \\
218 & Q235125.93$+$001306.6	      & Q2348$-$011  & 23:51:25.93& $+$00:13:06.6& 2.71 & 20.8 &n &     photo-$z=2.54$ \\
219 & Q235109.53$-$000101.6	      & Q2348$-$011  & 23:51:09.53& $-$00:01:01.6& 3.12 & 20.8 &n &  SDSS \\
220 & Q235142.12$+$000653.6	      & Q2348$-$011  & 23:51:42.12& $+$00:06:53.6& 3.04 & 21.3 &n &  SDSS \\
221 & Q235220.25$+$001735.9	      & Q2348$-$011  & 23:52:20.25& $+$00:17:35.9& 3.33 & 21.6 &n &   Stripe 82 \\
222 & SDSS J235251.66+001746.6        & Q2348$-$011  & 23:52:51.66& $+$00:17:46.6& 2.53 & 20.4 &n &   NED $z=2.33$ \\
223 & SDSS J235230.16+000552.0        & Q2348$-$011  & 23:52:30.16& $+$00:05:52.1& 3.14 & 20.1 &n &   NED $z=3.15$ \\
224 & SDSS J235219.08$-$000012.1      & Q2348$-$011  & 23:52:19.09& $-$00:00:12.1& 3.42 & 20.3 &n &   NED $z=3.43$ \\
225 & Q235259.02$+$000142.5	      & Q2348$-$011  & 23:52:59.02& $+$00:01:42.5& 2.18 & 20.8 &n &     SDSS \\
226 & SDSS J235224.13$-$000951.0    & Q2348$-$011  & 23:52:24.14& $-$00:09:51.0& 2.73 & 19.5 &n &   NED $z=2.74$ \\
227 & Q235136.89$-$002429.8	      & Q2348$-$011  & 23:51:36.89& $-$00:24:29.8& 2.33 & 21.9 &n &   Stripe 82 \\
228 & Q235416.52$-$001233.6	      & Q2348$-$011  & 23:54:16.52& $-$00:12:33.6& 3.21 & 20.5 &n &     SDSS \\
229 & Q235112.66$-$003124.8	      & Q2348$-$011  & 23:51:12.66& $-$00:31:24.8& 2.49 & 21.4 &y &         Central Imaging \\
230 & SDSS J235406.31$-$002110.5      & Q2348$-$011  & 23:54:06.31& $-$00:21:10.6& 3.52 & 20.4 &n &   NED $z=3.54$ \\
232 & Q000258.98$+$070318.4	      & Q2359+0653 & 00:02:58.98& $+$07:03:18.4& 2.72 & 20.9 &n &   Central Imaging \\
233 & Q2359+068	                      & Q2359+0653 & 00:01:40.60& $+$07:09:54.0& 3.23 & 18.4 &y &    NED $z=3.23$ \\
236 & Q000033.06$+$070716.1	      & Q2359+0653 & 00:00:33.06& $+$07:07:16.1& 2.86 & 19.6 &n &   Central Imaging \\
237 & Q000059.34$+$070900.3	      & Q2359+0653 & 00:00:59.34& $+$07:09:00.3& 2.42 & 19.7 &y &   Central Imaging \\
239 & Q000127.48$+$071911.8	      & Q2359+0653 & 00:01:27.48& $+$07:19:11.8& 2.87 & 20.7 &y &   Central Imaging \\
240 & Q000123.13$+$071559.1	      & Q2359+0653 & 00:01:23.13& $+$07:15:59.1& 2.33 & 22.0 &y &   Central Imaging \\
241 & Q000035.31$+$072503.1	      & Q2359+0653 & 00:00:35.31& $+$07:25:03.1& 2.54 & 20.6 &n &   Central Imaging \\
244 & Q000134.45$+$072312.9	      & Q2359+0653 & 00:01:34.45& $+$07:23:12.9& 2.40 & 19.4 &y &   Central Imaging \\
245 & Q000130.89$+$080536.8	      & Q2359+0653 & 00:01:30.89& $+$08:05:36.8& 2.18 & 18.9 &n &   Central Imaging \\
246 & Q000200.30$+$072636.8	      & Q2359+0653 & 00:02:00.30& $+$07:26:36.8& 2.88 & 20.0 &n &   Central Imaging \\
247 & Q000137.67$+$071412.2	      & Q2359+0653 & 00:01:37.67& $+$07:14:12.2& 2.99 & 20.8 &y &   Central Imaging \\
248 & Q000234.97$+$071349.3	      & Q2359+0653 & 00:02:34.97& $+$07:13:49.3& 2.60 & 20.6 &y &   Central Imaging \\
249 & Q030512.26$-$002727.8	      & Q0301$-$0035 & 03:05:12.26& $-$00:27:27.8& 2.58 & 21.5 &n &   Stripe 82 \\
250 & SDSS J030707.09$-$004901.5      & Q0301$-$0035 & 03:07:07.10& $-$00:49:01.5& 3.18 & 20.6 &n &   NED $z=3.18$ \\
251 & Q030622.41$-$004720.1	      & Q0301$-$0035 & 03:06:22.41& $-$00:47:20.1& 2.31 & 21.0 &n &     photo-$z=2.31$ \\
252 & Q030542.94$-$004058.5	      & Q0301$-$0035 & 03:05:42.94& $-$00:40:58.5& 2.71 & 21.3 &n &     SDSS \\
253 & Q030536.21$-$010218.0	      & Q0301$-$0035 & 03:05:36.21& $-$01:02:18.0& 2.96 & 21.8 &n &   Stripe 82 \\
254 & SDSS J030543.44$-$010622.1      & Q0301$-$0035 & 03:05:43.45& $-$01:06:22.2& 2.85 & 20.2 &n &   NED $z=2.85$ \\
255 & Q030335.03$-$003247.9	      & Q0301$-$0035 & 03:03:35.03& $-$00:32:47.9& 2.49 & 21.1 &n &    Central Imaging \\
256 & Q030504.18$-$010522.1	      & Q0301$-$0035 & 03:05:04.18& $-$01:05:22.1& 2.92 & 21.9 &n &   Stripe 82 \\
257 & SDSS J030437.56-011452.9        & Q0301$-$0035 & 03:04:37.57& $-$01:14:53.0& 2.42 & 19.8 &n &   NED $z=2.42$ \\
258 & Q030340.93$-$003540.7	      & Q0301$-$0035 & 03:03:40.93& $-$00:35:40.7& 2.62 & 21.7 &n &         Central Imaging \\
259 & Q030335.09$-$011235.7	      & Q0301$-$0035 & 03:03:35.09& $-$01:12:35.7& 3.78 & 21.4 &n &     photo-$z=3.10$ \\
261 & Q030244.18$-$003817.5	      & Q0301$-$0035 & 03:02:44.18& $-$00:38:17.5& 2.98 & 19.8 &n &     SDSS \\
262 & Q030210.01$-$005823.0	      & Q0301$-$0035 & 03:02:10.01& $-$00:58:23.0& 2.74 & 20.9 &n &     SDSS \\
263 & SDSS J030201.51$-$005833.5      & Q0301$-$0035 & 03:02:01.51& $-$00:58:33.5& 2.46 & 20.1 &n &   NED $z=2.46$ \\
264 & Q0301$-$0035		      & Q0301$-$0035 & 03:03:41.05& $-$00:23:21.9& 3.23 & 17.6 &y &   NED $z=3.23$ \\
265 & Q030252.20$-$002852.3	      & Q0301$-$0035 & 03:02:52.20& $-$00:28:52.3& 2.23 & 20.1 &n &     SDSS \\
266 & Q030241.61$-$002713.6	      & Q0301$-$0035 & 03:02:41.61& $-$00:27:13.6& 2.81 & 20.1 &n &         Central Imaging \\
267 & SDSS J030054.53$-$003403.3      & Q0301$-$0035 & 03:00:54.53& $-$00:34:03.4& 2.49 & 21.1 &n &   NED $z=2.49$ \\
268 & SDSS J030232.00$-$002734.1      & Q0301$-$0035 & 03:02:32.00& $-$00:27:34.2& 2.40 & 19.3 &n &   NED $z=2.40$ \\
269 & SDSS J030054.61$-$003051.8      & Q0301$-$0035 & 03:00:54.61& $-$00:30:51.8& 2.31 & 19.9 &n &   NED $z=2.31$ \\
270 & Q030111.16$-$002800.4	      & Q0301$-$0035 & 03:01:11.16& $-$00:28:00.4& 2.48 & 20.3 &n &     photo-$z=2.40$ \\
271 & Q030017.55$-$002258.7	      & Q0301$-$0035 & 03:00:17.55& $-$00:22:58.7& 2.16 & 21.4 &n &     SDSS \\
272 & Q025948.03$-$001749.6	      & Q0301$-$0035 & 02:59:48.03& $-$00:17:49.6& 2.56 & 21.9 &n &   Stripe 82 \\
273 & Q025959.18$-$001156.7	      & Q0301$-$0035 & 02:59:59.18& $-$00:11:56.7& 2.15 & 21.4 &n &   Stripe 82 \\
274 & SDSS J025949.14$-$000825.1      & Q0301$-$0035 & 02:59:49.15& $-$00:08:25.1& 2.65 & 20.7 &n &   NED $z=2.65$ \\
275 & Q030036.17$-$000549.4	      & Q0301$-$0035 & 03:00:36.17& $-$00:05:49.4& 2.60 & 21.1 &n &     photo-$z=2.31$ \\
276 & Q030254.04$-$001637.4	      & Q0301$-$0035 & 03:02:54.04& $-$00:16:37.4& 3.00 & 20.2 &n &     SDSS \\
277 & Q030127.86$-$000307.0	      & Q0301$-$0035 & 03:01:27.86& $-$00:03:07.0& 2.23 & 21.2 &n &   Stripe 82 \\
278 & Q030031.52$+$001143.8	      & Q0301$-$0035 & 03:00:31.52& $+$00:11:43.8& 2.20 & 21.2 &n &     photo-$z=2.31$ \\
279 & Q030124.12$+$000445.4	      & Q0301$-$0035 & 03:01:24.12& $+$00:04:45.4& 2.60 & 20.9 &n &     photo-$z=2.44$ \\
280 & Q030109.02$+$001013.8	      & Q0301$-$0035 & 03:01:09.02& $+$00:10:13.8& 3.01 & 20.5 &n &     photo-$z=2.90$ \\
281 & Q030151.92$+$000558.5	      & Q0301$-$0035 & 03:01:51.92& $+$00:05:58.5& 2.87 & 20.6 &n &     SDSS \\
282 & Q030131.79$+$001514.0	      & Q0301$-$0035 & 03:01:31.79& $+$00:15:14.0& 2.66 & 20.4 &n &     SDSS \\
283 & Q030259.87$-$000915.4	      & Q0301$-$0035 & 03:02:59.87& $-$00:09:15.4& 3.39 & 21.9 &n &         Central Imaging \\
284 & SDSS J030222.08+000631.0        & Q0301$-$0035 & 03:02:22.09& $+$00:06:31.1& 3.33 & 20.8 &n &   NED $z=3.33$ \\
285 & Q030324.69$-$001231.4	      & Q0301$-$0035 & 03:03:24.69& $-$00:12:31.4& 2.72 & 20.6 &n &         Central Imaging \\
286 & SDSS J030335.42$-$002001.1      & Q0301$-$0035 & 03:03:35.45& $-$00:20:01.1& 2.72 & 19.9 &n &   NED $z=2.72$ \\
287 & Q030211.38$+$002228.6	      & Q0301$-$0035 & 03:02:11.38& $+$00:22:28.6& 2.79 & 21.1 &n &     photo-$z=2.90$ \\
288 & Q030437.88$+$002100.8	      & Q0301$-$0035 & 03:04:37.88& $+$00:21:00.8& 2.93 & 20.3 &n &     SDSS \\
289 & Q030433.52$+$001601.1	      & Q0301$-$0035 & 03:04:33.52& $+$00:16:01.1& 2.36 & 21.7 &n &   Stripe 82 \\
290 & SDSS J030435.32-000251.0        & Q0301$-$0035 & 03:04:35.33& $-$00:02:51.0& 3.05 & 20.3 &n &   NED $z=3.05$ \\
291 & Q030558.72$+$002529.3	      & Q0301$-$0035 & 03:05:58.72& $+$00:25:29.3& 2.47 & 21.8 &n &   Stripe 82 \\
292 & WW2006  B		              & Q0301$-$0035 & 03:05:16.96& $+$00:00:43.5& 2.81 & 21.3 &n &   WW \\
293 & Q030506.79$-$000049.7	      & Q0301$-$0035 & 03:05:06.79& $-$00:00:49.7& 3.04 & 21.9 &n &   Stripe 82 \\
294 & LBQS 0302-0019		      & Q0301$-$0035 & 03:04:49.86& $-$00:08:13.5& 3.29 & 17.5 &y &   NED $z=3.29$ \\
295 & SDSS J030505.90$-$000616.5      & Q0301$-$0035 & 03:05:05.90& $-$00:06:16.5& 3.45 & 20.8 &n &   NED $z=3.45$ \\
296 & Q030515.60$-$001614.2	      & Q0301$-$0035 & 03:05:15.60& $-$00:16:14.2& 2.29 & 19.9 &n &   WW \\
297 & SDSS J030707.45$-$001601.4      & Q0301$-$0035 & 03:07:07.46& $-$00:16:01.4& 3.70 & 20.1 &n &   NED $z=3.70$ \\
\end{longtable}
\end{center}

\begin{table}
\centering
\begin{tabular}{clccccrr}
\hline
{Num}& {Name}           &  {$z_{\rm qso}$}&{Mag.}& {Max. \NHI} & {$z_{\rm abs}$} & Metals     \\ 
\hline
 16 &     $[$WHO91$]$ 0043$-$265 & 3.44 & 18.3 & 19.8 & 2.817 & \CIV, \FeII, \AlII         \\
 25 & SDSS J012642.91+000239.0   & 3.22 & 19.7 & 20.7 & 2.886 & \CIV, \FeII                \\
 35 &      Q012426.25$-$001708.1 & 2.67 & 21.2 & 19.3 & 2.682 & \CIV, \SiIV, \SiII         \\
 37 &      Q012355.95$-$001853.4 & 3.12 & 20.5 & 19.7 & 2.880 & \CIV, \CII                 \\
 38 &     Q012348.47$-$001538.8  & 2.88 & 21.1 & 19.8 & 2.822 & \CIV, \SiII, \SiIII        \\
 59 &    Q012351.00$+$005958.6   & 2.59 & 21.5 & 19.9 & 2.510 & \CIV, \SiIV                \\
 70 &    Q094224.73$-$120222.9   & 2.84 & 19.4 & 20.0 & 2.498 & \CIV, \SiIV, \AlII         \\
 72 &    Q094408.14$-$105040.0   & 2.68 & 20.8 & 20.0 & 2.255 & \CIV, \AlII                \\
 91 &    Q094357.66$-$105435.1   & 3.00 & 20.8 & 19.9 & 3.020 & \CIV, \SiIV, \SiII, \AlII  \\
 92 &    Q094400.94$-$114757.5   & 2.90 & 19.5 & 20.5 & 2.821 & \CII, \AlII                \\
108 &    Q120244.72$+$020528.5   & 3.52 & 20.3 & 19.2 & 3.537 & \CIV, \SiIV                \\
122 &    Q120001.29$+$003432.7   & 3.36 & 20.0 & 19.2 & 3.267 & \CIV, \SiIV                \\
131 & Q212904.90$-$160249.0      & 2.90 & 19.2 & 19.7 & 2.162 & \CIV, \SiIV, \FeII, \AlII  \\
141 &      Q213007.46$-$153320.9 & 3.46 & 21.9 & 19.4 & 3.267 & \CIV, \SiIV                \\
141 &      Q213007.46$-$153320.9 & 3.46 & 21.9 & 19.1 & 3.478 & \CIV, \SiIV                \\
142 &      Q213201.80$-$153256.4 & 2.74 & 17.8 & 19.8 & 2.338 & \CIV, \SiIV, \FeII         \\
\hline
\end{tabular}
\caption{DLA and LLS candidates towards low-resolution quasars in our
  first five LBG fields. Columns show the quasar number, name,
  emission redshift and magnitude from Table~\ref{tab:qsos}, a $1
  \sigma$ upper limit on the candidate system's column density
  (estimated using a Voigt profile with a single component of velocity
  width $b=50$~\kms), the absorber redshift and associated
  metals. Candidates were selected by eye, and were required to have
  strong \HI\ absorption with detectable absorption from two or more
  associated metal transitions.}
\label{tab:dlas}
\end{table}

\onecolumn

\begin{center}
\begin{longtable}{clcccccc}

\multicolumn{8}{c}{\sc \CIV\ Sample}\\
\\[-2ex]
\hline
\\[-2ex]
\multicolumn{1}{c}{{\#}} &
\multicolumn{1}{l}{{Field}} &
\multicolumn{1}{c}{{quasar \#}} &
\multicolumn{1}{c}{{$z_{qso}$}} &
\multicolumn{1}{c}{{$z_{abs}$}} &
\multicolumn{1}{c}{{$\sigma z_{abs}$}} &
\multicolumn{1}{c}{{EW}} &
\multicolumn{1}{c}{{$\sigma$EW}} \\
\\[-2ex]
\hline
\\[-2ex]
\endfirsthead
\multicolumn{8}{c}{{\tablename} \thetable{} -- Continued} \\
\\[-2ex]
\hline
\\[-2ex]
\multicolumn{1}{c}{{\#}} &
\multicolumn{1}{c}{{Field}} &
\multicolumn{1}{l}{{Quasar \#}} &
\multicolumn{1}{c}{{$z_{qso}$}} &
\multicolumn{1}{c}{{$z_{abs}$}} &
\multicolumn{1}{c}{{$\sigma z_{abs}$}} &
\multicolumn{1}{c}{{EW}} &
\multicolumn{1}{c}{{$\sigma$EW}} \\
\\[-2ex]
\hline
\\[-2ex]
\endhead

\\
  \multicolumn{7}{l}{{Continued on Next Page\ldots}} \\
\endfoot

\\[-2ex]
\hline \\
\caption{\CIV\ systems identified towards quasars in both our high and
  low resolution samples. The columns give the unique absorber number,
  the field for the absorber, the number from Table~\ref{tab:qsos} of
  the quasar toward which the absorber was seen, the quasar's
  redshift, the absorber redshift and one sigma error, and the
  observed EW in \AA\ and one sigma error.}
\label{tab:civ}
\endlastfoot

1 & HE0940$-$1050 & 95 & 2.22 & 1.58144 & 0.00022 & 2.157 & 0.065 \\
2 & HE0940$-$1050 & 95 & 2.22 & 2.19780 & 0.00026 & 0.952 & 0.029 \\
3 & HE0940$-$1050 & 98 & 2.93 & 2.42767 & 0.00016 & 1.401 & 0.010 \\
4 & HE0940$-$1050 & 70 & 2.84 & 2.49789 & 0.00009 & 5.073 & 0.029 \\
5 & HE0940$-$1050 & 79 & 2.96 & 2.48853 & 0.00021 & 2.670 & 0.060 \\
6 & HE0940$-$1050 & 79 & 2.96 & 2.89225 & 0.00009 & 2.587 & 0.006 \\
7 & HE0940$-$1050 & 92 & 2.90 & 2.21464 & 0.00020 & 1.335 & 0.019 \\
8 & HE0940$-$1050 & 81 & 3.48 & 2.56623 & 0.00014 & 4.665 & 0.077 \\
9 & HE0940$-$1050 & 81 & 3.48 & 2.57830 & 0.00022 & 2.652 & 0.079 \\
10 & HE0940$-$1050 & 81 & 3.48 & 3.13576 & 0.00026 & 2.440 & 0.045 \\
11 & HE0940$-$1050 & 96 & 2.33 & 1.85485 & 0.00028 & 0.923 & 0.030 \\
12 & HE0940$-$1050 & 96 & 2.33 & 2.32154 & 0.00040 & 0.510 & 0.013 \\
13 & HE0940$-$1050 & 94 & 2.56 & 1.86486 & 0.00033 & 0.344 & 0.005 \\
14 & HE0940$-$1050 & 94 & 2.56 & 1.88062 & 0.00032 & 0.658 & 0.006 \\
15 & HE0940$-$1050 & 94 & 2.56 & 1.94750 & 0.00015 & 1.322 & 0.009 \\
16 & HE0940$-$1050 & 94 & 2.56 & 2.18651 & 0.00016 & 0.932 & 0.006 \\
17 & HE0940$-$1050 & 94 & 2.56 & 2.34047 & 0.00017 & 0.841 & 0.007 \\
18 & HE0940$-$1050 & 72 & 2.68 & 2.25497 & 0.00037 & 2.033 & 0.118 \\
19 & HE0940$-$1050 & 93 & 3.15 & 2.97584 & 0.00014 & 3.904 & 0.030 \\
20 & HE0940$-$1050 & 93 & 3.15 & 3.12947 & 0.00026 & 1.035 & 0.010 \\
21 & HE0940$-$1050 & 89 & 2.83 & 2.81193 & 0.00048 & 1.717 & 0.023 \\
22 & HE0940$-$1050 & 89 & 2.83 & 2.82790 & 0.00011 & 0.968 & 0.002 \\
23 & HE0940$-$1050 & 80 & 2.45 & 2.34028 & 0.00024 & 1.389 & 0.046 \\
24 & HE0940$-$1050 & 101 & 3.05 & 2.22093 & 0.00005 & 0.992 & 0.005 \\
25 & HE0940$-$1050 & 101 & 3.05 & 2.33003 & 0.00005 & 3.062 & 0.006 \\
26 & HE0940$-$1050 & 101 & 3.05 & 2.40893 & 0.00005 & 0.358 & 0.004 \\
27 & HE0940$-$1050 & 101 & 3.05 & 2.51662 & 0.00005 & 0.083 & 0.007 \\
28 & HE0940$-$1050 & 101 & 3.05 & 2.61360 & 0.00005 & 0.057 & 0.005 \\
29 & HE0940$-$1050 & 101 & 3.05 & 2.64315 & 0.00005 & 0.218 & 0.007 \\
30 & HE0940$-$1050 & 101 & 3.05 & 2.66771 & 0.00005 & 0.641 & 0.006 \\
31 & HE0940$-$1050 & 101 & 3.05 & 2.82568 & 0.00005 & 0.679 & 0.011 \\
32 & HE0940$-$1050 & 101 & 3.05 & 2.82673 & 0.00005 & 1.629 & 0.013 \\
33 & HE0940$-$1050 & 101 & 3.05 & 2.83465 & 0.00005 & 1.621 & 0.007 \\
34 & HE0940$-$1050 & 101 & 3.05 & 2.86086 & 0.00005 & 0.085 & 0.007 \\
35 & HE0940$-$1050 & 101 & 3.05 & 2.91698 & 0.00005 & 0.346 & 0.007 \\
36 & HE0940$-$1050 & 101 & 3.05 & 2.93772 & 0.00005 & 0.133 & 0.009 \\
37 & HE0940$-$1050 & 101 & 3.05 & 3.03859 & 0.00005 & 0.091 & 0.006 \\
38 & PKS2126$-$158 & 138 & 2.43 & 2.32389 & 0.00027 & 4.056 & 0.106 \\
39 & PKS2126$-$158 & 138 & 2.43 & 2.40790 & 0.00017 & 0.543 & 0.003 \\
40 & PKS2126$-$158 & 130 & 2.58 & 2.23069 & 0.00010 & 5.385 & 0.085 \\
41 & PKS2126$-$158 & 131 & 2.90 & 2.14980 & 0.00005 & 0.092 & 0.012 \\
42 & PKS2126$-$158 & 131 & 2.90 & 2.16316 & 0.00005 & 3.748 & 0.031 \\
43 & PKS2126$-$158 & 131 & 2.90 & 2.35667 & 0.00005 & 0.474 & 0.035 \\
44 & PKS2126$-$158 & 131 & 2.90 & 2.43630 & 0.00005 & 0.250 & 0.020 \\
45 & PKS2126$-$158 & 131 & 2.90 & 2.82301 & 0.00005 & 0.376 & 0.023 \\
46 & PKS2126$-$158 & 131 & 2.90 & 2.84763 & 0.00005 & 0.653 & 0.025 \\
47 & PKS2126$-$158 & 134 & 3.27 & 2.39401 & 0.00005 & 1.662 & 0.003 \\
48 & PKS2126$-$158 & 134 & 3.27 & 2.45961 & 0.00005 & 0.317 & 0.003 \\
49 & PKS2126$-$158 & 134 & 3.27 & 2.48546 & 0.00005 & 0.086 & 0.003 \\
50 & PKS2126$-$158 & 134 & 3.27 & 2.55370 & 0.00005 & 0.162 & 0.003 \\
51 & PKS2126$-$158 & 134 & 3.27 & 2.63797 & 0.00005 & 2.277 & 0.003 \\
52 & PKS2126$-$158 & 134 & 3.27 & 2.67890 & 0.00005 & 0.695 & 0.003 \\
53 & PKS2126$-$158 & 134 & 3.27 & 2.72795 & 0.00005 & 0.371 & 0.004 \\
54 & PKS2126$-$158 & 134 & 3.27 & 2.76877 & 0.00005 & 3.366 & 0.006 \\
55 & PKS2126$-$158 & 134 & 3.27 & 2.81944 & 0.00005 & 0.590 & 0.006 \\
56 & PKS2126$-$158 & 134 & 3.27 & 2.90709 & 0.00005 & 0.543 & 0.004 \\
57 & PKS2126$-$158 & 134 & 3.27 & 2.96326 & 0.00005 & 0.314 & 0.004 \\
58 & PKS2126$-$158 & 134 & 3.27 & 2.96748 & 0.00005 & 0.355 & 0.004 \\
59 & PKS2126$-$158 & 134 & 3.27 & 3.09869 & 0.00005 & 0.021 & 0.003 \\
60 & PKS2126$-$158 & 134 & 3.27 & 3.21649 & 0.00005 & 0.299 & 0.003 \\
61 & PKS2126$-$158 & 128 & 3.86 & 3.08476 & 0.00031 & 0.476 & 0.003 \\
62 & PKS2126$-$158 & 128 & 3.86 & 3.35438 & 0.00038 & 0.480 & 0.002 \\
63 & PKS2126$-$158 & 128 & 3.86 & 3.40139 & 0.00024 & 0.480 & 0.002 \\
64 & PKS2126$-$158 & 128 & 3.86 & 3.57767 & 0.00012 & 1.370 & 0.003 \\
65 & PKS2126$-$158 & 128 & 3.86 & 3.73041 & 0.00030 & 0.606 & 0.003 \\
66 & PKS2126$-$158 & 132 & 2.54 & 1.95150 & 0.00007 & 3.289 & 0.009 \\
67 & PKS2126$-$158 & 132 & 2.54 & 2.29910 & 0.00025 & 0.996 & 0.007 \\
68 & PKS2126$-$158 & 140 & 2.40 & 1.70540 & 0.00014 & 0.657 & 0.003 \\
69 & PKS2126$-$158 & 140 & 2.40 & 1.91800 & 0.00006 & 2.013 & 0.003 \\
70 & PKS2126$-$158 & 142 & 2.74 & 2.33881 & 0.00010 & 0.972 & 0.002 \\
71 & PKS2126$-$158 & 129 & 2.14 & 1.96477 & 0.00006 & 1.668 & 0.002 \\
72 & PKS2126$-$158 & 129 & 2.14 & 1.99582 & 0.00004 & 3.478 & 0.004 \\
73 & PKS2126$-$158 & 129 & 2.14 & 2.03533 & 0.00010 & 0.868 & 0.002 \\
74 & PKS2126$-$158 & 141 & 3.46 & 2.54500 & 0.00025 & 0.964 & 0.008 \\
75 & PKS2126$-$158 & 141 & 3.46 & 3.11388 & 0.00021 & 1.853 & 0.015 \\
76 & PKS2126$-$158 & 141 & 3.46 & 3.26671 & 0.00035 & 1.499 & 0.016 \\
77 & J0124+0044 & 53 & 2.30 & 1.89338 & 0.00023 & 0.690 & 0.009 \\
78 & J0124+0044 & 26 & 2.77 & 2.50922 & 0.00020 & 2.016 & 0.039 \\
79 & J0124+0044 & 26 & 2.77 & 2.76856 & 0.00024 & 1.140 & 0.008 \\
80 & J0124+0044 & 49 & 4.07 & 3.11067 & 0.00024 & 1.628 & 0.018 \\
81 & J0124+0044 & 68 & 2.38 & 1.88617 & 0.00025 & 1.045 & 0.039 \\
82 & J0124+0044 & 39 & 2.48 & 2.17616 & 0.00033 & 2.415 & 0.079 \\
83 & J0124+0044 & 59 & 2.59 & 2.51074 & 0.00025 & 2.154 & 0.131 \\
84 & J0124+0044 & 48 & 2.94 & 2.14277 & 0.00014 & 1.075 & 0.005 \\
85 & J0124+0044 & 48 & 2.94 & 2.63615 & 0.00020 & 2.203 & 0.061 \\
86 & J0124+0044 & 52 & 2.97 & 2.14105 & 0.00013 & 1.419 & 0.009 \\
87 & J0124+0044 & 52 & 2.97 & 2.67890 & 0.00034 & 3.328 & 0.146 \\
88 & J0124+0044 & 62 & 3.00 & 2.19593 & 0.00042 & 1.516 & 0.086 \\
89 & J0124+0044 & 21 & 2.50 & 2.45733 & 0.00027 & 1.784 & 0.109 \\
90 & J0124+0044 & 21 & 2.50 & 2.49697 & 0.00005 & 1.912 & 0.003 \\
91 & J0124+0044 & 47 & 2.54 & 2.15139 & 0.00010 & 1.189 & 0.004 \\
92 & J0124+0044 & 37 & 3.12 & 2.88094 & 0.00027 & 1.715 & 0.039 \\
93 & J0124+0044 & 24 & 3.42 & 3.34344 & 0.00051 & 0.911 & 0.028 \\
94 & J0124+0044 & 43 & 2.47 & 1.80487 & 0.00022 & 1.270 & 0.020 \\
95 & J0124+0044 & 64 & 2.64 & 2.54622 & 0.00040 & 2.785 & 0.315 \\
96 & J0124+0044 & 46 & 3.11 & 2.23941 & 0.00020 & 2.293 & 0.068 \\
97 & J0124+0044 & 60 & 3.81 & 2.83356 & 0.00005 & 0.877 & 0.013 \\
98 & J0124+0044 & 60 & 3.81 & 2.86629 & 0.00005 & 0.082 & 0.006 \\
99 & J0124+0044 & 60 & 3.81 & 2.91041 & 0.00005 & 0.701 & 0.014 \\
100 & J0124+0044 & 60 & 3.81 & 2.94202 & 0.00005 & 0.111 & 0.009 \\
101 & J0124+0044 & 60 & 3.81 & 2.98670 & 0.00005 & 1.943 & 0.019 \\
102 & J0124+0044 & 60 & 3.81 & 3.06528 & 0.00005 & 1.500 & 0.019 \\
103 & J0124+0044 & 60 & 3.81 & 3.14814 & 0.00005 & 0.115 & 0.019 \\
104 & J0124+0044 & 60 & 3.81 & 3.18796 & 0.00005 & 0.135 & 0.018 \\
105 & J0124+0044 & 60 & 3.81 & 3.39219 & 0.00005 & 1.186 & 0.023 \\
106 & J0124+0044 & 60 & 3.81 & 3.54813 & 0.00007 & 1.794 & 0.002 \\
107 & J0124+0044 & 60 & 3.81 & 3.67356 & 0.00004 & 2.420 & 0.001 \\
108 & J0124+0044 & 60 & 3.81 & 3.76553 & 0.00011 & 0.661 & 0.000 \\
109 & J0124+0044 & 40 & 2.53 & 2.39121 & 0.00021 & 2.372 & 0.058 \\
110 & J0124+0044 & 40 & 2.53 & 2.43175 & 0.00022 & 2.345 & 0.058 \\
111 & J0124+0044 & 67 & 2.38 & 2.26440 & 0.00031 & 1.211 & 0.041 \\
112 & J0124+0044 & 66 & 2.51 & 2.28225 & 0.00024 & 2.219 & 0.062 \\
113 & J0124+0044 & 66 & 2.51 & 2.47245 & 0.00022 & 1.335 & 0.017 \\
114 & J0124+0044 & 63 & 2.98 & 2.31409 & 0.00028 & 1.085 & 0.042 \\
115 & J0124+0044 & 63 & 2.98 & 2.37429 & 0.00020 & 1.591 & 0.042 \\
116 & J0124+0044 & 63 & 2.98 & 2.47027 & 0.00028 & 1.155 & 0.039 \\
117 & J0124+0044 & 50 & 2.36 & 2.32717 & 0.00012 & 1.766 & 0.012 \\
118 & J1201+0116 & 122 & 3.36 & 2.43546 & 0.00006 & 3.754 & 0.007 \\
119 & J1201+0116 & 122 & 3.36 & 2.46989 & 0.00011 & 2.434 & 0.011 \\
120 & J1201+0116 & 122 & 3.36 & 3.26698 & 0.00015 & 2.168 & 0.016 \\
121 & J1201+0116 & 112 & 3.84 & 3.79466 & 0.00016 & 5.871 & 0.067 \\
122 & J1201+0116 & 104 & 2.38 & 1.65636 & 0.00008 & 2.675 & 0.017 \\
123 & J1201+0116 & 104 & 2.38 & 2.04113 & 0.00027 & 1.126 & 0.056 \\
124 & J1201+0116 & 108 & 3.52 & 3.42823 & 0.00039 & 8.836 & 0.666 \\
125 & J1201+0116 & 121 & 2.23 & 1.77878 & 0.00026 & 2.348 & 0.197 \\
126 & J1201+0116 & 110 & 3.20 & 2.37602 & 0.00005 & 0.101 & 0.006 \\
127 & J1201+0116 & 110 & 3.20 & 2.47715 & 0.00005 & 0.449 & 0.009 \\
128 & J1201+0116 & 110 & 3.20 & 2.68444 & 0.00005 & 1.451 & 0.022 \\
129 & J1201+0116 & 110 & 3.20 & 2.78999 & 0.00005 & 0.396 & 0.023 \\
130 & J1201+0116 & 110 & 3.20 & 3.07951 & 0.00010 & 0.734 & 0.032 \\
131 & Q0042$-$2627 & 5 & 3.29 & 2.47481 & 0.00005 & 0.998 & 0.031 \\
132 & Q0042$-$2627 & 5 & 3.29 & 2.50670 & 0.00005 & 0.281 & 0.052 \\
133 & Q0042$-$2627 & 5 & 3.29 & 2.72816 & 0.00005 & 0.421 & 0.037 \\
134 & Q0042$-$2627 & 5 & 3.29 & 2.77891 & 0.00005 & 0.108 & 0.035 \\
135 & Q0042$-$2627 & 5 & 3.29 & 2.82754 & 0.00005 & 0.288 & 0.037 \\
136 & Q0042$-$2627 & 5 & 3.29 & 3.10189 & 0.00005 & 0.538 & 0.039 \\
137 & Q0042$-$2627 & 5 & 3.29 & 3.14458 & 0.00005 & 0.389 & 0.071 \\
138 & Q0042$-$2627 & 5 & 3.29 & 3.21252 & 0.00005 & 0.272 & 0.065 \\
139 & Q0042$-$2627 & 5 & 3.29 & 3.23594 & 0.00010 & 0.843 & 0.003 \\
140 & Q0042$-$2627 & 7 & 2.79 & 2.33870 & 0.00010 & 0.620 & 0.130 \\
141 & Q0042$-$2627 & 7 & 2.79 & 2.34175 & 0.00007 & 6.356 & 0.050 \\
142 & Q0042$-$2627 & 7 & 2.79 & 2.59946 & 0.00009 & 1.145 & 0.007 \\
143 & Q0042$-$2627 & 12 & 2.90 & 2.49148 & 0.00017 & 1.475 & 0.019 \\
144 & Q0042$-$2627 & 12 & 2.90 & 2.82448 & 0.00009 & 1.256 & 0.012 \\
145 & Q0042$-$2627 & 14 & 3.05 & 2.26461 & 0.00009 & 1.037 & 0.008 \\
146 & Q0042$-$2627 & 14 & 3.05 & 2.33871 & 0.00008 & 0.619 & 0.004 \\
147 & Q0042$-$2627 & 14 & 3.05 & 2.56880 & 0.00010 & 0.800 & 0.140 \\
148 & Q0042$-$2627 & 14 & 3.05 & 2.73960 & 0.00010 & 1.410 & 0.250 \\
149 & Q0042$-$2627 & 14 & 3.05 & 2.75630 & 0.00010 & 0.730 & 0.180 \\
150 & Q0042$-$2627 & 8 & 2.48 & 2.14446 & 0.00012 & 1.905 & 0.035 \\
151 & Q0042$-$2627 & 9 & 2.46 & 1.75183 & 0.00006 & 0.734 & 0.002 \\
152 & Q0042$-$2627 & 9 & 2.46 & 1.86964 & 0.00006 & 2.220 & 0.027 \\
153 & Q0042$-$2627 & 9 & 2.46 & 2.02084 & 0.00007 & 0.818 & 0.006 \\
154 & Q0042$-$2627 & 9 & 2.46 & 2.15417 & 0.00023 & 0.514 & 0.009 \\
155 & Q0042$-$2627 & 9 & 2.46 & 2.27132 & 0.00005 & 1.710 & 0.019 \\
156 & Q0042$-$2627 & 9 & 2.46 & 2.43659 & 0.00003 & 1.127 & 0.003 \\
157 & Q0042$-$2627 & 9 & 2.46 & 2.43907 & 0.00004 & 0.946 & 0.003 \\
158 & Q0042$-$2627 & 1 & 3.33 & 2.49787 & 0.00011 & 0.561 & 0.006 \\
159 & Q0042$-$2627 & 1 & 3.33 & 2.90194 & 0.00006 & 1.212 & 0.004 \\
160 & Q0042$-$2627 & 1 & 3.33 & 3.24812 & 0.00011 & 0.691 & 0.012 \\
161 & Q0042$-$2627 & 1 & 3.33 & 3.25543 & 0.00012 & 0.755 & 0.005 \\
162 & Q0042$-$2627 & 16 & 3.44 & 2.71935 & 0.00005 & 0.202 & 0.015 \\
163 & Q0042$-$2627 & 16 & 3.44 & 2.81853 & 0.00005 & 2.061 & 0.038 \\
164 & Q0042$-$2627 & 16 & 3.44 & 3.04437 & 0.00005 & 3.086 & 0.024 \\
165 & Q0042$-$2627 & 16 & 3.44 & 3.12003 & 0.00005 & 0.882 & 0.014 \\
166 & Q0042$-$2627 & 16 & 3.44 & 3.12944 & 0.00005 & 0.402 & 0.017 \\
167 & Q0042$-$2627 & 16 & 3.44 & 3.15350 & 0.00005 & 0.180 & 0.020 \\
168 & Q0042$-$2627 & 16 & 3.44 & 3.25486 & 0.00005 & 1.255 & 0.016 \\
169 & Q0042$-$2627 & 16 & 3.44 & 3.37157 & 0.00005 & 0.453 & 0.016 \\
170 & Q0042$-$2627 & 6 & 2.50 & 1.88975 & 0.00011 & 1.213 & 0.012 \\
171 & Q0042$-$2627 & 6 & 2.50 & 1.95550 & 0.00010 & 0.710 & 0.080 \\
172 & Q0042$-$2627 & 6 & 2.50 & 2.40002 & 0.00003 & 0.781 & 0.001 \\
173 & Q0042$-$2627 & 6 & 2.50 & 2.50560 & 0.00010 & 0.450 & 0.060 \\
174 & Q0042$-$2627 & 4 & 3.31 & 2.29549 & 0.00009 & 1.082 & 0.010 \\
175 & Q0042$-$2627 & 2 & 2.98 & 2.12850 & 0.00010 & 1.790 & 0.180 \\
176 & Q0042$-$2627 & 2 & 2.98 & 2.22710 & 0.00010 & 2.770 & 0.420 \\

\end{longtable}
\end{center}

\end{document}